\documentclass[opre,nonblindrev]{informs3}

\OneAndAHalfSpacedXI % current default line spacing
\usepackage{endnotes}
\let\footnote=\endnote
\let\enotesize=\normalsize
\def\notesname{Endnotes}%
\def\makeenmark{\hbox to1.275em{\theenmark.\enskip\hss}}
\def\enoteformat{\rightskip0pt\leftskip0pt\parindent=1.275em
  \leavevmode\llap{\makeenmark}}

\usepackage[english]{babel}
\usepackage[utf8]{inputenc}  
\usepackage[babel=true]{csquotes}
\usepackage{wrapfig}
\usepackage{float}
\usepackage{mathtools}
\usepackage{subfig}
\usepackage{lmodern}
\usepackage{amsfonts}
\usepackage{amsmath}
\usepackage{enumerate}
\usepackage{amssymb}

\newcommand{\preal}{ p^{\mathrm{real}} }
\newcommand{\deltainf}{\delta^{\mathrm{inf}}}
\newcommand{\deltasup}{\delta^{\mathrm{sup}}}
\DeclareMathOperator\supp{supp}
\DeclarePairedDelimiter\floor{\lfloor}{\rfloor}
\DeclarePairedDelimiter\ceil{\lceil}{\rceil}

\usepackage{natbib}
 \bibpunct[, ]{(}{)}{,}{a}{}{,}%
 \def\bibfont{\small}%
 \def\bibsep{\smallskipamount}%
 \def\bibhang{24pt}%
\TheoremsNumberedThrough     % Preferred (Theorem 1, Lemma 1, Theorem 2)
\ECRepeatTheorems

\EquationsNumberedThrough    % Default: (1), (2), ...
\begin{document}
%%%%%%%%%%%%%%%%

%\setcounter{page}{1}
%\DOI{10.1287/xxxx.0000.0000}%

\RUNAUTHOR{Flajolet, Blandin and Jaillet}

\RUNTITLE{Robust Adaptive Routing Under Uncertainty}

% \TITLE{Bundling Information Goods of Decreasing Value}
\TITLE{Robust Adaptive Routing Under Uncertainty}

\ARTICLEAUTHORS{%
\AUTHOR{Arthur Flajolet}
\AFF{Operations Research Center, Massachusetts Institute of Technology, Cambridge, MA 02139, flajolet@mit.edu \\
	 Ecole Polytechnique, Route de Saclay, 91120 Palaiseau, France}
\AUTHOR{S\'{e}bastien Blandin}
\AFF{IBM Research Collaboratory, 9 Changi Business Park Central 1, Singapore 486048, Singapore, sblandin@sg.ibm.com}
\AUTHOR{Patrick Jaillet}
\AFF{Department of Electrical Engineering and Computer Science, Operations Research Center, Massachusetts Institute of
Technology, Cambridge, Massachusetts 02139, jaillet@mit.edu}

} % end of the block

\ABSTRACT{%
We consider the problem of finding an optimal history-dependent routing strategy on a directed graph weighted by stochastic arc costs when the objective is to minimize the risk of spending more than a prescribed budget. To help mitigate the impact of the lack of information on the arc cost probability distributions, we introduce a robust counterpart where the distributions are only known through confidence intervals on some statistics such as the mean, the mean absolute deviation, and any quantile. Leveraging recent results in distributionally robust optimization, we develop a general-purpose algorithm to compute an approximate optimal strategy. To illustrate the benefits of the robust approach, we run numerical experiments with field data from the Singapore road network.
}%

%\KEYWORDS{deterministic inventory theory; infinite linear programming duality;
%  existence of optimal policies; semi-Markov decision process; cyclic schedule}

 %\HISTORY{ .}

\maketitle
%%%%%%%%%%%%%%%%%%%%%%%%%%%%%%%%%%%%%%%%%%%%%%%%%%%%%%%%%%%%%%%%%%%%%%

% Samples of sectioning (and labeling) in OPRE
% NOTE: (1) \section and \subsection do NOT end with a period
%       (2) \subsubsection and lower need end punctuation
%       (3) capitalization is as shown (title style).
%
%\section{Introduction.}\label{intro} %%1.
%\subsection{Duality and the Classical EOQ Problem.}\label{class-EOQ} %% 1.1.
%\subsection{Outline.}\label{outline1} %% 1.2.
%\subsubsection{Cyclic Schedules for the General Deterministic SMDP.}
%  \label{cyclic-schedules} %% 1.2.1
%\section{Problem Description.}\label{problemdescription} %% 2.

% Text of your paper here

\section{Introduction}

\subsection{Motivation}
Stochastic Shortest Path (SSP) problems have emerged as natural extensions to the classical shortest path problem when arc costs are uncertain and modeled as outcomes of random variables. In particular, we consider in this paper the class of adaptive SSPs, which can be formulated as \emph{Markov Decision Processes} (MDPs), where we optimize over all history-dependent strategies. As standard with MDPs, optimal policies are characterized by dynamic programming equations involving expected values (e.g. \cite{bertsekas1991analysis}). Yet, computing the expected value of a function of a random variable generally requires a full description of its probability distribution, and this can be hard to obtain accurately due to errors and sparsity of measurements. In practice, only finite samples are available and an optimal strategy based on approximated arc cost probability distributions may be suboptimal with respect to the real arc cost probability distributions.
\\  
\indent One of the most common applications of SSPs deals with the problem of routing vehicles in transportation networks. Providing driving itineraries is a challenging task as suppliers have to cope simultaneously with limited knowledge about random fluctuations in traffic congestion (e.g. caused by traffic incidents, variability of travel demand) and users’ desire to arrive on time. These considerations have led to the definition of the Stochastic On-Time Arrival (SOTA) problem, an adaptive SSP problem with the objective of maximizing the probability of on-time arrival, and formulated using dynamic programming in \cite{nie2006arriving}. The algorithm proposed in \cite{samitha10} to solve this problem assumes the knowledge of the complete arc travel-time distributions. Yet, in practice, such distributions tend to be estimated from samples which are sparse and error-prone.
\\
\indent In recent years, \emph{Distributionally Robust Optimization} (DRO) has emerged as a new framework for decision-making under uncertainty when the underlying distributions are only known through some statistics or from collections of samples. DRO was put forth in an effort to capture both risk (uncertainty on the outcomes) and ambiguity (uncertainty on the probabilities of the outcomes) when optimizing over a set of alternatives, thus lying at the crossroad between stochastic and robust optimization. The computational complexity of this approach can vary greatly, depending on the nature of the ambiguity sets and on the structure of the optimization problem, see \cite{wiesemann2014distributionally} and \cite{delage2010distributionally} for convex problems, and \cite{calafiore2006distributionally} for chance-constraint problems. Even in the absence of decision variables, the theory proves useful in order to derive either numerical or closed form bounds on expected values using tools drawn from linear programming, e.g. tailored dual simplex algorithm in \cite{prekopa1990discrete}, and from semidefinite programming as in \cite{bertsimas2005optimal} and \cite{vandenberghe2007generalized}.
\\
\indent In the case of limited knowledge of the arc cost probability distributions, we propose to bring DRO to bear on adaptive SSP problems to help mitigate the impact of the lack of information and introduce Distributionally Robust Adaptive Stochastic Shortest Path problems. Our work fits into the literature on distributionally robust MDPs where the transition probabilities are only known to lie in prescribed ambiguity sets (e.g. \cite{RobustMDPElGahoui}, \cite{iyengar2005robust}, \cite{xu2010distributionally}, and \cite{wiesemann2013robust}). While some of the methods developed in the aforementioned literature can be shown to carry over, adaptive SSPs exhibit a particular structure that allows for a large variety of ambiguity sets and enables the development of faster solution procedures. Specifically, optimal strategies for finite-horizon distributionally robust MDPs are characterized by a Bellman recursion on the worst-case expected reward-to-go. While standard approaches focus on computing this last quantity for each state independently from one another, closely related problems (e.g. estimating an expected value $\mathbb{E}[f(t - X)]$ where the random variable $X$ is fixed but $t$ varies depending on the state) carry across states for adaptive SSPs, and, as a result, making the most of previous computations becomes crucial to achieve computational tractability. 

\subsection{Related Work}

Extending the shortest path problem by assigning random, as opposed to deterministic, costs to arcs requires some additional modeling assumptions. Over the years, many formulations have been proposed which differ along three main features:
\begin{itemize}
	\item{The specific objective function to optimize: in the presence of uncertainty, the most natural approach is to minimize the total expected costs, see \cite{bertsekas1991analysis}, and \cite{miller2000least} for time-dependent random costs. However, this approach is oblivious to risk. In an attempt to take that factor into account, \cite{prescott1983optimal} proposed earlier to rely on utility functions of moments (e.g. mean costs and variances) involving an inherent trade-off, and considered multi-objective criteria. However, Bellman’s principle of optimality no longer holds for arcs weighted by multidimensional costs, giving rise to computational hardness. A different approach consists of introducing a budget, set by the user, corresponding to the maximum total cost he is willing to pay to reach his terminal node. Such approaches have been considered along several different directions. Research efforts have considered either minimizing the probability of budget overrun (see \cite{frank1969shortest}, \cite{nikolova2006stochastic}, and also \cite{xu2011probabilistic} for probabilistic goal MDPs), minimizing more general functions of the budget overrun as in \cite{nikolova2006optimal}, minimizing refined satisficing measures in order to guarantee good performances with respect to several other objectives as in \cite{jailletroutingopti2013}, and constraining the probability of over-spending while optimizing the expected costs as in \cite{xu2012optimization}.} 
	\item{The admissible set of strategies over which we are free to optimize: incorporating uncertainty may cause history-dependent strategies to significantly outperform a priori paths depending on the chosen performance index. This is the case for the SOTA problem where two types of formulations have been considered: (i) an a-priori formulation which consists in finding a path before taking any actions, see \cite{nikolova2006stochastic} and \cite{nie2009shortest}; and (ii) an adaptive formulation which allows to update
the path to go based on the remaining budget, see \cite{nie2006arriving}, \cite{samitha10}, and \cite{parmentier2014stochastic}.}
	\item{The knowledge on the random arc costs taken as an input: it can range from the full knowledge of the probability distributions to having access to only a few samples drawn from them. In practical settings, the problem of estimating accurately some statistics (e.g. mean cost and variance) seems more reasonable than retrieving the full probability distribution. For instance, \cite{jailletroutingopti2013} consider lower-order statistics (minimum, average and maximum costs) and make use of closed form bounds derived in the DRO theory. These considerations were extensively investigated in the context of distributionally robust MDPs. From a theoretical standpoint, \cite{wiesemann2013robust} show that a property coined as rectangularity has to be satisfied by the ambiguity sets for computational tractability, while \cite{iyengar2005robust} characterizes the optimal policies with a dynamic programming equation for general rectangular ambiguity sets. The ambiguity sets are parametric in \cite{wiesemann2013robust}, where the parameter lies in the intersection of finitely many ellipsoids, are based on likelihood measures in \cite{RobustMDPElGahoui}, and are defined by linear inequalities in \cite{white1994markov}.}

\end{itemize}
We give an overview of prior formulations in Table \ref{table-litteraturereview}.

{\def\arraystretch{1.3}
\begin{table}[H]
   \TABLE
   {Literature review. \label{table-litteraturereview}}
   {\begin{tabular}{|c|c|c|c|c|}
  \hline
   Author(s) & Objective function & Strategy & Uncertainty description & Approach  \\
   \hline
  \cite{prescott1983optimal} & utility function  & a priori & moments & \begin{tabular}{c} dominated \\ paths \end{tabular} \\
  \hline
  \cite{nikolova2006stochastic} & \begin{tabular}{c} probability of \\ budget overrun \end{tabular} & a priori & normal distributions & \begin{tabular}{c} convex \\ optimization \end{tabular} \\
  \hline
  \begin{tabular}{c} \cite{nie2006arriving} \\ \cite{samitha10} \end{tabular} & \begin{tabular}{c} probability of \\ budget overrun \end{tabular} & adaptive &  distributions & \begin{tabular}{c} dynamic \\ programming \end{tabular} \\
  \hline
  \cite{RobustMDPElGahoui} & expected cost & adaptive & \begin{tabular}{c} maximum-likelihood \\ ambiguity sets \end{tabular} & \begin{tabular}{c} dynamic \\ programming \end{tabular} \\
  \hline
  \begin{tabular}{c} \cite{jailletroutingopti2013} \\ \cite{jailletroutingext2014} \end{tabular} & \begin{tabular}{c} requirements \\ violation \end{tabular} & a priori & \begin{tabular}{c} distributions or \\ moments \end{tabular} & \begin{tabular}{c} iterative \\ procedure \end{tabular} \\
  \hline
  \cite{gabrel2011new} & worst-case cost & a priori & \begin{tabular}{c} intervals or \\ discrete scenarios \end{tabular} & \begin{tabular}{c} integer \\ programming \end{tabular} \\
  \hline
  \cite{parmentier2014stochastic} & \begin{tabular}{c} monotone \\ risk measure \end{tabular} & a priori & distributions & \begin{tabular}{c} labeling \\ algorithm \end{tabular} \\
  \hline
  Our work & \begin{tabular}{c} risk function \\ of the budget \\ overrun \end{tabular} & adaptive &  \begin{tabular}{c} distributions or \\ confidence intervals \\ on statistics \end{tabular} & \begin{tabular}{c} dynamic \\ programming \end{tabular} \\
  \hline
   \end{tabular}}
   {}
\end{table}
}
   
\subsection{Contributions}
The main contributions of this paper can be summarized as follows:
\begin{enumerate}
	\item{We extend the class of adaptive SSP problems, first introduced in \cite{fan2005arriving} when the objective is to minimize the probability of budget overrun, to general risk functions of the budget overrun. We characterize optimal strategies and identify conditions on the risk function under which infinite cycling is provably suboptimal. For any risk function satisfying these conditions, we provide an efficient solution procedure to compute an $\epsilon$-approximate optimal strategy for any $\epsilon > 0$.}
	\item{We introduce the distributionally robust version of this general problem, under rectangular ambiguity sets. We characterize optimal robust strategies and extend the conditions ruling out infinite cycling. For any risk function satisfying these conditions, we provide efficient solution procedures to compute an $\epsilon$-approximate optimal strategy when the arc cost distributions are only known through confidence intervals on piecewise affine statistics (e.g. the mean, the mean absolute deviation, any quantile...) for any $\epsilon > 0$. }
\end{enumerate}
Special cases where the objective is to minimize the probability of budget overrun and the arc costs are independent and take on values that are multiple of a unit cost can serve as a basis for comparison with prior work on distributionally robust MDPs. For this subclass of problems, our formulation can be interpreted as a distributionally robust MDP with finite horizon $N$, finitely many states $n$ (resp. actions $m$), and a rectangular ambiguity set. Using the solution methodology developed in this paper, we can compute an $\epsilon$-optimal strategy with complexity $O(m \cdot n \cdot \log(\frac{N}{\epsilon}) \cdot \log(n) )$.
\\
\indent The remainder of the paper is organized as follows. In Section \ref{section-problemformulation}, we introduce the adaptive SSP problem and its distributionally robust counterpart. Section \ref{section-analysis-nominal-problem} (resp. Section \ref{section-analysis-robust-problem}) is devoted to the theoretical and computational analysis of the nominal (resp. robust) problem. In Section \ref{section-numericalresults}, we consider a vehicle routing application and present results of numerical experiments run with field data from the Singapore road network. In Section \ref{sec-extensions}, we relax some of the assumptions made in Section \ref{section-problemformulation} and extend the results presented in Sections \ref{section-analysis-nominal-problem} and \ref{section-analysis-robust-problem}.

\paragraph{Notations}
For a function $g(\cdot)$ and a random variable $X$ distributed according to $p$, we denote by $\mathbb{E}_{X \sim p}[g(X)]$ the expected value of $g(X)$. For a set $S \subset \mathbb{R}^n$, $\bar{S}$ is the closure of $S$ in the standard topology of $\mathbb{R}^n$, $\mathrm{conv}(S)$ denotes the convex hull generated by $S$ and $|S|$ denotes the cardinality of $S$. For a set $S \subset \mathbb{R}^2$, $\hat{S}$ denotes the upper convex hull of $S$, i.e. $\hat{S} = \{ (x, y) \in \mathbb{R}^2 \; : \; \exists (a, b) \in \mathrm{conv}(S)$ such that $x = a$ and $y \geq b \}$.

\section{Problem Formulation}
\label{section-problemformulation}
In this section, we formulate the adaptive SSP problem for any risk function of the budget overrun. Then, we introduce the distributionally robust approach based on ambiguity sets to tackle the situation of limited knowledge of the arc cost probability distributions.

\subsection{Nominal problem}
\label{section-nominal-problem}
Let $\mathcal{G} = (\mathcal{V}, \mathcal{A})$ be a finite directed graph where each arc $(i,j) \in \mathcal{A}$ is assigned a collection of non-negative random costs $(c^\tau_{ij})_{\tau \geq 0}$. We consider a user traveling through $\mathcal{G}$ leaving from $s$ and wishing to reach $d$ within a total prescribed budget $T$. Having already spent a total cost $\tau$ and being at node $i$, choosing to cross arc $(i, j)$ would incur an additional cost $c^\tau_{ij}$, whose value becomes known after the arc is crossed. In vehicle routing applications, $c^\tau_{ij}$ typically models the travel time along arc $(i, j)$ at time $\tau$ and $T$ is the deadline imposed at the destination. The objective is to find a strategy to reach $d$ maximizing a risk function of the budget overrun, denoted by $f(\cdot)$. Mathematically, this corresponds to solving:
\begin{equation}
	\label{eq-nominal-problem}
	\sup\limits_{\pi \in \Pi} \mathbb{E}[f( T - X_\pi )],
\end{equation}
where $\Pi$ is the set of all history-dependent randomized strategies, i.e. mappings from the past realizations of the costs and the previously visited nodes to probability distributions over the set of neighboring nodes, and $X_\pi$ is the random cost associated with strategy $\pi$ when leaving from node $s$ with budget $T$. We denote by $\mathcal{H}$ the set of all possible histories of the previously experienced costs and previously visited nodes. Examples of natural risk functions include $f(t) = t \cdot 1_{t \leq 0}$, $f(t) = 1_{t \geq 0}$, and $f(t) = - |t|$ which translate into, respectively, minimizing the expected budget overrun, maximizing the probability of completion within budget, and penalizing the expected deviation from the target budget. When $f(t) = 1_{t \geq 0}$, we recover the adaptive SOTA problem introduced in \cite{fan2005arriving}. We will restrict our attention to risk functions satisfying natural properties meant to prevent infinite cycling in Theorem \ref{lemma-LET-is-opt-nominal} of Section \ref{section-characterization-policy-nominal}, e.g. maximizing the expected budget overrun is not allowed. Without any additional assumption on the random costs, \eqref{eq-nominal-problem} is computationally intractable and characterizing an optimal solution is theoretically hard. To simplify the problem, a common approach in the literature is to assume independence of the arc costs, see for example \cite{fan2005arriving} and \cite{jailletroutingopti2013}. 
\begin{assumption}
	\label{assumption-independence}
	$(c^\tau_{ij})_{(i, j) \in \mathcal{A}, \tau \geq 0}$ are independent random variables.
\end{assumption}
In practice, the costs of neighboring arcs can be highly correlated for some applications and Assumption \ref{assumption-independence} may then appear unreasonable. It turns out that most of the results derived in this paper can be extended to the case where the dependence can be modeled by \emph{Markov} chains of finite order, i.e., where the cost of an arc depends on the past $m \in \mathbb{N}$ experienced costs. This is of course at the price of more technicalities and an increased complexity both in terms of modeling and of computational requirements. To simplify the presentation, Assumption \ref{assumption-independence} is used throughout most of the paper and the extension to \emph{Markov} chains is discussed in Section \ref{sec-extensions-independence}. For the same reason, we further assume that the random costs are identically distributed across $\tau$. 
\begin{assumption}
	\label{assumption-time-dependence}
	For each arc $(i, j) \in \mathcal{A}$, the distribution of $c^\tau_{ij}$ does not depend on $\tau$.
\end{assumption}
The extension to $\tau$-dependent arc cost distributions is detailed in Section \ref{sec-extensions-time-dependent-cost}. For clarity of the exposition, we omit the superscript $\tau$ in the notations when it is unnecessary and simply denote the costs by $(c_{ij})_{(i, j) \in \mathcal{A}}$, even though the cost of an arc corresponds to an independent realization of its corresponding random variable each time it is crossed. We denote the probability distribution of $c_{ij}$ by $p_{ij}$. Throughout the paper, we also assume that the arc cost distributions have compact supports. This is a perfectly reasonable assumption in many practical settings, such as in transportation networks.
\begin{assumption}
	\label{assumption-compactsupport}
	$\forall (i,j) \in \mathcal{A}$, $p_\mathrm{ij}$ has compact support included in $[\deltainf_{ij}, \deltasup_{ij}]$ with $\deltainf_{ij}>0$ and $\deltasup_{ij} < \infty$. Thus $\deltainf = \min\limits_{(i,j) \in \mathcal{A}} \deltainf_{ij} > 0$ and $\deltasup = \max\limits_{(i,j) \in \mathcal{A}} \deltasup_{ij} < \infty$.
\end{assumption}
Assumption \ref{assumption-compactsupport} is motivated by computational considerations, see Section \ref{section-algorithm-nominal}, but is also substantially needed when proving theoretical properties satisfied by optimal solutions to \eqref{eq-nominal-problem}, and when analyzing the complexity of the proposed algorithms.

\subsection{Distributionally robust problem}
\label{section-distributionally-robust-problem}
One of the major limitations of the approach described in Section \ref{section-nominal-problem} is that it requires a full description of the uncertainty. Under Assumptions \ref{assumption-independence} and \ref{assumption-time-dependence}, this is equivalent to having access to the exact arc cost probability distributions. Yet, in practice, we often only have access to a limited number of realizations of the random variables $c_{ij}$. In these circumstances, it is tempting to estimate empirical arc cost distributions and to take them as input to problem \eqref{eq-nominal-problem}. However, estimating accurately a distribution with samples drawn from realizations usually requires a very large sample size, and our experimental evidence suggests that, as a result, the corresponding solutions may perform poorly when only few samples are available, as we will see in Section \ref{section-numericalresults}. To mitigate the impact of the lack of information on the arc cost distributions, we adopt a distributionally robust point of view where, for each arc $(i, j) \in \mathcal{A}$, we assume that $p_{ij}$ is only known to lie in an ambiguity set $\mathcal{P}_{ij}$. We make the following assumption on these ambiguity sets throughout the paper.

\begin{assumption}
	\label{assumption-compact-ambiguity-set}
	$\forall (i,j) \in \mathcal{A}$, $\mathcal{P}_{ij}$ is not empty, closed for the weak topology, and a subset of $\mathcal{P}([\deltainf_{ij}, \deltasup_{ij}])$, the set of probability measures on $[\deltainf_{ij}, \deltasup_{ij}]$.
\end{assumption}
The last part of Assumption \ref{assumption-compact-ambiguity-set} is a natural extension of Assumption \ref{assumption-compactsupport}, and is essential for computational tractability, see Section \ref{section-analysis-robust-problem}. The robust counterpart of \eqref{eq-nominal-problem} for an ambiguity-averse user is then given by:
\begin{equation}
	\label{eq-ideal-robust-problem}
	\sup\limits_{\pi \in \Pi} \; \inf\limits_{\forall (i, j) \in \mathcal{A}, \; p_{ij} \in \mathcal{P}_{ij}} \; \mathbb{E}_{ \mathbf{p} }[f(T - X_\pi)],
\end{equation}
where the notation $\mathbf{p}$ refers to the fact that the costs $(c_{ij})_{(i, j) \in \mathcal{A}}$ are independent and distributed according to $(p_{ij})_{(i, j) \in \mathcal{A}}$. 
\\
\indent As a byproduct of the results obtained for the nominal problem in Section \ref{section-characterization-policy-nominal}, \eqref{eq-ideal-robust-problem} can be equivalently viewed as a distributionally robust MDP in the extended space state $(i, \tau) \in \mathcal{V} \times \mathbb{R}_+$ where $i$ is the current location and $\tau$ is the total cost spent so far and where the transition probabilities from any state $(i, \tau)$ to any state $(j, \tau')$, for $j \in \mathcal{V}(i)$ and $\tau' \geq \tau$, are only known to jointly lie in a global ambiguity set. As shown in \cite{wiesemann2013robust}, the tractability of a distributionally robust MDP hinges on the decomposability of the global ambiguity set as a Cartesian product over the space state of individual ambiguity sets, a property coined as \emph{rectangularity}. While the global ambiguity set of \eqref{eq-ideal-robust-problem} is rectangular with respect to our original state space $\mathcal{V}$, it is not with respect to the extended space space $\mathcal{V} \times \mathbb{R}_+$. Thus, we are led to enlarge our ambiguity set to make it rectangular and consider a robust relaxation of \eqref{eq-ideal-robust-problem}. This boils down to allowing the arc cost distributions to vary in their respective ambiguity sets as a function of the total cost spent so far. This approach leads to the following choice for our robust formulation associated with an ambiguity-averse user:

\begin{equation}
	\label{eq-robust-problem}
	\sup\limits_{\pi \in \Pi} \; \inf\limits_{\forall \tau, \forall (i, j) \in \mathcal{A}, \; p^\tau_{ij} \in \mathcal{P}_{ij}} \; \mathbb{E}_{ \mathbf{p^\tau} }[f(T - X_\pi)],
\end{equation}
where the notation $\mathbf{p^\tau}$ refers to the fact that, for any arc $(i, j) \in \mathcal{A}$, the costs $(c^\tau_{ij})_{\tau \geq 0}$ are independent and distributed according to $(p^\tau_{ij})_{\tau \geq 0}$. Note that when Assumption \ref{assumption-time-dependence} is relaxed, we have a different ambiguity set for each pair $((i, j), \tau) \in \mathcal{A} \times \mathbb{R}_+$, which is denoted in this case by $\mathcal{P}^\tau_{ij}$, and \eqref{eq-robust-problem} is precisely the robust counterpart of \eqref{eq-nominal-problem} as opposed to a robust relaxation, see Section \ref{sec-extensions-time-dependent-cost}. Also observe that \eqref{eq-robust-problem} reduces to \eqref{eq-nominal-problem} when the ambiguity sets are singleton, i.e. $\mathcal{P}_{ij} = \{ p_{ij} \}$. In the sequel, we focus on \eqref{eq-robust-problem} and refer to this optimization problem as the robust problem. But we will also investigate the performance of an optimal solution to \eqref{eq-robust-problem} with respect to the optimization problem \eqref{eq-ideal-robust-problem}, both from a theoretical standpoint in Section \ref{section-ambiguity-set}, and from a practical standpoint in Section \ref{section-numericalresults}. Finally note that we consider general ambiguity sets satisfying Assumption \ref{assumption-compact-ambiguity-set} when we study the theoretical properties of \eqref{eq-robust-problem}. However, for tractability purposes, the solution procedure that we develop in Section \ref{section-algorithm-robust} only applies to ambiguity sets defined by confidence intervals on piecewise affine statistics, such as the mean, the absolute mean deviation, or any quantile. We refer to Section \ref{section-ambiguity-set} for a discussion on the modeling power of these ambiguity sets and on how to build them with samples. Similarly as for the nominal problem, we will also restrict our attention to risk functions satisfying natural properties meant to prevent infinite cycling in Theorem \ref{lemma-LET-is-opt-robust} of Section \ref{section-characterization-policy-robust}.

\section{Theoretical and computational analysis of the nominal problem}
\label{section-analysis-nominal-problem}

\subsection{Characterization of optimal policies}
\label{section-characterization-policy-nominal}

\begin{figure}[t]
	\centering
	\includegraphics[scale=0.3]{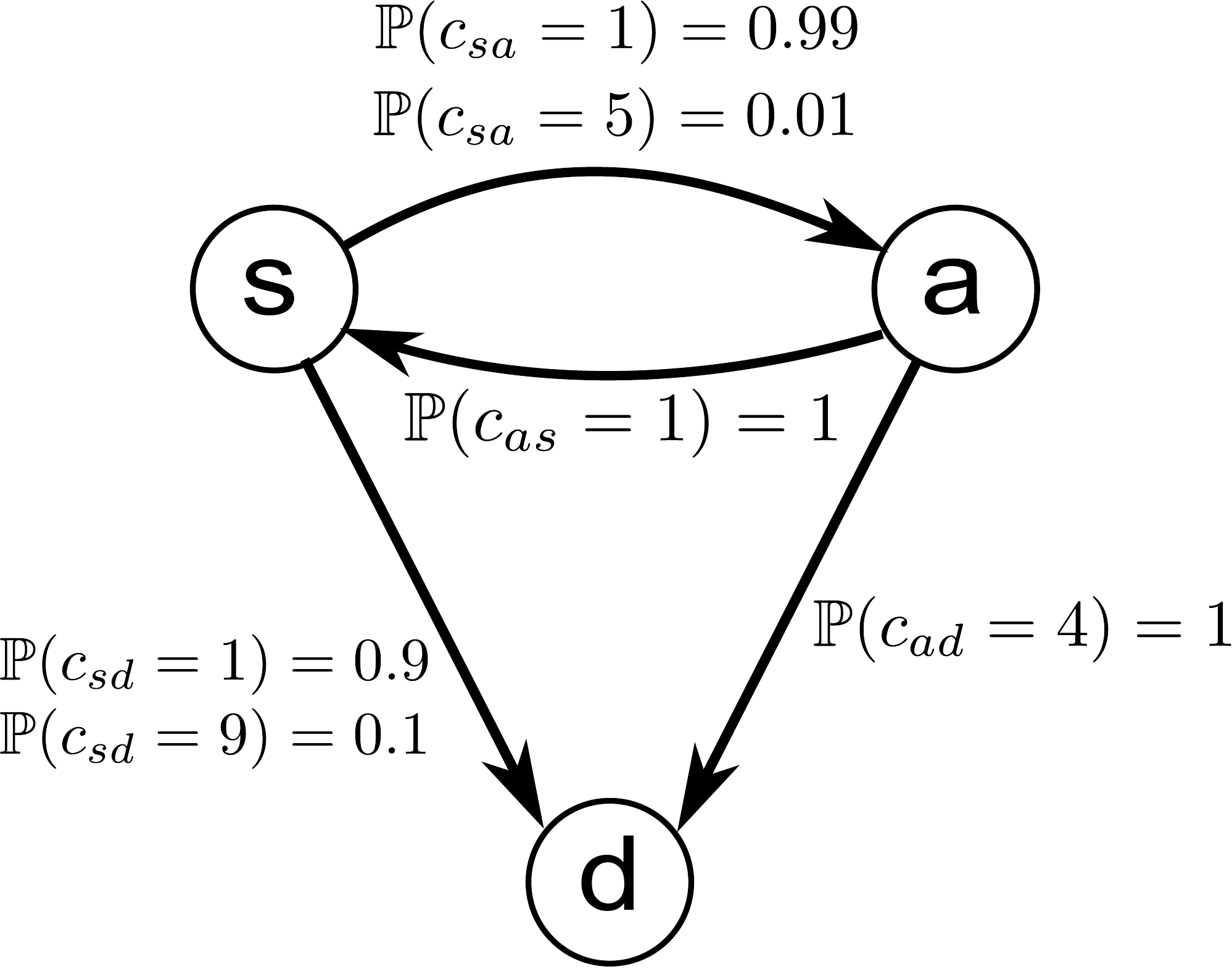}
	\caption{Existence of loops. If the initial budget is $T=8$ and the risk function is $f(t) = t \cdot 1_{t \leq 0}$, the optimal strategy to travel from $s$ to $d$ is to go to $a$ first. This is because going to $d$ directly incurs an expected delay of $0.1$, while going to $a$ first and then planning to go to $d$ incurs an expected delay of $0.01$. If we end up getting a cost $c_{sa} = 5$ on the way to $a$, then, performing a similar analysis, the optimal strategy is to go back to $s$.} 
	\label{fig-counterexampleloop}
\end{figure}

Perhaps the most important property of \eqref{eq-nominal-problem} is that \emph{Bellman's Principle of Optimality} can be shown to hold. Specifically, for any history of the process $h \in \mathcal{H}$, an optimal strategy to \eqref{eq-nominal-problem} must also be an optimal strategy to the subproblem of minimizing the risk function given this history. Otherwise, we could modify this strategy for this particular history and take it to be an optimal strategy for this subproblem. This operation could only increase the objective function of the optimization problem \eqref{eq-nominal-problem}, which would contradict the optimality of the strategy. 
\\
\indent Another, less obvious, interesting feature of \eqref{eq-nominal-problem} is that, even for perfectly natural risk functions $f(\cdot)$, making decisions according to an optimal strategy may lead to cycle back to a previously visited location. This may happen, for instance, when the objective is to maximize the probability of completion within budget, see \cite{samitha10}, and their example can be adapted when the objective is to minimize the expected budget overrun, see Figure \ref{fig-counterexampleloop}. 
While counter-intuitive at first, the existence of loops is a direct consequence of the stochasticity of the costs when the decision maker is concerned about the risk of going over budget, as illustrated in Figure \ref{fig-counterexampleloop}. On the other hand, the existence of infinitely many loops is particularly troublesome from a modeling perspective as it would imply that a user traveling through $\mathcal{V}$ following the optimal strategy may get at a location $i \neq d$ having already spent an arbitrarily large budget with positive probability. Furthermore, infinite cycling is also problematic from a computational standpoint because describing an optimal strategy would require unlimited storage capacity. We argue that infinite cycling arises only when the risk function is poorly chosen. This is obvious when $f(t) = - t \cdot 1_{t \leq 0}$, which corresponds to maximizing the expected budget overrun, but we stress that it is not merely a matter of monotonicity. Infinite cycling may occur even if $f(\cdot)$ is increasing as we highlight in Example \ref{example-counterxamplefnotconcavenotdecaying}. 

\begin{figure}[t]
	\begin{center}
		\subfloat[Graph, $s$ and $d$ are respectively the source and the destination.]{\label{fig-counterexamplegraph}\includegraphics[scale=0.3]{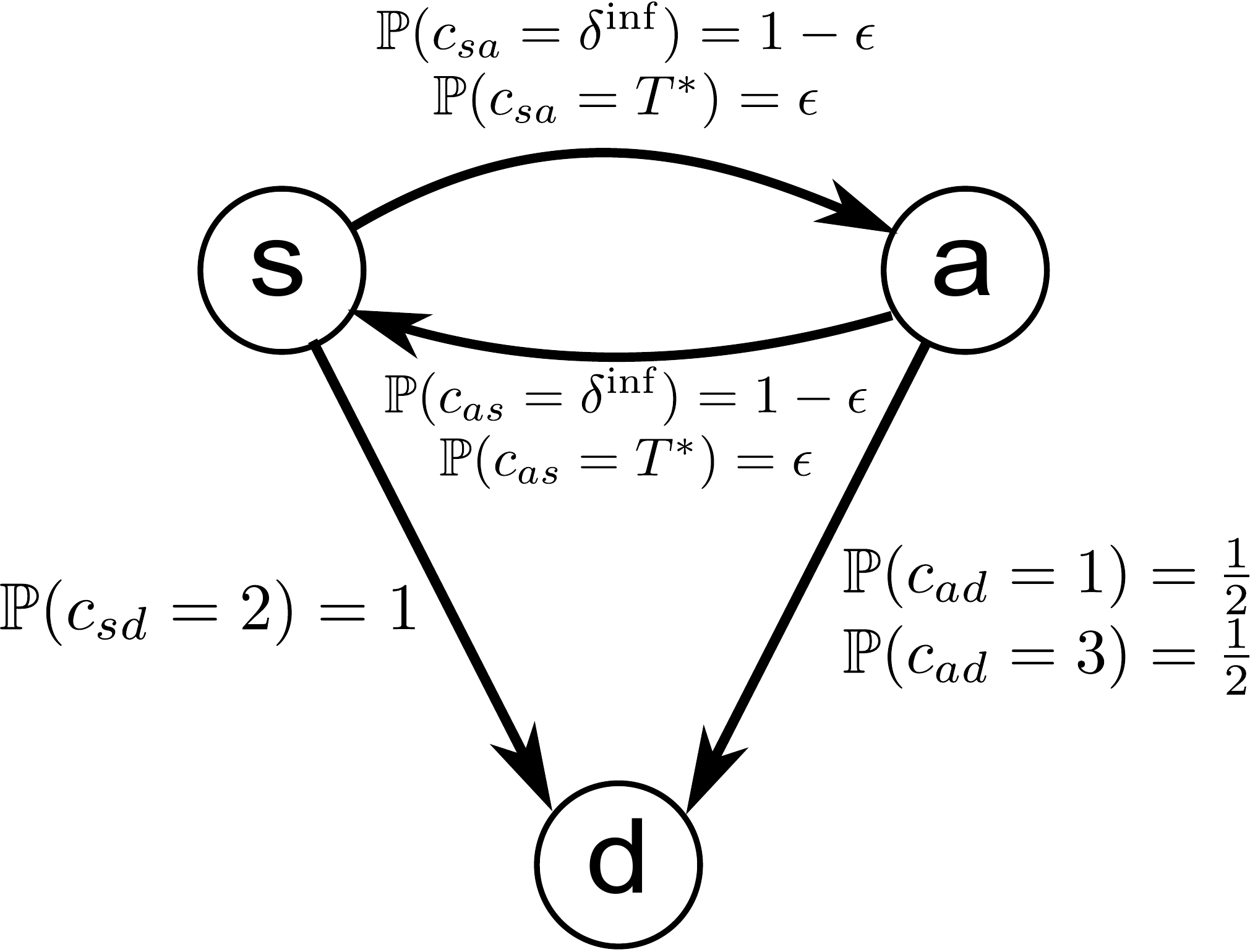}} 
		\hspace{1cm}
		\subfloat[risk function. $T$ is the initial budget, $2 T^{*}$ is the period of $f'(\cdot)$.]{\label{fig-counterexampleobjectivefunction}\includegraphics[scale=0.4]{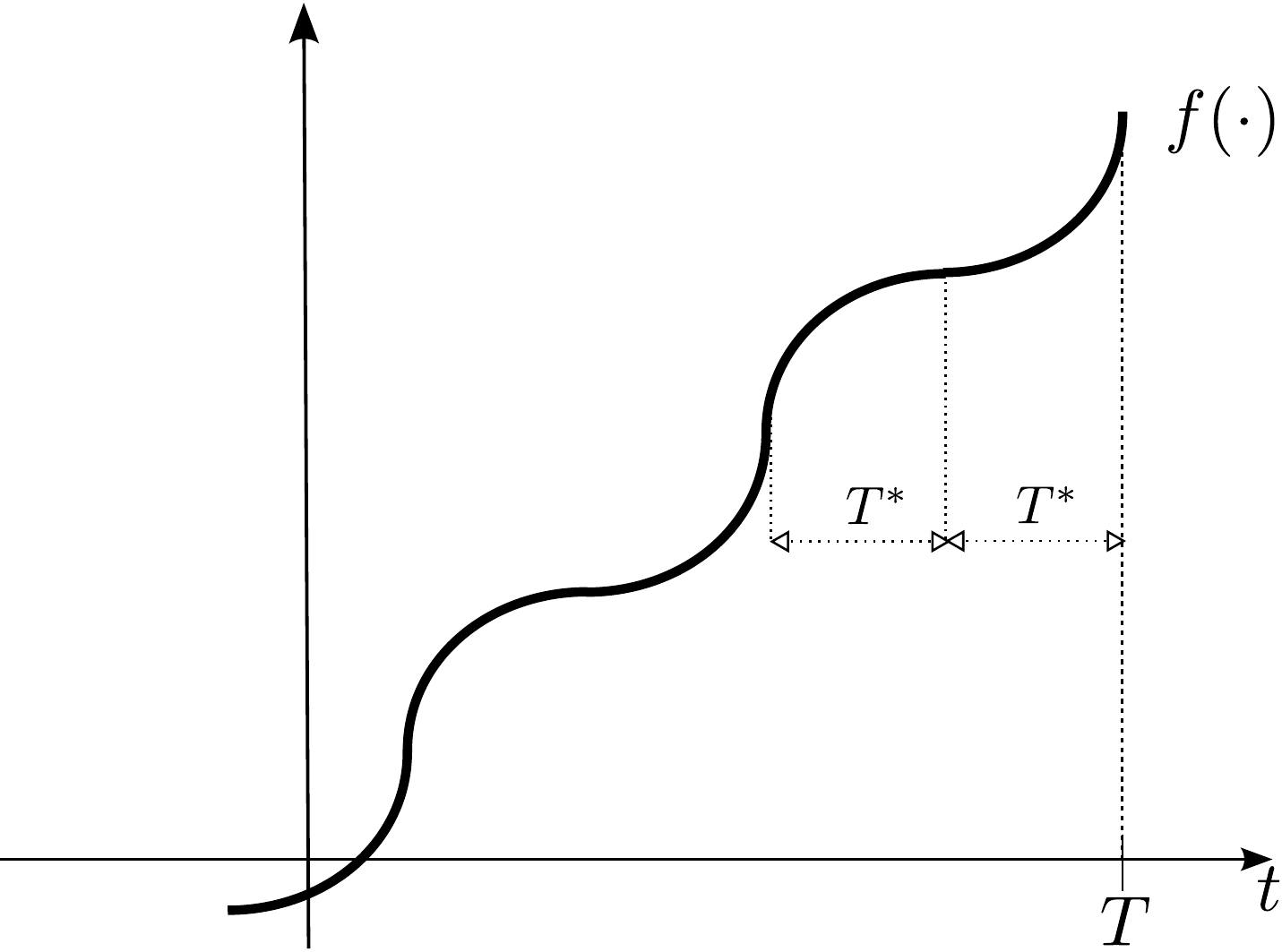}}\hspace{1cm} 
	\end{center}
	
	\caption{Existence of infinite cycling from Example \ref{example-counterxamplefnotconcavenotdecaying}.}	
	\label{fig-generalcounterexample}
\end{figure}

\begin{example}
	\label{example-counterxamplefnotconcavenotdecaying} 
	
	Consider the simple directed graph of Figure \ref{fig-counterexamplegraph} and the risk function $f(\cdot)$ illustrated in Figure \ref{fig-counterexampleobjectivefunction}. $f(\cdot)$ is defined piecewise, alternating between concavity and convexity on intervals of size $T^*$ and the same pattern is repeated every $2 T^{*}$. This means that, for this particular objective, the attitude towards risk keeps fluctuating as the budget decreases, from being risk-averse when $f(\cdot)$ is locally concave to being risk-seeking when $f(\cdot)$ is locally convex. Now take $\deltainf << 1$, $\epsilon << 1$ and $T^{*} > 3$ and consider finding a strategy to get to $d$ starting from $s$ with initial budget $T$ which we choose to take at a point where $f(\cdot)$ switches from being concave to being convex, see Figure \ref{fig-counterexampleobjectivefunction}. Going straight to $d$ incurs an expected objective value of $f(T-2) < \frac{1}{2} f(T-1) + \frac{1}{2} f(T-3)$ and we can make this gap arbitrarily large by properly defining $f(\cdot)$. Therefore, by taking $\epsilon$ and $\deltainf$ small enough, going to $a$ first is optimal. With probability $\epsilon>0$, we arrive at $a$ with a remaining budget of $T-T^{*}$. Afterwards, the situation is reversed as we are willing to take as little risk as possible and the corresponding optimal solution is to go back to $s$. With probability $\epsilon$, we arrive at $s$ with a budget of $T-2 T^{*}$ and we are back in the initial situation, showing the existence of infinite cycling.
\end{example}

In light of Example \ref{example-counterxamplefnotconcavenotdecaying}, we identify a set of sufficient asymptotic conditions on $f(\cdot)$ ruling out the possibility of infinite cycling.

\begin{theorem}
\label{lemma-LET-is-opt-nominal}
Case 1: If there exists $T_1$ such that either:
\begin{enumerate}[(a)]
	\item $f(\cdot)$ is increasing, concave, and $C^2$ on $(-\infty, T_1)$ and such that $\frac{f''}{f'} \rightarrow_{-\infty} 0$,
	\item $f(\cdot)$ is $C^1$ on $(-\infty, T_1)$ and $\lim_{ - \infty} f'$ exists, is positive, and is finite,
\end{enumerate}
then there exists $T_f$ such that, for any $T \geq 0$ and as soon as the total cost spent so far is larger than $T - T_f$, any optimal policy to \eqref{eq-nominal-problem} follows the shortest-path tree rooted at $d$ with respect to the mean arc costs, which we denote by $\mathcal{T}$. 
\\
Case 2: If there exists $T_f$ such that the support of $f(\cdot)$ is included in $[T_f, \infty)$, then following $\mathcal{T}$ is optimal as soon as the total cost spent so far is larger than $T - T_f$.
\end{theorem}
For a node $i$, $\mathcal{T}(i)$ refers to the set of immediate successors of $i$ in $\mathcal{T}$. The proof is deferred to the online supplement, Section \ref{section-proofextendsota}.
\\
\indent Observe that, in addition to not being concave, the choice of $f(\cdot)$ in Example \ref{example-counterxamplefnotconcavenotdecaying} does not satisfy property (b) as $f'(\cdot)$ is $2 T^{*}$-periodic. An immediate consequence of Theorem \ref{lemma-LET-is-opt-nominal} is that an optimal strategy to \eqref{eq-nominal-problem} does not include any loop as soon as the total cost spent so far is larger than $T - T_f$. Since each arc has a positive minimum cost, this rules out infinite cycling. The parameter $T_f$ can be computed through direct reasoning on the risk function $f(\cdot)$ or by inspecting the proof of Theorem \ref{lemma-LET-is-opt-nominal}. Remark that any polynomial of even degree with a negative leading coefficient satisfies condition (a) of Theorem \ref{lemma-LET-is-opt-nominal}. Examples of valid objectives include maximization of the probability of completion within budget $f(t) = 1_{t \geq 0}$ with $T_f = 0$, minimization of the budget overrun $f(t) = t \cdot 1_{t \leq 0}$ with $T_f = 0$, and minimization of the squared budget overrun $f(t) = - t^2 \cdot 1_{t \leq 0}$ with 
$$
	T_f = - \frac{ |\mathcal{V}| \cdot \deltasup \cdot \max\limits_{i \in \mathcal{V} } M_i }{ 2 \cdot \min\limits_{i \neq d} \min\limits_{j \in \mathcal{V}(i), j \notin \mathcal{T}(i) } \{ \mathbb{E}[c_{ij}] + M_j - M_i \}  },
$$
where $M_i$ is the minimum expected cost to go from $i$ to $d$ and with the convention that the minimum of an empty set is equal to $\infty$. When $f(\cdot)$ is increasing but does not satisfy condition (a) or (b), the optimal strategy may follow a different shortest-path tree. For instance, if $f(t) = - \exp(-t)$, the optimal policy is to follow the shortest path to $d$ with respect to $(\log( \mathbb{E}[ \exp(c_{ij}) ] ))_{(i,j) \in \mathcal{A}}$. Conversely, if $f(t) = \exp(t)$, the optimal policy is to follow the shortest path to $d$ with respect to $(- \log( \mathbb{E}[ \exp(- c_{ij}) ] ))_{(i,j) \in \mathcal{A}}$. For these reasons, proving that an optimal strategy to \eqref{eq-nominal-problem} does not include infinitely many loops when $f(\cdot)$ does not satisfy the assumptions of Theorem \ref{lemma-LET-is-opt-nominal} requires objective-specific (and possibly graph-specific) arguments. To illustrate this last point, observe that the conclusion of Theorem \ref{lemma-LET-is-opt-nominal} always holds for a graph consisted of a single simple path regardless of the definition of $f(\cdot)$, even if this function is decreasing. Hence, the assumptions of Theorem \ref{lemma-LET-is-opt-nominal} are not necessary in general to prevent infinite cycling but restricting our attention to this class of risk functions enables us to study the problem in a generic fashion and to develop a general-purpose algorithm in Section \ref{section-solution-methodology-nominal}. 
\\
\indent Another remarkable property of \eqref{eq-nominal-problem} is that it can be equivalently formulated as a MDP in the extended space state $(i, t) \in \mathcal{V} \times (-\infty, T]$ where $i$ is the current location and $t$ is the remaining budget. As a result, standard techniques for MDPs can be applied to show that there exists an optimal \emph{Markov} policy $\pi^*_f$ which is a mapping from the current location and the remaining budget to the next node to visit. Furthermore, the optimal \emph{Markov} policies are characterized by the dynamic programming equation:
\begin{equation}
	\label{eq-dp-nominal}
	\begin{aligned}
	 & u_d(t) = f(t) \quad  && t \leq T \\
	 & u_i(t) = \max\limits_{j \in \mathcal{V}(i)} \int_0^{\infty} p_{ij}(\omega) \cdot u_j(t-\omega) \mathrm{d}\omega \quad && i \neq d, t \leq T \\
	 & \pi^*_f(i, t) \in \argmax\limits_{j \in \mathcal{V}(i)} \int_0^{\infty} p_{ij}(\omega) \cdot u_j(t-\omega) \mathrm{d}\omega \quad && i \neq d, t \leq T,
	\end{aligned}
\end{equation}
where $\mathcal{V}(i) = \{ j \in \mathcal{V} \; | \; (i,j) \in \mathcal{A} \}$ refers to the set of immediate successors of $i$ in $\mathcal{G}$ and $u_i(t)$ is the expected objective-to-go when leaving $i \in \mathcal{V}$ with remaining budget $t$. The interpretation of \eqref{eq-dp-nominal} is simple. At each node $i \in \mathcal{V}$, and for each potential remaining budget $t$, the decision maker should pick the outgoing edge $(i,j)$ that yields the maximum expected objective-to-go if acting optimally thereafter. 
\begin{proposition}
	\label{lemma-dp-optimal-nominal}
	Under the same assumptions as in Theorem \ref{lemma-LET-is-opt-nominal}, any Markov policy solution to \eqref{eq-dp-nominal} is an optimal strategy for \eqref{eq-nominal-problem}.
\end{proposition}
The proof is deferred to the online supplement, Section \ref{section-proof-dp-optimal-nominal}.

\subsection{Solution methodology}
\label{section-solution-methodology-nominal}
In order to solve \eqref{eq-nominal-problem}, we use Proposition \ref{lemma-dp-optimal-nominal} and compute a \emph{Markov} policy solution to the dynamic program \eqref{eq-dp-nominal}. We face two main challenges when we carry out this task. First, \eqref{eq-dp-nominal} is a continuous dynamic program. To solve this program numerically, we approximate the functions $(u_i(\cdot))_{i \in \mathcal{V}}$ by piecewise constant functions, as detailed in Section \ref{section-discretization-nominal}. Second, as illustrated in Figure \ref{fig-counterexampleloop} of Section \ref{section-characterization-policy-nominal}, an optimal \emph{Markov} strategy solution to \eqref{eq-dp-nominal} may contain loops. Hence, in the presence of a cycle in $\mathcal{G}$, say $i \rightarrow j \rightarrow i$, observe that computing $u_i(t)$ requires to know the value of $u_j(t)$ which in turns depends on $u_i(t)$. As a result, it is a-priori unclear how to solve \eqref{eq-dp-nominal} without resorting to value or policy iteration. We explain how to sidestep this difficulty and construct efficient label-setting algorithms in Section \ref{section-algorithm-nominal}. In particular, using these algorithms, we can compute:
\begin{itemize}
	\item{an optimal solution to \eqref{eq-nominal-problem} in $O( |\mathcal{A}| \cdot  \frac{ T - T_f }{\Delta t} \cdot \log^2( \frac{\deltasup}{\Delta t}) + |\mathcal{V}|^2 \cdot \frac{\deltasup}{\Delta t} \cdot \log( |\mathcal{V}| \cdot \frac{\deltasup}{\Delta t}  ) )$ computation time when the arc costs only take on values that are multiple of $\Delta t > 0$ and for any risk function $f(\cdot)$ satisfying Theorem \ref{lemma-LET-is-opt-nominal}. This simplifies to $O( |\mathcal{A}| \cdot  \frac{ T }{\Delta t} \cdot \log^2( \frac{\deltasup}{\Delta t} ) )$ when the objective is to maximize the probability of completion within budget,}
	\item an $\epsilon$-approximate solution to \eqref{eq-nominal-problem} in 
	$$
		O( \frac{ (|\mathcal{V}| + \frac{T - T_f}{\deltainf} )^2}{\epsilon} \cdot [ \; |\mathcal{A}| \cdot  (T - T_f) \cdot \log^2( \frac{(|\mathcal{V}| + \frac{T - T_f}{\deltainf}  ) \cdot \deltasup}{\epsilon}) + |\mathcal{V}|^2 \cdot \deltasup \cdot \log( \frac{(|\mathcal{V}| + \frac{T - T_f}{\deltainf} ) \cdot |\mathcal{V}|  \cdot \deltasup}{\epsilon}  ) \; ] )		
	$$   
	computation time when the risk function is Lipschitz on compact sets.
\end{itemize}

\subsubsection{Discretization scheme}
\label{section-discretization-nominal}
For each node $i \in \mathcal{V}$, we approximate $u_i(\cdot)$ by a piecewise constant function $u^{\Delta t}_i(\cdot)$ of uniform stepsize $\Delta t$. Under the conditions of Theorem \ref{lemma-LET-is-opt-nominal}, we only need to approximate $u_i(\cdot)$ for a remaining budget larger than $k^{\text{min}}_i \cdot \Delta t$, for $k^{\text{min}}_i = \floor*{ \frac{T_f - (|\mathcal{V}| - \text{level}(i, \mathcal{T}) + 1) \cdot \deltasup}{\Delta t} }$, where $\text{level}(i, \mathcal{T})$ is defined as the level of node $i$ in the rooted tree $\mathcal{T}$, i.e. the number of parent nodes of $i$ in $\mathcal{T}$ plus one. This is because, following the shortest path tree $\mathcal{T}$ once the remaining budget drops below $T_f$, we can never get to state $i$ with remaining budget less than $k^{\text{min}}_i \cdot \Delta t$. We use the approximation:
\begin{equation}
	\label{eq-discretize-approximation-nominal}
	\begin{aligned}	 
	 & u^{\Delta t}_i(t) = u^{\Delta t}_i( \floor*{ \frac{t}{\Delta t} } \cdot \Delta t ) \quad && i \in \mathcal{V}, t \in [k^{\text{min}}_i \cdot \Delta t, T] \\
	 & \pi^{\Delta t}(i, t) = \pi^{\Delta t}(i, \floor*{ \frac{t}{\Delta t} } \cdot \Delta t ) \quad && i \neq d, t \in [k^{\text{min}}_i \cdot \Delta t, T], \\
	 \end{aligned}
\end{equation}
and the values at the mesh points are determined by the set of equalities:
\begin{equation}
	\label{eq-discretization-scheme-nominal}
	\begin{aligned}
	 & u^{\Delta t}_d( k \cdot \Delta t) = f( k \cdot \Delta t ) \quad && k = k^{\text{min}}_d, \cdots, \floor*{ \frac{T}{\Delta t} } \\
	 & u^{\Delta t}_i( k \cdot \Delta t ) =  \max\limits_{j \in \mathcal{V}(i)} \int_0^{\infty} p_{ij}(\omega) \cdot u^{\Delta t}_j( k \cdot \Delta t - \omega) \mathrm{d}\omega \quad && i \neq d, k = \floor*{ \frac{T_f}{\Delta t} }, \cdots, \floor*{ \frac{T}{\Delta t} }\\
	 & \pi^{\Delta t}(i, k \cdot \Delta t ) \in \argmax\limits_{j \in \mathcal{V}(i)} \int_0^{\infty} p_{ij}(\omega) \cdot u^{\Delta t}_j( k \cdot \Delta t - \omega) \mathrm{d}\omega \quad && i \neq d, k = \floor*{ \frac{T_f}{\Delta t} }, \cdots, \floor*{ \frac{T}{\Delta t} } \\
	& u^{\Delta t}_i( k \cdot \Delta t ) =  \max\limits_{j \in \mathcal{T}(i)} \int_0^{\infty} p_{ij}(\omega) \cdot u^{\Delta t}_j( k \cdot \Delta t - \omega) \mathrm{d}\omega \quad && i \neq d, k = k^{\text{min}}_i, \cdots, \floor*{ \frac{T_f}{\Delta t} } - 1 \\
	& \pi^{\Delta t}(i, k \cdot \Delta t ) \in \argmax\limits_{j \in \mathcal{T}(i)} \int_0^{\infty} p_{ij}(\omega) \cdot u^{\Delta t}_j( k \cdot \Delta t - \omega) \mathrm{d}\omega \quad && i \neq d, k = k^{\text{min}}_i, \cdots, \floor*{ \frac{T_f}{\Delta t} } - 1.
	\end{aligned}
\end{equation}
Notice that for $t \leq T_f$, we rely on Theorem \ref{lemma-LET-is-opt-nominal} and only consider, for each node $i \neq d$, the immediate neighbors of $i$ in $\mathcal{T}$. This is of critical importance to be able to solve \eqref{eq-discretization-scheme-nominal} with a label-setting algorithm, see Section \ref{section-algorithm-nominal}. The next result provides insight into the quality of the policy $\pi^{\Delta t}$ as an approximate solution to \eqref{eq-nominal-problem}.
\begin{proposition}
	\label{lemma-quality-approx-nominal}
	Consider a solution to the global discretization scheme \eqref{eq-discretize-approximation-nominal} and \eqref{eq-discretization-scheme-nominal}, $(\pi^{\Delta t}, (u^{\Delta t}_i(\cdot))_{i \in \mathcal{V}})$. We have:
	\begin{enumerate}
		\item{If $f(\cdot)$ is non-decreasing, the functions $(u^{\Delta t}_i(\cdot))_{i \in \mathcal{V}}$ converge pointwise almost everywhere to $(u_i(\cdot))_{i \in \mathcal{V}}$ as $\Delta t \rightarrow 0$,}
		\item{If $f(\cdot)$ is continuous, the functions $(u^{\Delta t}_i(\cdot))_{i \in \mathcal{V}}$ converge uniformly to $(u_i(\cdot))_{i \in \mathcal{V}}$ and $\pi^{\Delta t}$ is a $o(1)$-approximate optimal solution to \eqref{eq-nominal-problem} as $\Delta t \rightarrow 0$,}
		\item{If $f(\cdot)$ is Lipschitz on compact sets (e.g. if $f(\cdot)$ is $C^1$), the functions $(u^{\Delta t}_i(\cdot))_{i \in \mathcal{V}}$ converge uniformly to $(u_i(\cdot))_{i \in \mathcal{V}}$ at speed $\Delta t$ and $\pi^{\Delta t}$ is a $O(\Delta t)$-approximate optimal solution to \eqref{eq-nominal-problem} as $\Delta t \rightarrow 0$,}	
		\item{If $f(t) = 1_{t \geq 0}$ and the distributions $(p_{ij})_{(i, j) \in \mathcal{A}}$ are continuous, the functions $(u^{\Delta t}_i(\cdot))_{i \in \mathcal{V}}$ converge uniformly to $(u_i(\cdot))_{i \in \mathcal{V}}$ and $\pi^{\Delta t}$ is a $o(1)$-approximate optimal solution to \eqref{eq-nominal-problem} as $\Delta t \rightarrow 0$.}
	\end{enumerate}	
\end{proposition}
The proof is deferred to the online supplement, Section \ref{section-proof-quality-approx-nominal}. 
\\
\indent If the distributions $(p_{ij})_{(i, j) \in \mathcal{A}}$ are discrete and $f(\cdot)$ is piecewise constant, an exact optimal solution to \eqref{eq-nominal-problem} can be computed by appropriately choosing a different discretization length for each node. In this paper, we focus on discretization schemes with a uniform stepsize $\Delta t$ for mathematical convenience. We stress that choosing adaptively the discretization length can improve the quality of the approximation for the same number of computations, see \cite{hoy2015approximately}.

\subsubsection{Solution procedures}
\label{section-algorithm-nominal}
The key observation enabling the development of label-setting algorithms to solve \eqref{eq-dp-nominal} is made by \cite{samitha10}. They note that, when the risk function is the probability of completion within budget, $u_i(t)$ can be computed for $i \in \mathcal{V}$ and $t \leq T$ as soon as the values taken by $u_j(\cdot)$ on $(-\infty, t - \deltainf]$ are available for all neighboring nodes $j \in \mathcal{V}(i)$ since $p_{ij}(\omega) = 0$ for $\omega \leq \deltainf$ under Assumption \ref{assumption-compactsupport}. They propose a label-setting algorithm which consists in computing the functions $(u_i(\cdot))_{i \in \mathcal{V}}$ block by block, by interval increments of size $\deltainf$. After the following straightforward initialization step: $u_i(t) = 0$ for $t \leq 0$ and $i \in \mathcal{V}$, they first compute $(u_i(\cdot)_{[0, \deltainf]})_{i \in \mathcal{V}}$, then $(u_i(\cdot)_{[0, 2 \cdot \deltainf]})_{i \in \mathcal{V}}$ and so on to eventually derive $(u_i(\cdot)_{[0, T]})_{i \in \mathcal{V}}$. While this incremental procedure can still be applied for general risk functions, the initialization step gets tricky if $f(\cdot)$ does not have a one-sided compact support of the type $[a, \infty)$. Theorem \ref{lemma-LET-is-opt-nominal} is crucial in this respect because the shortest-path tree $\mathcal{T}$ induces an ordering of the nodes to initialize the collection of functions $(u_i(\cdot))_{i \in \mathcal{V}}$ for remaining budgets smaller than $T_f$. The functions can subsequently be computed for larger budgets using the incremental procedure outlined above. To be specific, we solve \eqref{eq-discretization-scheme-nominal} in three steps. First, we compute $T_f$ (defined in Theorem \ref{lemma-LET-is-opt-nominal}). Inspecting the proof of Theorem \ref{lemma-LET-is-opt-nominal}, observe that $T_f$ only depends on few parameters, namely the risk function $f(\cdot)$, the expected arc costs, and the maximum arc costs. Next, we compute the values $u^{\Delta t}_i( k \cdot \Delta t)$ for $k \in \{k^{\text{min}}_i, \cdots, \floor*{ \frac{T_f}{\Delta t} } - 1 \}$ starting at node $i = d$ and traversing the tree $\mathcal{T}$ in a breadth-first fashion using fast Fourier transforms with complexity $O(|\mathcal{V}|^2 \cdot \frac{\deltasup}{\Delta t} \cdot \log( |\mathcal{V}| \cdot \frac{\deltasup}{\Delta t}  )  )$. Note that this step can be made to run significantly faster for specific risk functions, e.g. for the probability of completion within budget where $u^{\Delta t}_i( k \cdot \Delta t) = 0$ for $k < \floor*{\frac{T_f}{\Delta t}}$ and any $i \in \mathcal{V}$. Finally, we compute the values $u^{\Delta t}_i( k \cdot \Delta t)$ for $k \in \{ \floor*{ \frac{T_f}{\Delta t} } + m \cdot \floor*{ \frac{\deltainf}{ \Delta t } }, \cdots, \floor*{ \frac{T_f}{\Delta t} } + (m+1) \cdot \floor*{ \frac{\deltainf}{ \Delta t } } \}$ for all nodes $i \in \mathcal{V}$ by induction on $m$. 
\paragraph{Complexity analysis.}
The description of the last step of the label-setting approach leaves out one detail that has a dramatic impact on the runtime complexity. We need to specify how to compute the convolution products arising in \eqref{eq-discretization-scheme-nominal} for $k \geq \floor*{ \frac{T_f}{\Delta t} }$, keeping in mind that, for any node $i \in \mathcal{V}$, the values $u^{\Delta t}_i(k \cdot \Delta t)$ for $k \in \{ \floor*{ \frac{T_f}{\Delta t} }, \cdots, \floor*{ \frac{T}{\Delta t} } \}$ become available online by chunks of length $\floor*{\frac{\deltainf}{\Delta t}}$ as the label-setting algorithm progresses. A naive implementation consisting in applying the pointwise definition of convolution products has a runtime complexity $O( |\mathcal{A}| \cdot \frac{ (T - T_f) \cdot ( \deltasup - \deltainf )}{(\Delta t)^2})$. Using fast Fourier transforms for each chunk brings down the complexity to $O( |\mathcal{A}| \cdot  \frac{ (T - T_f) }{\Delta t} \cdot  \frac{\deltasup}{\deltainf} \cdot \log(\frac{\deltasup}{\Delta t}))$. Applying another online scheme developed in \cite{dean2010speeding} and \cite{samitha2012atmos}, based on the idea of zero-delay convolution, leads to a worst-case complexity $O( |\mathcal{A}| \cdot  \frac{ (T - T_f)}{\Delta t} \cdot \log^2( \frac{\deltasup}{\Delta t} ) )$. Numerical evidence suggest that this last implementation significantly speeds up the computations, see \cite{samitha2012atmos}.

\section{Theoretical and computational analysis of the robust problem}
\label{section-analysis-robust-problem}

\subsection{Characterization of optimal policies}
\label{section-characterization-policy-robust}
The properties satisfied by optimal solutions to the nominal problem naturally extend to their robust counterparts, which we recall are defined as optimal solutions to \eqref{eq-robust-problem}. In fact, all the results derived in this section are strict generalizations of those obtained in Section \ref{section-characterization-policy-nominal} for singleton ambiguity sets. We point out that the rectangularity of the global ambiguity set is essential for the results to carry over to the robust setting as it guarantees that \emph{Bellman's Principle of Optimality} continue to hold, which is an absolute prerequisite for computational tractability.
\\
\indent Similarly as what we have seen for the nominal problem, infinite cycling might occur in the robust setting, depending on the risk function at hand. This difficulty can be shown not to arise under the same conditions on $f(\cdot)$ as for the nominal problem.

\begin{theorem}
\label{lemma-LET-is-opt-robust}
Case 1: If there exists $T_1$ such that either:
\begin{enumerate}[(a)]
	\item $f(\cdot)$ is increasing, concave, and $C^2$ on $(-\infty, T_1)$ and such that $\frac{f''}{f'} \rightarrow_{-\infty} 0$,
	\item $f(\cdot)$ is $C^1$ on $(-\infty, T_1)$ and $\lim_{ - \infty} f'$ exists, is positive, and is finite,
\end{enumerate}
then there exists $T^r_f$ such that, for any $T \geq 0$ and as soon as the total cost spent so far is larger than $T - T^r_f$, any optimal policy solution to \eqref{eq-robust-problem} follows the shortest-path tree rooted at $d$ with respect to the worst-case mean arc costs, i.e. $(\max_{ p_{ij} \in \mathcal{P}_{ij} } \mathbb{E}_{X \sim p_{ij}}[X] )_{(i, j) \in \mathcal{A}}$, which we denote by $\mathcal{T}^r$.
\\
Case 2: If there exists $T_f$ such that the support of $f(\cdot)$ is included in $[T_f, \infty)$, then following $\mathcal{T}^r$ is optimal as soon as the total cost spent so far is larger than $T - T^r_f$.
\end{theorem}
For a node $i$, $\mathcal{T}^r(i)$ refers to the set of immediate successors of node $i$ in $\mathcal{T}^r$. The proof is deferred to the online supplement, Section \ref{section-proof-lemma-LET-is-opt-robust}. 
\\
\indent Interestingly, $T^r_f$ is determined by the exact same procedure as $T_f$ provided the expected arc costs are substituted with the worst-case expected costs. For instance, when $f(t) = - t^2 \cdot 1_{t \leq 0}$, we may take:
$$
	T^r_f = - \frac{ |\mathcal{V}| \cdot \deltasup \cdot \max\limits_{i \in \mathcal{V}} M_i }{ 2 \cdot \min\limits_{i \neq d} \min\limits_{j \in \mathcal{V}(i), j \notin \mathcal{T}^r(i) } \{ \max_{ p_{ij} \in \mathcal{P}_{ij} } \mathbb{E}_{X \sim p_{ij}}[X] + M_j - M_i \}  },
$$
where $M_i$ is the worst-case minimum expected cost to go from $i$ to $d$. 
\\
\indent Last but not least, problem \eqref{eq-robust-problem} can be formulated as a distributionally robust MDP in the extended space state $(i, t) \in \mathcal{V} \times (-\infty, T]$. As a result, one can show that there exists an optimal \emph{Markov} policy $\pi^*_{f, \mathcal{P}}$ characterized by the dynamic programming equation:
\begin{equation}
	\label{eq-dp-robust}
	\begin{aligned}
     & u_d(t) = f(t) \quad  && t \leq T \\
	 & u_i(t) = \max\limits_{j \in \mathcal{V}(i)} \inf\limits_{p_{ij} \in \mathcal{P}_{ij}} \int_0^{\infty} p_{ij}(\omega) \cdot u_j(t-\omega) \mathrm{d}\omega \quad && i \neq d, t \leq T \\
	 & \pi^*_{f, \mathcal{P}}(i, t) \in \argmax\limits_{j \in \mathcal{V}(i)} \inf\limits_{p_{ij} \in \mathcal{P}_{ij}} \int_0^{\infty} p_{ij}(\omega) \cdot u_j(t-\omega) \mathrm{d}\omega \quad && i \neq d, t \leq T,
	\end{aligned}
\end{equation}
where $u_i(t)$ is the worst-case expected objective-to-go when leaving $i \in \mathcal{V}$ with remaining budget $t$. Observe that  \eqref{eq-dp-robust} only differs from \eqref{eq-dp-nominal} through the presence of the infimum over $\mathcal{P}_{ij}$.
\begin{proposition}
	\label{lemma-dp-optimal-robust}
	Any Markov policy solution to \eqref{eq-dp-robust} is an optimal strategy for \eqref{eq-robust-problem}.
\end{proposition}
The proof is deferred to the online supplement, Section \ref{section-proof-lemma-dp-optimal-robust}. 

\subsection{Tightness of the robust problem}
\label{section-tightness-robust-relaxation}
The optimization problem \eqref{eq-robust-problem} is a robust relaxation of \eqref{eq-ideal-robust-problem} in the sense that, for any strategy $\pi \in \Pi$, we have:
$$
	\inf\limits_{\forall (i, j) \in \mathcal{A}, \; p_{ij} \in \mathcal{P}_{ij}} \; \mathbb{E}_{ \mathbf{p} }[f(T - X_\pi)] \geq \inf\limits_{\forall \tau, \forall (i, j) \in \mathcal{A}, \; p^\tau_{ij} \in \mathcal{P}_{ij}} \; \mathbb{E}_{ \mathbf{p^\tau} }[f(T - X_\pi)].
$$
We say that \eqref{eq-ideal-robust-problem} and \eqref{eq-robust-problem} are equivalent if they share the same optimal value and if there exists a common optimal strategy. For general risk functions, ambiguity sets, and graphs, \eqref{eq-ideal-robust-problem} and \eqref{eq-robust-problem} are not equivalent. In this section, we highlight several situations of interest for which \eqref{eq-ideal-robust-problem} and \eqref{eq-robust-problem} happen to be equivalent and we bound the gap between the optimal values of \eqref{eq-ideal-robust-problem} and \eqref{eq-robust-problem} for a subclass of risk functions. In this paper, we solve \eqref{eq-robust-problem} instead of \eqref{eq-ideal-robust-problem} for computational tractability, irrespective of whether or not \eqref{eq-ideal-robust-problem} and \eqref{eq-robust-problem} are equivalent. Hence, the results presented in this section are included mainly for illustrative purposes, i.e. we do not impose further restrictions on the risk function or the ambiguity sets here.

\paragraph{Equivalence of \eqref{eq-ideal-robust-problem} and \eqref{eq-robust-problem}.}
As a simple first example, observe that when $f(\cdot)$ is non-decreasing and $\mathcal{P}_{ij} = \mathcal{P}([\deltainf_{ij}, \deltasup_{ij}])$, both \eqref{eq-ideal-robust-problem} and \eqref{eq-robust-problem} reduce to a standard robust approach where the goal is to find a path minimizing the sum of the worst-case arc costs. The following result identifies conditions of broader applicability when the decision maker is risk-seeking.

\begin{lemma}
\label{lemma-tight-relaxation-if-convex}
Suppose that $f(\cdot)$ is convex and satisfies property (b) in Case 1 of Theorem \ref{lemma-LET-is-opt-robust} and that, for any arc $(i, j) \in \mathcal{V}$, the Dirac distribution supported at $\max_{ p_{ij} \in \mathcal{P}_{ij} } \mathbb{E}_{X \sim p_{ij}}[X]$ belongs to $\mathcal{P}_{ij}$. Then, \eqref{eq-ideal-robust-problem} and \eqref{eq-robust-problem} are equivalent.
\end{lemma}
The proof is deferred to the online supplement, Section \ref{section-proof-lemma-tight-relaxation-if-convex}.
\\
\indent To illustrate Lemma \ref{lemma-tight-relaxation-if-convex}, observe that the assumptions are satisfied for $f(t) = \exp(a \cdot t) + b \cdot t$, with $a$ and $b$ taken as positive values, and when the ambiguity sets are defined through confidence intervals on the expected costs, i.e. for any arc $(i, j) \in \mathcal{A}$:
\begin{equation*}
	\mathcal{P}_{ij} = \{ p \in \mathcal{P}([\deltainf_{ij}, \deltasup_{ij}]) : \hspace{0.2cm} \mathbb{E}_{X \sim p}[ X ] \in [\alpha_{ij}, \beta_{ij}] \},
\end{equation*}
with $\alpha_{ij} \leq \beta_{ij}$. Further note that adding upper bounds on the mean deviation or on higher order moments in the definition of the ambiguity sets does not alter the conclusion of Lemma \ref{lemma-tight-relaxation-if-convex}. We move on to another situation of interest where \eqref{eq-ideal-robust-problem} and \eqref{eq-robust-problem} can be shown to be equivalent.
\begin{lemma}
\label{lemma-tight-relaxation-if-higher-order-convex}
	Take $K \in \mathbb{N}$. Suppose that:
	\begin{itemize}
		\item{$\mathcal{G}$ is a single-path graph,}
		\item{$f(\cdot)$ is $C^{K+1}$ and $f^{(K+1)}(t) > 0 \; \forall t$ or $f^{(K +1)}(t) < 0 \; \forall t$,}
		\item{For any arc $(i, j) \in \mathcal{A}$:
		\begin{equation*}
			\mathcal{P}_{ij} = \{ p \in \mathcal{P}([\deltainf_{ij}, \deltasup_{ij}]) : \hspace{0.2cm} \mathbb{E}_{X \sim p}(X) = m_{\mathrm{ij}}^1, \cdots, \mathbb{E}_{X \sim p}(X^K) = m_{\mathrm{ij}}^K \},
		\end{equation*}	
		where $m_{\mathrm{ij}}^1, \cdots, m_{\mathrm{ij}}^K$ are non-negative.}
	\end{itemize}		
	Then \eqref{eq-ideal-robust-problem} and \eqref{eq-robust-problem} are equivalent.
\end{lemma}
The proof is deferred to the online supplement, Section \ref{section-proof-lemma-tight-relaxation-if-higher-order-convex}.
\\
\indent When $\mathcal{G}$ is a single-path graph, the optimal value of \eqref{eq-ideal-robust-problem} corresponds to the worst-case risk function when following this path, given that the arc cost distributions are only known to lie in the ambiguity sets. While it is a priori unclear how to compute this quantity, Proposition \ref{lemma-quality-approx-robust} of Section \ref{section-discretization-robust} establishes that the optimal value of \eqref{eq-robust-problem} can be determined with arbitrary precision provided the inner optimization problems appearing in the discretization scheme of Section \ref{section-discretization-robust} can be computed numerically. Hence, even in this seemingly simplistic situation, the equivalence between \eqref{eq-ideal-robust-problem} and \eqref{eq-robust-problem} is an important fact to know as it has significant computational implications. Lemma \ref{lemma-tight-relaxation-if-higher-order-convex} shows that, when the risk function is $(K+1)$th order convex or concave and when the arc cost distributions are only known through the first $K$-order moments, \eqref{eq-ideal-robust-problem} and \eqref{eq-robust-problem} are in fact equivalent. For this particular class of ambiguity sets, the inner optimization problems of the discretization scheme of Section \ref{section-discretization-robust} can be solved using semidefinite programming, see \cite{bertsimas2005optimal}.

\paragraph{Bounding the gap between the optimal values of \eqref{eq-ideal-robust-problem} and \eqref{eq-robust-problem}.}
It turns out that, for a particular subclass of risk functions, we can bound the gap between the optimal values of \eqref{eq-ideal-robust-problem} and \eqref{eq-robust-problem} uniformly over all graphs and ambiguity sets.

\begin{lemma}
\label{lemma-bound-gap-robust-relaxation}
	Denote the optimal value of \eqref{eq-ideal-robust-problem} (resp. \eqref{eq-robust-problem}) by $v^*$ (resp. $v$). 
	\\
	If there exists $\gamma, a >0$ and $\beta, b$ such that one of the following conditions holds:
	\begin{itemize}
		\item $\gamma \cdot t + \beta \geq f(t) \geq a \cdot t + b \quad \forall t \leq T$,
		\item $\gamma \cdot \exp(t) + \beta \geq f(t) \geq a \cdot \exp(t) + b \quad \forall t \leq T$,
		\item $- \gamma \cdot \exp(- t) + \beta \geq f(t) \geq - a \cdot \exp(- t) + b \quad \forall t \leq T$,
	\end{itemize}		
	then $v^* \geq v \geq \frac{a}{\gamma} \cdot (v^* - \beta) + b$.
\end{lemma}
The proof is deferred to the online supplement, Section \ref{section-proof-lemma-bound-gap-robust-relaxation}.

\subsection{Solution methodology}
\label{section-solution-methodology-robust}
We proceed as in Section \ref{section-solution-methodology-nominal} and compute an approximate \emph{Markov} policy solution to \eqref{eq-dp-robust}. The computational challenges faced when solving the nominal problem carry over to the robust counterpart, but with additional difficulties to overcome. Specifically, the continuity of the problem leads us to build a discrete approximation in Section \ref{section-discretization-robust} similar to the one developed for the nominal approach. We also extend the label-setting algorithm of Section \ref{section-algorithm-nominal} to tackle the potential existence of cycles at the beginning of Section \ref{section-algorithm-robust}. However, the presence of an inner optimization problem in \eqref{eq-dp-robust} is a distinctive feature of the robust problem which poses a new computational challenge. As a result, and in contrast with the situation for the nominal problem where this optimization problem reduces to a convolution product, it is not a priori obvious how to solve the discretization scheme numerically, let alone efficiently. As can be expected, the exact form taken by the ambiguity sets has a major impact on the computational complexity of the inner optimization problem. In an effort to mitigate the computational burden, we restrict our attention to a subclass of ambiguity sets defined by confidence intervals on piecewise affine statistics in Section \ref{section-ambiguity-set}. While this simplification might seem restrictive, we show that this subclass displays significant modeling power. Finally, we develop two general-purpose algorithms in Section \ref{section-algorithm-robust} for this particular subclass of ambiguity sets. The computational attractiveness of these approaches hinges on the existence of a data structure, presented in Section \ref{section-dynamic-convex-hull}, maintaining the convex hull of a dynamic set of points efficiently. The mechanism behind this data structure can be regarded as the counterpart of the online fast Fourier scheme for the nominal approach. In particular, using the algorithms developed in this section, we can compute:
\begin{itemize}
	\item an $\epsilon$-approximate solution to \eqref{eq-robust-problem} in 
	$$
		O(  \frac{ |\mathcal{A}| \cdot (T - T^r_f) + |\mathcal{V}|^2 \cdot \deltasup }{ \Delta t} \cdot \log( \frac{ \deltasup - \deltainf }{ \Delta t }  ) \cdot \log(  \frac{ |\mathcal{V}| + \frac{T - T^r_f}{\deltainf}  }{ \epsilon }   )  )
	$$
	computation time when the arc costs only take on values that are multiple of $\Delta t > 0$ and for any continuous risk function $f(\cdot)$ satisfying Theorem \ref{lemma-LET-is-opt-robust}. This also applies when the objective is to maximize the probability of completion within budget and even simplifies to $O( |\mathcal{A}| \cdot \frac{  T }{ \Delta t} \cdot \log( \frac{ \deltasup - \deltainf }{ \Delta t }  ) \cdot \log(  \frac{ T }{ \epsilon \cdot \deltainf } ) )$,
	\item an $\epsilon$-approximate solution to \eqref{eq-robust-problem} in 
	$$
		O( \frac{ (|\mathcal{V}| + \frac{T - T^r_f}{\deltainf} )^2 \cdot ( |\mathcal{A}| \cdot (T - T^r_f) + |\mathcal{V}|^2 \cdot \deltasup  ) }{\epsilon} \cdot \log( \frac{ (|\mathcal{V}| + \frac{T - T^r_f}{\deltainf}) \cdot (\deltasup - \deltainf) }{ \epsilon }  ) \cdot \log(  \frac{ |\mathcal{V}| + \frac{T - T^r_f}{\deltainf}  }{ \epsilon }   ) )
	$$
	computation time when the risk function is Lipschitz on compact sets.
\end{itemize}

\subsubsection{Discretization scheme}
\label{section-discretization-robust}
For each node $i \in \mathcal{V}$, we approximate $u_i(\cdot)$ by a piecewise affine continuous function $u^{\Delta t}_i(\cdot)$ of uniform stepsize $\Delta t$. This is in contrast with Section \ref{section-discretization-nominal} where we use a piecewise constant approximation. This change is motivated by computational considerations. Essentially, the continuity of $u^{\Delta t}_i(\cdot)$ guarantees strong duality for the inner optimization problem appearing in \eqref{eq-dp-robust}. Similarly as for the nominal problem, we only need to approximate $u_i(\cdot)$ for a remaining budget larger than $k^{r, \text{min}}_i \cdot \Delta t$, for $k^{r, \text{min}}_i = \floor*{ \frac{T^r_f - (|\mathcal{V}| - \text{level}(i, \mathcal{T}^r) + 1) \cdot \deltasup}{\Delta t} }$, where $\text{level}(i, \mathcal{T}^r)$ is the level of node $i$ in $\mathcal{T}^r$. Specifically, we use the approximation:
\begin{equation}
	\label{eq-discretize-approximation-robust}
	\begin{aligned}
	 & u^{\Delta t}_i(t) = (1 - \frac{t}{\Delta t} + \floor*{ \frac{t}{\Delta t} } ) \cdot u^{\Delta t}_i( \floor*{ \frac{t}{\Delta t} } \cdot \Delta t ) + (\frac{t}{\Delta t} - \floor*{ \frac{t}{\Delta t} } ) \cdot u^{\Delta t}_i(\ceil*{ \frac{t}{\Delta t} } \cdot \Delta t) && i \in \mathcal{V}, t \in [k^{r, \text{min}}_i \cdot \Delta t, T] \\
	 & \pi^{\Delta t}(i, t) = \pi^{\Delta t}(i, \floor*{ \frac{t}{\Delta t} } \cdot \Delta t ) && i \neq d, t \in [k^{r, \text{min}}_i \cdot \Delta t, T],
	\end{aligned}
\end{equation}
and the values at the mesh points are determined by the set of equalities:
\begin{equation}
	\label{eq-discretization-scheme-robust}
	\begin{aligned}
	 & u^{\Delta t}_d( k \cdot \Delta t) = f( k \cdot \Delta t ) \quad && k = k^{r, \text{min}}_d, \cdots, \floor*{ \frac{T}{\Delta t} } \\
	 & u^{\Delta t}_i( k \cdot \Delta t ) =  \max\limits_{j \in \mathcal{V}(i)} \inf\limits_{p_{ij} \in \mathcal{P}_{ij}} \int_0^{\infty} p_{ij}(\omega) \cdot u^{\Delta t}_j( k \cdot \Delta t - \omega) \mathrm{d}\omega \quad && i \neq d, k = \floor*{ \frac{T^r_f}{\Delta t} }, \cdots, \floor*{ \frac{T}{\Delta t} }\\
	 & \pi^{\Delta t}(i, k \cdot \Delta t ) \in \argmax\limits_{j \in \mathcal{V}(i)} \inf\limits_{p_{ij} \in \mathcal{P}_{ij}} \int_0^{\infty} p_{ij}(\omega) \cdot u^{\Delta t}_j( k \cdot \Delta t - \omega) \mathrm{d}\omega \quad && i \neq d, k = \floor*{ \frac{T^r_f}{\Delta t} }, \cdots, \floor*{ \frac{T}{\Delta t} } \\
	& u^{\Delta t}_i( k \cdot \Delta t ) =  \max\limits_{j \in \mathcal{T}^r(i)} \inf\limits_{p_{ij} \in \mathcal{P}_{ij}} \int_0^{\infty} p_{ij}(\omega) \cdot u^{\Delta t}_j( k \cdot \Delta t - \omega) \mathrm{d}\omega \quad && i \neq d, k = k^{r, \text{min}}_i, \cdots, \floor*{ \frac{T^r_f}{\Delta t} } - 1 \\
	& \pi^{\Delta t}(i, k \cdot \Delta t ) \in \argmax\limits_{j \in \mathcal{T}^r(i)} \inf\limits_{p_{ij} \in \mathcal{P}_{ij}} \int_0^{\infty} p_{ij}(\omega) \cdot u^{\Delta t}_j( k \cdot \Delta t - \omega) \mathrm{d}\omega \quad && i \neq d, k = k^{r, \text{min}}_i, \cdots, \floor*{ \frac{T^r_f}{\Delta t} } - 1.
	\end{aligned}
\end{equation} 
As we did for the nominal problem, we can quantify the quality of $\pi^{\Delta t}$ as an approximate solution to \eqref{eq-robust-problem} as a function of the regularity of the risk function.

\begin{proposition}
	\label{lemma-quality-approx-robust}
	Consider a solution to the global discretization scheme \eqref{eq-discretize-approximation-robust} and \eqref{eq-discretization-scheme-robust}, $(\pi^{\Delta t}, (u^{\Delta t}_i(\cdot))_{i \in \mathcal{V}})$. We have:
	\begin{enumerate}
		\item{If $f(\cdot)$ is non-decreasing, the functions $(u^{\Delta t}_i(\cdot))_{i \in \mathcal{V}}$ converge pointwise almost everywhere to $(u_i(\cdot))_{i \in \mathcal{V}}$ as $\Delta t \rightarrow 0$.}
		\item{If $f(\cdot)$ is continuous, the functions $(u^{\Delta t}_i(\cdot))_{i \in \mathcal{V}}$ converge uniformly to $(u_i(\cdot))_{i \in \mathcal{V}}$ and $\pi^{\Delta t}$ is a $o(1)$-approximate optimal solution to \eqref{eq-robust-problem} as $\Delta t \rightarrow 0$.}
		\item{If $f(\cdot)$ is Lipschitz on compact sets (e.g. if $f(\cdot)$ is $C^1$), the functions $(u^{\Delta t}_i(\cdot))_{i \in \mathcal{V}}$ converge uniformly to $(u_i(\cdot))_{i \in \mathcal{V}}$ at speed $\Delta t$ and $\pi^{\Delta t}$ is a $O(\Delta t)$-approximate optimal solution to \eqref{eq-robust-problem}
		 as $\Delta t \rightarrow 0$.}	
	\end{enumerate}	
\end{proposition}
The proof is deferred to the online supplement, Section \ref{section-proof-lemma-quality-approx-robust}.

\subsubsection{Ambiguity sets}
\label{section-ambiguity-set}
For computational tractability, we restrict our attention to the following subclass of ambiguity sets.

\begin{definition}
\label{def-setprobability}
 For any arc $(i,j) \in \mathcal{A}$:
\begin{equation*}
	\mathcal{P}_{ij} = \{ p \in \mathcal{P}([\deltainf_{ij}, \deltasup_{ij}]) : \hspace{0.2cm} \mathbb{E}_{X \sim p}[ g^{ij}_q(X) ] \in [\alpha^{ij}_q, \beta^{ij}_q], \; q = 1, \cdots, Q_{ij}  \},
\end{equation*}
where:
\begin{itemize}
	\item{the functions $(g^{ij}_q(\cdot))_{q = 1, \cdots, Q_{ij}}$ are piecewise affine with a finite number of pieces on $[\deltainf_{ij}, \deltasup_{ij}]$ and such that $\mathcal{P}_{ij}$ is closed for the weak topology,}
	\item{$ -\infty \leq \alpha^{ij}_q \leq \beta^{ij}_q \leq \infty$ for $q = 1, \cdots, Q_{ij}$.}
\end{itemize}
\end{definition}
Note that Definition \ref{def-setprobability} allows to model one-sided constraints by either taking $\alpha^{ij}_q  = - \infty$ or $\beta^{ij}_q  = \infty$. Moreover, we point out that the functions $(g^{ij}_q(\cdot))_{q = 1, \cdots, Q_{ij}}$ need not be continuous to guarantee closeness of $\mathcal{P}_{ij}$. For instance, the constraints $\mathbb{E}_{X \sim p}[ 1_{X \in S} ] \leq \beta$ and $\mathbb{E}_{X \sim p}[ 1_{X \in S'} ] \geq \beta$, for $S$ (resp. $S'$) an open (resp. a closed) set, are perfectly valid. In terms of modeling power, Definition \ref{def-setprobability} allows to have constraints on standard statistics, such as the mean value, the mean absolute deviation, and the median, but also to capture distributional asymmetry, through constraints on any quantile or of the type $\mathbb{E}_{X \sim p}[ X \cdot 1_{X > \theta} ] \leq \beta$, and to incorporate higher-order information, e.g. the variance or the skewness, since continuous functions can be approximated arbitrarily well by piecewise affine functions on a compact set. Finally, observe that Definition \ref{def-setprobability} also allows to model the situation where $c_{ij}$ only takes values in a prescribed finite set $S$ through the constraint $\mathbb{E}_{X \sim p}[ 1_{X \in S} ] \geq 1$.

\paragraph{Data-driven ambiguity sets.} Ambiguity sets of the form introduced in Definition \ref{def-setprobability} can be built using a combination of prior knowledge and historical data. To illustrate, suppose that, for any arc $(i, j) \in \mathcal{A}$, we have observed $n_{ij}$ samples drawn from the corresponding arc cost distribution. Setting aside computational aspects, there is an inherent trade-off at play when designing ambiguity sets with this empirical data: using more statistics and/or narrowing the confidence intervals will improve the quality of the guarantee on the risk function provided by the robust approach, but will, on the other hand, deteriorate the probability that this guarantee holds. Assuming we are set on which statistics to use, the trade-off is simple to resolve as far as confidence intervals are concerned. Using Hoeffding's and Boole's inequalities, the confidence interval for statistics $q$ of arc $(i, j)$ should be centered at the empirical average and have width $\epsilon^{ij}_q$ determined by:
$$
	\frac{ \epsilon^{ij}_q }{ \max\limits_{[\deltainf_{ij}, \deltasup_{ij}]} g^{ij}_q - \min\limits_{[\deltainf_{ij}, \deltasup_{ij}]} g^{ij}_q } = \sqrt{ \frac{ \log( \frac{2}{\epsilon} \cdot \sum\limits_{(i, j) \in \mathcal{A}} Q_{ij}  ) }{  2 n_{ij} } }
$$
in order to achieve a probability $1 - \epsilon$ that the guarantee holds. Choosing which statistics to use is a more complex endeavor. It is not even clear whether using more statistics is beneficial since the confidence intervals jointly expand with $\sum\limits_{(i, j) \in \mathcal{A}} Q_{ij}$. Numerical evidence presented in Section \ref{section-numericalresults} suggests that low-order statistics, such as the mean, tend to be more informative when only few samples are available. Conversely, as sample sizes get very large, incorporating higher-order information seem to improve the quality of the strategy derived. In the limit where the statistics can be computed exactly, we should use as many statistics as possible. This observation is supported by the following lemma.

\begin{lemma}
\label{lemma-convergence-more-moments}
For any arc $(i, j) \in \mathcal{A}$, consider $(\mathcal{P}^k_{ij})_{k \in \mathbb{N}}$, a sequence of nested ambiguity sets satisfying Assumption \ref{assumption-compact-ambiguity-set}. If $f(\cdot)$ is continuous, then the optimal value of the robust problem \eqref{eq-robust-problem} when the uncertainty sets are taken as $( \mathcal{P}^k_{ij} )_{(i, j) \in \mathcal{A}} $ monotonically converges to the optimal value of \eqref{eq-robust-problem} when the uncertainty sets are taken as $( \cap_{ k \in \mathbb{N} } \mathcal{P}^k_{ij} )_{(i, j) \in \mathcal{A}}$ as $k \rightarrow \infty$.
\\
In particular, if $\cap_{ k \in \mathbb{N} } \mathcal{P}^k_{ij}$ is a singleton for all arcs $(i, j) \in \mathcal{A}$, then the optimal value of the robust problem converges to the value of the nominal problem \eqref{eq-nominal-problem}.
\end{lemma}
The proof is deferred to the online supplement, Section \ref{section-proof-lemma-convergence-more-moments}.
\\
\indent Using the Weierstrass approximation theorem, observe that the second part of Lemma \ref{lemma-convergence-more-moments} applies in particular when the ambiguity sets $\mathcal{P}^k_{ij}$ are defined by the first $k$-order moments.

\subsubsection{Solution procedures}
\label{section-algorithm-robust}
We develop two general-purpose methods to compute a solution to the discretization scheme \eqref{eq-discretization-scheme-robust} for the class of ambiguity sets identified in Section \ref{section-ambiguity-set}. The first method, based on the ellipsoid algorithm, computes an $\epsilon-$approximate solution to \eqref{eq-discretization-scheme-robust} with worst-case complexity:
$$
	O(  \frac{ |\mathcal{A}| \cdot (T - T^r_f) + |\mathcal{V}|^2 \cdot \deltasup }{ \Delta t} \cdot \log( \frac{ \deltasup - \deltainf }{ \Delta t }  ) \cdot \log(  \frac{ |\mathcal{V}| + \frac{T - T^r_f}{\deltainf}  }{ \epsilon }   )  ),
$$
provided $f(\cdot)$ is continuous and where the hidden factors are linear in the number of pieces of each statistic and polynomial in the number of statistics. We remind the reader that the complexity of solving the discretization scheme \eqref{eq-discretization-scheme-nominal} for the nominal problem is $O( |\mathcal{A}| \cdot  \frac{ T - T_f }{\Delta t} \cdot \log^2( \frac{\deltasup}{\Delta t}) + |\mathcal{V}|^2 \cdot \frac{\deltasup}{\Delta t} \cdot \log( |\mathcal{V}| \cdot \frac{\deltasup}{\Delta t}  ) )$ when using zero-delay convolution. While these bounds are not directly comparable because some of the parameters required to specify a robust instance are not relevant for a nominal instance and vice versa, we point out that they share many similarities, including the almost linear dependence on $\frac{1}{\Delta t}$. The second method, based on delayed column generation and warm starting techniques, is more practical but has worst-case complexity exponential in $\frac{1}{ \Delta t }$. We stress that none of these approaches can be used to solve the nominal problem as the latter is not a particular case of the robust problem for the restricted class of ambiguity sets defined in Section \ref{section-ambiguity-set}. Indeed, characterizing a single distribution generally requires infinitely many moment constraints.

\paragraph{Label-setting approach.}
To cope with the potential existence of cycles, we remark that the label-setting approach developed for the nominal approach trivially extends to the robust setting. Similarly as for the nominal problem, we proceed in three steps to solve \eqref{eq-discretization-scheme-robust}. First, we compute $T^r_f$. Next, we compute the values $u^{\Delta t}_i( k \cdot \Delta t)$ for $k \in \{k^{r, \text{min}}_i, \cdots, \floor*{ \frac{T^r_f}{\Delta t} } - 1 \}$ starting at node $i = d$ and traversing the tree $\mathcal{T}^r$ in a breadth-first fashion. Finally, we compute the values $u^{\Delta t}_i( k \cdot \Delta t)$ for $k \in \{ \floor*{ \frac{T^r_f}{\Delta t} } + m \cdot \floor*{ \frac{\deltainf}{ \Delta t } }, \cdots, \floor*{ \frac{T^r_f}{\Delta t} } + (m+1) \cdot \floor*{ \frac{\deltainf}{ \Delta t } } \}$ for all nodes $i \in \mathcal{V}$ by induction on $m$. Of course, an efficient procedure solving the inner optimization problem of \eqref{eq-discretization-scheme-robust} is a prerequisite for carrying out the last two steps. This will be our focus in the remainder of this section.

\paragraph{Solving the Inner Optimization Problem.}
Consider any arc $(i, j) \in \mathcal{A}$. We need to solve, at each step $k \in \{ k^{r, \text{min}}_i, \cdots, \floor*{ \frac{T}{\Delta t} } \}$, the optimization problem:
\begin{equation}
	\label{eq-primal-inner-problem}
	\begin{aligned}
	& \inf_{ p \in \mathcal{P}([\deltainf_{ij}, \deltasup_{ij}]) } 
	& & \mathbb{E}_{X \sim p}[ u^{\Delta t}_j(k \cdot \Delta t - X) ] \\
	& \text{subject to}
	& & \mathbb{E}_{X \sim p}[ g^{ij}_q(X) ] \in [\alpha^{ij}_q, \beta^{ij}_q] \quad  q = 1, \cdots, Q_{ij}.  \\
	\end{aligned}
\end{equation}	
Since the set of non-negative measures on $[\deltainf_{ij}, \deltasup_{ij}]$ is a cone, \eqref{eq-primal-inner-problem} can be cast as a conic linear problem. As a result, standard conic duality theory applies and the optimal value of \eqref{eq-primal-inner-problem} can be equivalently computed by solving a dual optimization problem which turns out to be easier to study. For a thorough exposition of the duality theory of general conic linear problems, the reader is referred to \cite{Shapiro00onduality}. To simplify the presentation, we assume that $(\alpha^{ij}_q)_{q = 1, \cdots, Q_{ij}}$ and $(\beta^{ij}_q)_{q = 1, \cdots, Q_{ij}}$ are all finite quantities but this is by no means a limitation of our approach. 

\begin{lemma}
\label{lemma-dual-inner-problem}
The optimization problem \eqref{eq-primal-inner-problem} has the same optimal value as the semi-infinite linear program:
\begin{equation}
	\label{eq-dual-inner-problem}
	\begin{aligned} 
	& \sup\limits_{\substack{ z \in \mathbb{R} \\ y_1, \cdots, y_{Q_{ij}} \in \mathbb{R} \\ x_1, \cdots, x_{Q_{ij}} \in \mathbb{R} }  }
	& &  z + \sum_{q = 1}^{ Q_{ij} } ( \alpha^{ij}_q \cdot x_q  - \beta^{ij}_q \cdot y_q ) \\
	& \text{subject to}
	& & z + \sum_{q = 1}^{ Q_{ij} } ( x_q  - y_q ) \cdot g^{ij}_q(\omega) \leq u^{\Delta t}_j(k \cdot \Delta t - \omega) \quad \forall \omega \in [\deltainf_{ij}, \deltasup_{ij}] \\
	&
	& & y_q, x_q \geq 0 \quad q = 1, \cdots, Q_{ij}.
	\end{aligned}
\end{equation}	
	\proof{Proof}
	We take the Lagrangian dual of \eqref{eq-primal-inner-problem}. Since $u^{\Delta t}_j(\cdot)$ is continuous by construction and $\mathcal{P}_{ij}$ is not empty and compact by assumption, Proposition 3.1 in \cite{Shapiro00onduality} shows that strong duality holds.
	\Halmos
	\endproof
\end{lemma}
Because the functions $(g^{ij}_q(\cdot))_{q = 1, \cdots, Q_{ij}}$ are all piecewise affine, we can partition $[\deltainf_{ij}, \deltasup_{ij}]$ into $R_{ij}$ non-overlapping intervals $(I_r)_{r = 1, \cdots, R_{ij}}$ such that the functions $(g^{ij}_q(\cdot))_{q = 1, \cdots, Q_{ij}}$ are all affine on $I_r$ for any $r \in \{ 1, \cdots, R_{ij} \}$, i.e.:
$$
	g^{ij}_q( \omega ) =  a^{ij}_{q, r} \cdot \omega + b^{ij}_{q, r} \; \text{ if } \; \omega \in I_r
$$
for any $q \in \{1, \cdots, Q_{ij} \}$ and $\omega \in [\deltainf_{ij}, \deltasup_{ij}]$. This decomposition enables us to show that the feasible region of \eqref{eq-dual-inner-problem} can be described with finitely many inequalities.

\begin{lemma}
\label{lemma-dual-inner-problem-finite-lp}
The semi-infinite linear program \eqref{eq-dual-inner-problem} can be reformulated as the following finite linear program:
\begin{equation}
	\label{eq-dual-inner-problem-finite-lp}
  \begin{alignedat}{3}
    \sup\limits_{\substack{ z \in \mathbb{R} \\ y_1, \cdots, y_{Q_{ij}} \in \mathbb{R} \\ x_1, \cdots, x_{Q_{ij}} \in \mathbb{R} }  } & \; \; z + \sum_{q = 1}^{ Q_{ij} } ( \alpha^{ij}_q \cdot x_q  - \beta^{ij}_q \cdot y_q ) &  \\
    \text{subject to} & \; \; z + \sum_{q = 1}^{ Q_{ij} } ( x_q  -  y_q) \cdot (a^{ij}_{q, r} \cdot l \cdot \Delta t + b^{ij}_{q, r}) \leq u^{\Delta t}_j( (k - l) \cdot \Delta t) \quad  & l = \ceil*{ \frac{\inf(I_r)}{\Delta t} }, \cdots, \floor*{ \frac{\sup( I_r )}{\Delta t} }  \\
     & \; \; &  r = 1, \cdots, R_{ij} \\
     & \; \;  z + \sum_{q = 1}^{ Q_{ij} } ( x_q - y_q  ) \cdot (a^{ij}_{q, r} \cdot \sup(I_r) + b^{ij}_{q, r}) \leq u^{\Delta t}_j( k  \cdot \Delta t - \sup(I_r)) &  r = 1, \cdots, R_{ij}  \\
     & \; \;  z + \sum_{q = 1}^{ Q_{ij} } ( x_q - y_q ) \cdot (a^{ij}_{q, r} \cdot \inf(I_r) + b^{ij}_{q, r}) \leq u^{\Delta t}_j( k  \cdot \Delta t - \inf(I_r)) &  r = 1, \cdots, R_{ij} \\
     & \; \; y_q, x_q \geq 0 \quad q = 1, \cdots, Q_{ij}. &
  \end{alignedat}
\end{equation}
\proof{Proof}
	Take $z, y_1, \cdots, y_{Q_{ij}}, x_1, \cdots, x_{Q_{ij}} \in \mathbb{R}$ and $r \in \{ 1, \cdots, R_{ij} \}$. Since the function $\omega \rightarrow z + \sum_{q = 1}^{ Q_{ij} } ( x_q - y_q ) \cdot g^{ij}_q(\omega)$ is affine on $I_r$, this function lies below the continuous piecewise affine function $u^{\Delta t}_j( k  \cdot \Delta t - \cdot)$ on $I_r$ if and only if it lies below $u^{\Delta t}_j( k  \cdot \Delta t - \cdot)$ at every breakpoint of $u^{\Delta t}_j( k  \cdot \Delta t - \cdot)$ on $\bar{I_r}$ and at the boundary points of $\bar{I_r}$. Since the collection of intervals $(I_r)_{r = 1, \cdots, R_{ij}}$ forms a partition of $[\deltainf_{ij}, \deltasup_{ij}]$, this establishes the claim.
\Halmos
\endproof
\end{lemma}
While \eqref{eq-dual-inner-problem-finite-lp} is a finite linear program and can thus be solved with an interior point algorithm, the large number of constraints calls for an efficient separation oracle, which we develop next, and the use of the ellipsoid algorithm. The key is to refine the idea of Lemma \ref{lemma-dual-inner-problem-finite-lp}. Specifically, for any $r \in \{1, \cdots, R_{ij} \}$ and $l \in \{ \ceil*{ \frac{\inf(I_r)}{\Delta t} }, \cdots, \floor*{ \frac{\sup( I_r )}{\Delta t} } \}$, the constraint
$$
	z + \sum_{q = 1}^{ Q_{ij} } ( x_q  -  y_q) \cdot (a^{ij}_{q, r} \cdot l \cdot \Delta t + b^{ij}_{q, r}) \leq u^{\Delta t}_j( (k - l) \cdot \Delta t)
$$
does not limit the feasible region if $(l \cdot \Delta t, u^{\Delta t}_j( (k - l) \cdot \Delta t) )$ is not an extreme point of the upper convex hull of $\{ ( m \cdot \Delta t, u^{\Delta t}_j( (k - m) \cdot \Delta t)), \; m = \floor*{ \frac{\inf(I_r)}{\Delta t} }, \cdots, \ceil*{ \frac{\sup(I_r)}{\Delta t} } \}$. Denote by $\mathcal{L}^{k, r}_{ij}$ the subset of integers $l$ such that $(l \cdot \Delta t, u^{\Delta t}_j( (k - l) \cdot \Delta t) )$ is such an extreme point. Observe that the function
$$
	l \rightarrow u^{\Delta t}_j( (k - l) \cdot \Delta t) - [z + \sum_{q = 1}^{ Q_{ij} } ( x_q  -  y_q) \cdot (a^{ij}_{q, r} \cdot l \cdot \Delta t + b^{ij}_{q, r}) ] 
$$
is convex on $\mathcal{L}^{k, r}_{ij}$, therefore a minimizer of this function can be found by binary search. As a result, all we need to be able to separate efficiently for the subset of constraints:
$$
	z + \sum_{q = 1}^{ Q_{ij} } ( x_q  -  y_q) \cdot (a^{ij}_{q, r} \cdot l \cdot \Delta t + b^{ij}_{q, r}) \leq u^{\Delta t}_j( (k - l) \cdot \Delta t) \quad  l = \ceil*{ \frac{\inf(I_r)}{\Delta t} }, \cdots, \floor*{ \frac{\sup( I_r )}{\Delta t} } 
$$
is a means to perform binary search on $\mathcal{L}^{k, r}_{ij}$ efficiently. We defer the presentation of a data structure designed for this purpose to Section \ref{section-dynamic-convex-hull} and make the following assumption to conclude the computational study.

\begin{assumption} 
	\label{assumption-data-structure}
	For any two integers $L, L'$ such that $\ceil*{ \frac{ \deltainf_{ij} }{ \Delta t } } \leq L < L' \leq \floor*{ \frac{ \deltasup_{ij} }{ \Delta t } }$, there exists a data structure that can maintain, dynamically as $k$ increases from $k = k^{r, \text{min}}_i$ to $k = \floor*{ \frac{T}{\Delta t} }$, a description of the upper convex hull of $\{ ( l \cdot \Delta t, u^{\Delta t}_j( (k -l) \cdot \Delta t)), \; l = L, \cdots, L' \}$ allowing to perform binary search on the first coordinate of the extreme points with a global complexity $O( (\frac{T}{\Delta t} - k^{r, \text{min}}_i) \cdot \log( \frac{\deltasup - \deltainf}{\Delta t}  ) )$.
\end{assumption}
Equipped with a data structure satisfying Assumption \ref{assumption-data-structure}, the separation oracle has runtime complexity $O( \log( \frac{ \deltasup - \deltainf }{\Delta t} ) )$  given that there are at most $\floor*{ \frac{ \deltasup_{ij} }{ \Delta t } } - \ceil*{ \frac{ \deltainf_{ij} }{ \Delta t } }$ extreme points at any step $k$. Using the ellipsoid algorithm, we can compute the optimal value of \eqref{eq-primal-inner-problem} with precision $\epsilon$ in $O( \log( \frac{ \deltasup - \deltainf }{\Delta t} ) \cdot \log( \frac{1}{\epsilon} ) )$ running time, where the hidden factors are polynomial in $Q_{ij}$ and linear in $R_{ij}$. We point out that relying on a data structure satisfying Assumption \ref{assumption-data-structure} is critical to achieve this complexity: recomputing the upper convex hull from scratch at every time step $k$ would increase the complexity to $O( \frac{ \deltasup - \deltainf }{\Delta t}  \cdot \log( \frac{1}{\epsilon} ) )$ (achieved using, for instance, Andrew's monotone chain convex hull algorithm).

\subparagraph{Practical general purpose method.}
Due to the limited practicability of the ellipsoid algorithm, we have developed another method based on delayed column generation to solve the inner optimization problem. To simplify the presentation, we assume that $(\inf(I_r))_{r = 1, \cdots, R_{ij}}$ and $(\sup(I_r))_{r = 1, \cdots, R_{ij}}$ are all multiples of $\Delta t$. Since \eqref{eq-dual-inner-problem-finite-lp} is a linear program with a non-empty feasible set, we can equivalently compute its value by solving the dual optimization problem given by:
\begin{equation}
	\label{eq-primal-after-manipulations-inner-problem}
	\begin{aligned}
	& \inf\limits_{ p_0, \cdots, p_L \in \mathbb{R} } 
	& & \sum_{l = 0, \cdots, L} p_l \cdot u^{\Delta t}_j( (k - l) \cdot \Delta t - \deltainf_{ij}) \\
	& \text{subject to}
	& & \sum_{l = 0, \cdots, L} p_l \cdot g^{ij}_q( l \cdot \Delta t + \deltainf_{ij}) \in [\alpha^{ij}_q, \beta^{ij}_q] \quad  q = 1, \cdots, Q_{ij}  \\
	&
	& & \sum_{l = 0, \cdots, L} p_l  = 1 \\
	&
	& & p_l \geq 0 \quad l = 0, \cdots, L, 
	\end{aligned}
\end{equation}	
where $L = \frac{ \deltasup_{ij} - \deltainf_{ij} }{ \Delta t }$. Observe that the feasible set of the linear program \eqref{eq-primal-after-manipulations-inner-problem} does not change across steps $k = k^{r, \text{min}}_i, \cdots, \floor*{ \frac{T}{\Delta t} }$. Hence, we can warm start the primal simplex algorithm with the optimal solution found at the previous step. Furthermore, the separation oracle developed for the dual optimization problem can also be used as a subroutine for delayed column generation.

\subparagraph{Faster procedure when the mean is the only statistics.}
If the ambiguity sets are only defined through a confidence interval on the mean value, i.e.:
\begin{equation*}
	\mathcal{P}_{ij} = \{ p \in \mathcal{P}([\deltainf_{ij}, \deltasup_{ij}]) : \hspace{0.2cm} \mathbb{E}_{X \sim p}[ X ] \in [\alpha^{ij}, \beta^{ij}] \},
\end{equation*}
then \eqref{eq-dual-inner-problem-finite-lp} can be solved to optimality in $O(\log(\frac{\deltasup_{ij} - \deltainf_{ij}}{\Delta t}))$ computation time without resorting to the ellipsoid algorithm. First observe that \eqref{eq-dual-inner-problem-finite-lp} simplifies to:
\begin{equation}
	\label{eq-primal-inner-problem-only-mean}
	\begin{aligned}
	& \sup\limits_{ z, y, x \in \mathbb{R} } 
	& & z + \alpha^{ij} \cdot x - \beta^{ij} \cdot y \\
	& \text{subject to}
	& & z + (x - y) \cdot l \cdot \Delta t \leq u^{\Delta t}_j( (k - l) \cdot \Delta t ), \quad l = \ceil*{ \frac{ \deltainf_{ij} }{\Delta t} }, \cdots, \floor*{ \frac{ \deltasup_{ij} }{\Delta t} } \\
	& 
	& & z + (x - y) \cdot \deltasup_{ij} \leq u^{\Delta t}_j( k \cdot \Delta t - \deltasup_{ij}) \\
	& 
	& & z + (x - y) \cdot \deltainf_{ij} \leq u^{\Delta t}_j( k \cdot \Delta t - \deltainf_{ij}) \\
	&
	& & y, x \geq 0.
	\end{aligned}
\end{equation}	
As it turns out, we can identify an optimal feasible basis to \eqref{eq-primal-inner-problem-only-mean} by direct reasoning.
\begin{lemma} 
	\label{lemma-optimal-basis-inner-problem-only-mean}
	An optimal solution to \eqref{eq-primal-inner-problem-only-mean} can be found by performing three binary searches on the first coordinate of the extreme points of the upper convex hull of 
	$$
		\{ ( l \cdot \Delta t, u^{\Delta t}_j( (k-l) \cdot \Delta t)), \; l = \floor*{ \frac{ \deltainf_{ij} }{\Delta t} }, \cdots, \ceil*{ \frac{ \deltasup_{ij} }{\Delta t} } \} \cup \{ (\deltasup_{ij}, u^{\Delta t}_j( k \cdot \Delta t - \deltasup_{ij})), (\deltainf_{ij}, u^{\Delta t}_j( k \cdot \Delta t - \deltainf_{ij}))  \}.
	$$
\end{lemma}
The proof is deferred to the online supplement, Section \ref{section-proof-lemma-optimal-basis-inner-problem-only-mean}.
\\
\indent Hence, \eqref{eq-primal-inner-problem-only-mean} can be solved to optimality in $O(\log(\frac{\deltasup_{ij} - \deltainf_{ij}}{\Delta t}))$ running time provided that the extreme points are stored in a data structure satisfying Assumption \ref{assumption-data-structure}.

\subparagraph{Faster procedure when the statistics are piecewise constant.}
When the statistics are piecewise constant, we have:
$$
	a^{ij}_{q, r} = 0 \quad q = 1, \cdots, Q_{ij}, \; r= 1, \cdots, R_{ij}.
$$
Hence, for any $r \in \{1, \cdots, R_{ij} \}$, the set of constraints
$$
	z + \sum_{q = 1}^{ Q_{ij} } ( x_q  -  y_q) \cdot (a^{ij}_{q, r} \cdot l \cdot \Delta t + b^{ij}_{q, r}) \leq u^{\Delta t}_j( (k - l) \cdot \Delta t) \quad l = \ceil*{ \frac{\inf(I_r)}{\Delta t} }, \cdots, \floor*{ \frac{\sup( I_r )}{\Delta t} }
$$
is equivalent to the single constraint:
$$
	z + \sum_{q = 1}^{ Q_{ij} } ( x_q  -  y_q) \cdot b^{ij}_{q, r} \leq \min\limits_{l = \ceil*{ \frac{\inf(I_r)}{\Delta t} }, \cdots, \floor*{ \frac{\sup( I_r )}{\Delta t} }} u^{\Delta t}_j( (k - l) \cdot \Delta t),
$$
whose right-hand side can be computed by binary search on $\mathcal{L}^{k, r}_{ij}$. As a result, the linear program \eqref{eq-dual-inner-problem-finite-lp} has $2 \cdot Q_{ij} + 1$ variables and $2 \cdot Q_{ij} + 3 \cdot R_{ij}$ constraints and can be solved to precision $\epsilon$ with an interior-point algorithm in $O( \log(\frac{1}{\epsilon}) )$ computation time. Typically, piecewise constant statistics can be used to bound the probability that a given event occurs, see Section \ref{section-ambiguity-set}.

\subsubsection{Dynamic convex hull algorithm}
\label{section-dynamic-convex-hull}
\begin{figure}[t]
	\centering
	\subfloat[$\hat{\mathcal{C}}_{k}$ is the hatched area.]{\label{fig-original-convex-hull}\includegraphics[scale=0.4]{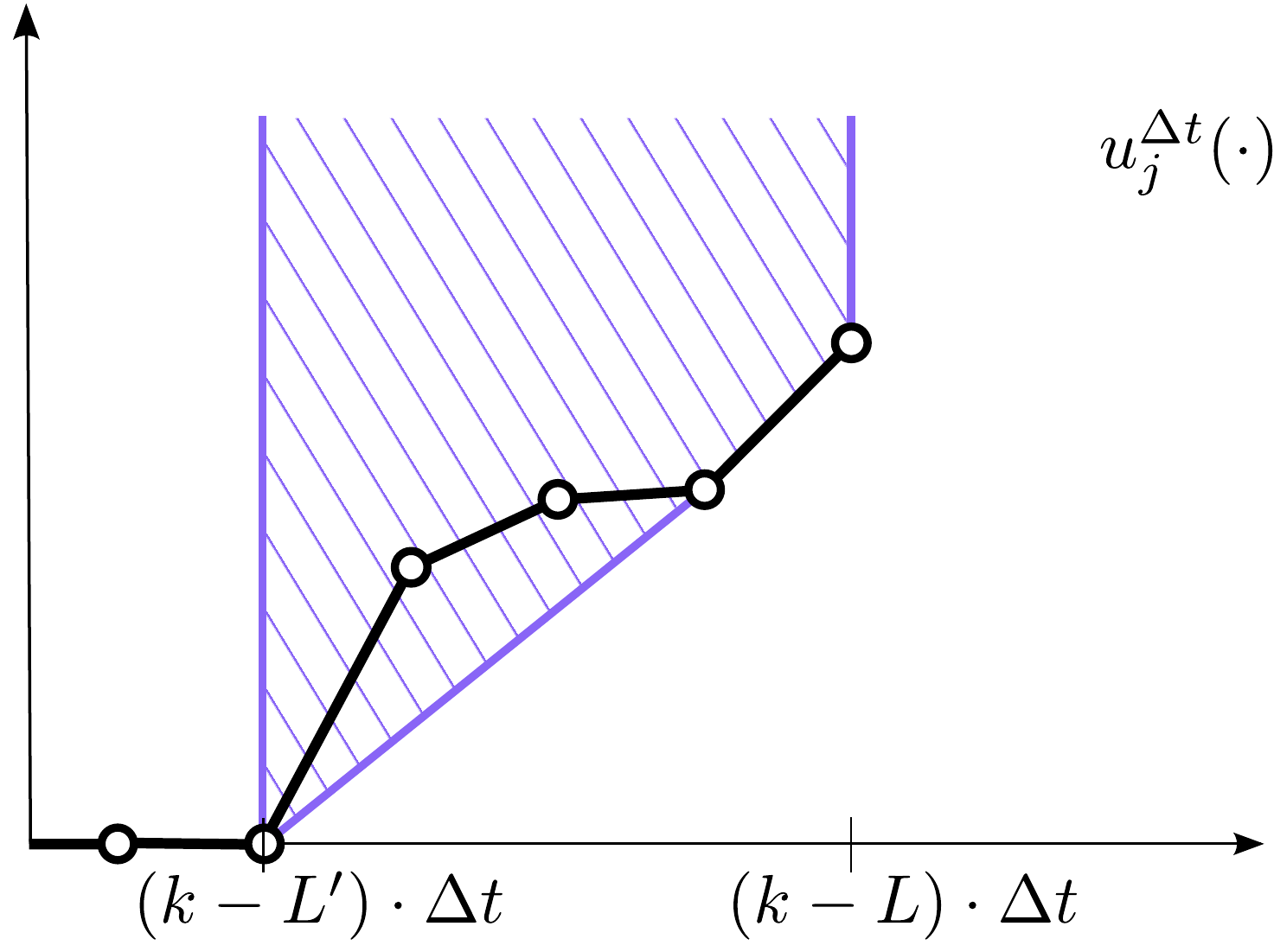}} \hspace{0.5cm}
	\subfloat[$\hat{\mathcal{C}}_{k+1}$ is the hatched area.]{\label{fig-update-convex-hull}\includegraphics[scale=0.4]{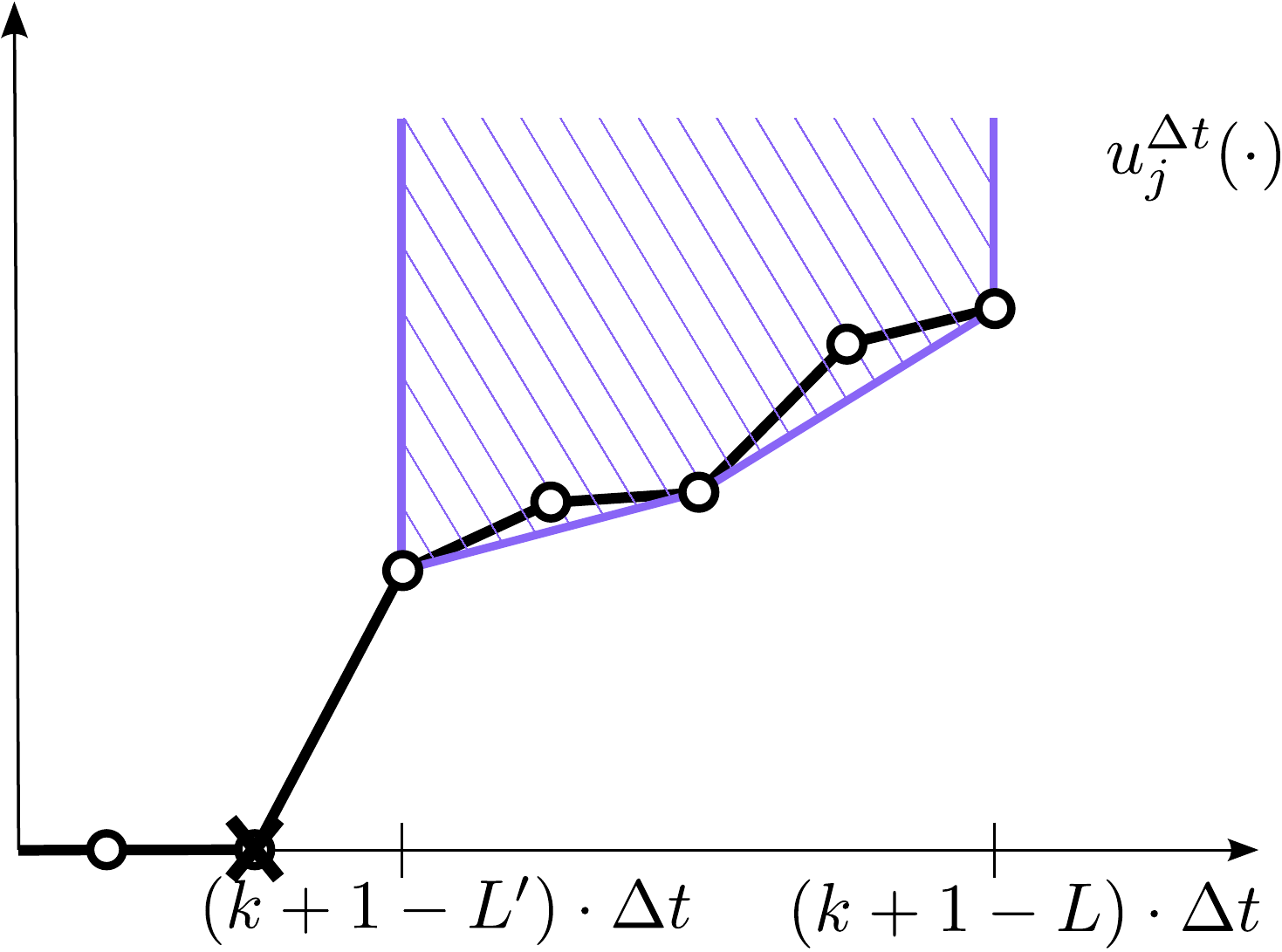}}
	\caption{The graph of $u^{\Delta t}_j(\cdot)$ is plotted in black. The dot points represent the breakpoints of $u^{\Delta t}_j(\cdot)$.}	\label{fig-extremepoints}
\end{figure}

Fix an arc $(i, j) \in \mathcal{A}$ and two integers $L < L'$ in $\{ \ceil*{ \frac{ \deltainf_{ij} }{ \Delta t } }, \cdots, \floor*{ \frac{ \deltasup_{ij} }{ \Delta t } } \}$. We are interested in the extreme points of the upper convex hull of $\{ (l \cdot \Delta t, u^{\Delta t}_j( (k - l) \cdot \Delta t)), \; l = L, \cdots, L' \}$ for $k \in \{ k^{r, \text{min}}_i, \cdots, \floor*{ \frac{T}{\Delta t} } \}$. To simplify the notations, it is convenient to reverse the x-axis and shift the x-coordinate by $k \cdot \Delta t$ which leads us to equivalently look at the extreme points of the upper convex hull of:
$$
	\mathcal{C}_k = \{ (l \cdot \Delta t, u^{\Delta t}_j( l \cdot \Delta t)), \; l = k - L', \cdots, k - L \},
$$
for $k \in \{ k^{r, \text{min}}_i, \cdots, \floor*{ \frac{T}{\Delta t} } \}$. There is a one-to-one mapping between the extreme points of these two sets which consists in applying the reverse transformation. For any $k$, $\hat{\mathcal{C}}_k$ denotes the upper convex hull of $\mathcal{C}_k$. Note that $\hat{\mathcal{C}}_k$ is a convex set and has a finitely many extreme points, all of which are in $\mathcal{C}_k$. Since the values $(u^{\Delta t}_j(l \cdot \Delta t))_{l = k^{r, \text{min}}_j, \cdots, \floor*{ \frac{T}{\Delta t} }}$ become sequentially available in ascending order of $l$ by chunks of size $\floor*{ \frac{\deltainf }{\Delta t} }$ as the label-setting algorithm progresses, a search for the extreme points of $\hat{\mathcal{C}}_{k+1}$ begins upon identification of the extreme points of $\hat{\mathcal{C}}_k$. Observe that $\hat{\mathcal{C}}_k$ updates to $\hat{\mathcal{C}}_{k+1}$ by removing the leftmost point $( (k-L') \cdot \Delta t, u^{\Delta t}_j( (k-L') \cdot \Delta t ) )$ and appending $( (k + 1 - L ) \cdot \Delta t, u^{\Delta t}_j( (k + 1 - L) \cdot \Delta t ) ) $ to the right, see Figure \ref{fig-extremepoints} for an illustration. In this process, deleting a point is arguably the most challenging operation because it might turn a formerly non-extreme point into one, see Figure \ref{fig-update-convex-hull} where this happens to be the case for the third leftmost point. In contrast, inserting a new point can only turn a formerly extreme point into a non-extreme one. Hence, deletions require us to do some bookkeeping other than simply keeping track of the extreme points of $\hat{\mathcal{C}}_k$ as $k$ increases. 
\\
\indent Maintaining the extreme points of a dynamically changing set is a well-studied class of problems in computational geometry known as \emph{Dynamic Convex Hull} problems. Specific instances from this class differ along the operations to be performed on the set (e.g. insertions, deletions), the queries to be answered on the extreme points, and the dimensionality of the input data. \cite{brodal2002dynamic} design a data structure maintaining a description of the upper convex hull of a finite set of $N$ points in $\mathbb{R}^2$. This data structure satisfies Assumption \ref{assumption-data-structure} as it allows to insert points, to delete points, and to perform binary search on the first coordinate of the extreme points, all in amortized time $O(\log(N))$ and with $O(N)$ space usage. For the purpose of being self-contained, we design our own data structure in the online supplement Section \ref{section-tailored-dynamic-algorithm} to tackle the particular dynamic convex hull problem at hand. Our approach is based on Andrew's monotone chain convex hull algorithm, see \cite{andrew1979another}, and only uses two arrays and a stack. The data structure developed in \cite{brodal2002dynamic} is more complex than ours but can handle arbitrary dynamic convex hull problems.

%%%%%%%%%%%%%%%%%%%%%%%%%%%%%%%%%%%%%%%%%%%%%%%%%

\section{Numerical experiments}
\label{section-numericalresults}
In this section, we compare, using a real-world application with field data from the Singapore road network, the performance of the nominal and robust approaches to vehicle routing when traffic measurements are scarce and uncertain. To benchmark the performance of the robust approach, we propose a realistic framework where both the nominal and robust approaches can be efficiently computed and for which it is up to the user to pick one.

\subsection{Framework}
\label{section-numexpframework}

\begin{figure}[t]
	\centering	
	\includegraphics[scale=0.24]{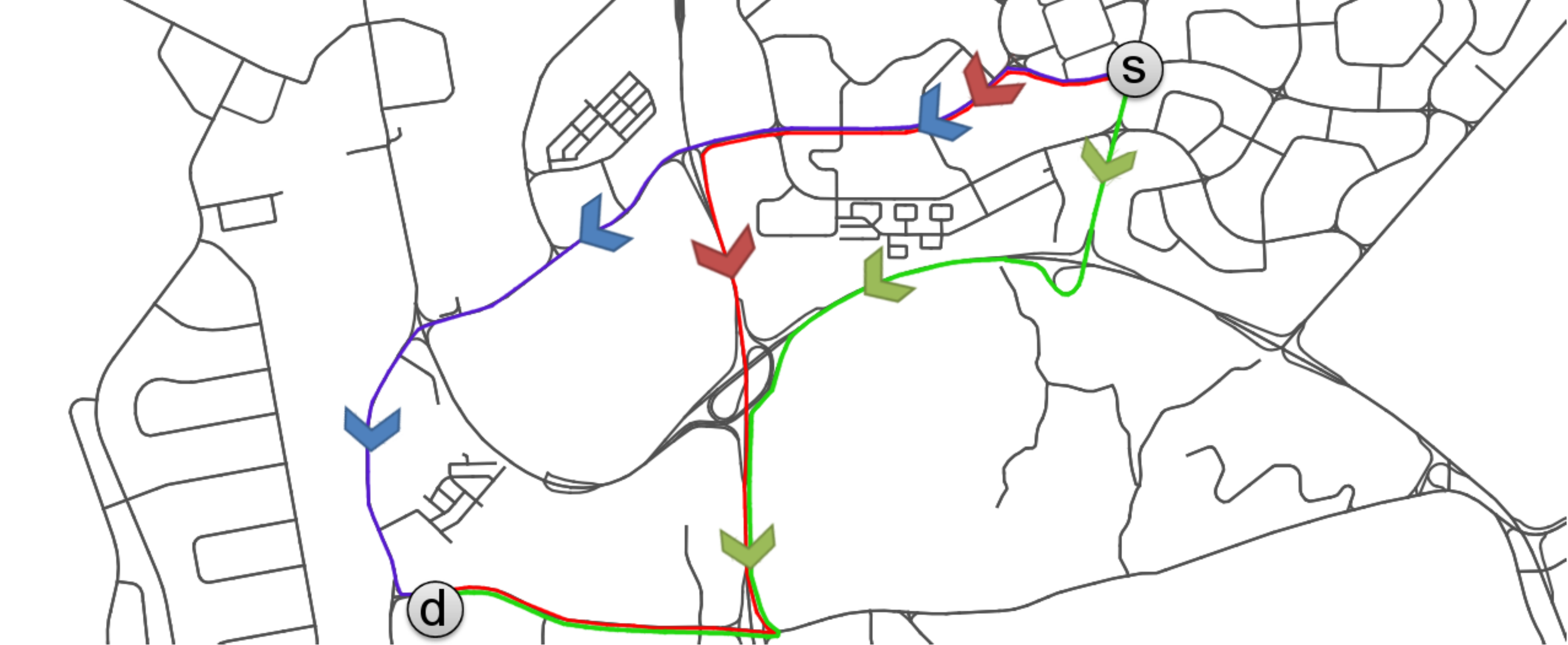}
	\caption{Local map. $s$ and $d$ locate the departure and arrival nodes. Three paths are highlighted. The left one (blue) is 5.3-km long and takes 9 minutes to travel. The middle one (red) is 6.4-km long and takes 8 minutes to travel. The rightmost one (green) is 6.1-km long and takes 10 minutes to travel.} \label{fig-newodbasic4}
\end{figure}

We work on a network composed of the main roads of Singapore with 20,221 arcs and 11,018 nodes for a total length of 1131 kilometers of roads. The data consists of a 15-day recording of GPS probe vehicle speed samples coming from a combined fleet of over 15,000 taxis. Features of each recording include current location, speed and status (free, waiting for a customer, occupied). We denote by $s$ and $d$ the departure and arrival nodes. Because there is usually only one reasonable route to get from $s$ to $d$ for most pairs $(s, d)$ in our network, the benefits of using one vehicle routing approach over another would not be apparent if we were to pick $(s, d)$ uniformly at random over $\mathcal{V}^2$. Instead, we choose to hand-pick a pair $(s, d)$ with at least two reasonable routes to get from $s$ to $d$ with similar travel times so that the best driving itinerary depends on the actual traffic conditions. We choose $s=$ \enquote{Woodlands avenue 2} and $d=$ \enquote{Mandai link}, see Figure \ref{fig-newodbasic4}, but the results would be similar for other pairs satisfying this property. 

\paragraph{Method of performance evaluation.}
Consider the following real-world situation. A user has to find an itinerary to get from $s$ to $d$ within a given budget $T$ (the deadline) and with an objective to maximize the probability of on-time arrival, but when only a few vehicle speed samples are available in order to assess arc travel time uncertainty.
\\
\indent To model this real-world situation, we assume that the full set of samples of vehicle speed measurement available in our dataset in fact represents the real traffic conditions, characterized by the corresponding travel-time distributions $\preal_{ij}$'s, which are obtained from the full set of samples. Mimicking the fact that the $\preal_{ij}$'s are actually not fully available, we then consider the case where only a fraction of the full set of samples, say $\lambda \in [0, 1]$, is available. Based on this limited data, the challenge is to select an itinerary with a probability of on-time arrival with respect to the real traffic conditions $\preal_{ij}$'s as high as possible. We propose to use the methods listed in Table \ref{tab-allmethods} to choose such an itinerary. For each of these methods, the process goes as follows:

\begin{enumerate}
	\item{Estimate the arc-based travel-time parameters required to run the method using the fraction of data available.}
	\item{Run the corresponding algorithm to find an itinerary, depending on the chosen method.}
	\item{Compute the probability of on-time arrival of this itinerary for the \textbf{real} traffic conditions ($\lambda = 1$).}
\end{enumerate}

The result obtained depends on both $\lambda$ and the available samples as there are many ways to pick a fraction $\lambda$ out of the entire dataset. Hence, for each $\lambda$ in a set $\Lambda$, we randomly pick $\lambda \cdot N_{ij}$ samples for each arc $(i,j)$, where $N_{ij}$ is the number of samples collected in the entire dataset for that particular arc. For each $\lambda \in \Lambda$, and for each method, we store the calculated probability of on-time arrival. We repeat this procedure 100 times.  

{\def\arraystretch{1.2}
\begin{table}[t]
   \centering
   \caption{\label{tab-allmethods} Methods considered. $I^{\text{m}}_{ij}$ and $I^{\text{md}}_{ij}$ are confidence intervals.} 
   \begin{tabular}{|c|c|c|}
  \hline
   Method & \begin{tabular}{c} Travel-time parameters to \\ estimate from samples \end{tabular} & Approach  \\
  \hline
  RobustM & $\deltainf_{ij}$, $\deltasup_{ij}$, $I^{\text{m}}_{ij}$ & \begin{tabular}{c} \eqref{eq-robust-problem} with \\ $\mathcal{P}_{ij} = \{ p \in \mathcal{P}([\deltainf_{ij}, \deltasup_{ij}]) :  \mathbb{E}_{X \sim p}[ X ] \in I^{\text{m}}_{ij} \}$ \end{tabular} \\
  \hline
  RobustMD & \begin{tabular}{c} $\deltainf_{ij}$, $\deltasup_{ij}$, $I^{\text{m}}_{ij}$, $I^{\text{md}}_{ij}$, \\ $m_{ij} = \frac{ \max(I^{\text{m}}_{ij}) + \min(I^{\text{m}}_{ij}) }{2}$ \end{tabular}& \begin{tabular}{c} 
  \eqref{eq-robust-problem} with \\ $\mathcal{P}_{ij} = \{ p \in \mathcal{P}([\deltainf_{ij}, \deltasup_{ij}]) :$ \begin{tabular}{c} $\mathbb{E}_{X \sim p}[ X ] \in I^{\text{m}}_{ij}$ \\ $\mathbb{E}_{X \sim p}[ | X  - m_{ij} | ] \in I^{\text{md}}_{ij}$ \end{tabular} $\}$ \end{tabular}  \\
  \hline
  Empirical & empirical distributions $p_{ij}$ & \eqref{eq-nominal-problem} with $p_{ij}$ \\  
  \hline
  LET & empirical mean $m_{ij}$ & standard shortest path \\
  \hline
	\end{tabular}
\end{table}
}
A few remarks are in order. We choose $\Lambda = \{0.001, 0.002, 0.005\}$, this corresponds to an average number of samples per arc of $[5.5, 9.4, 25.1]$ respectively (we take at least one sample per arc). The average arc length is 163 meters, hence we set $\Delta t = 0.02$ second to get a good accuracy. This parameter has a significant impact on the running time and it could also be optimized. We include the LET method as it is a reasonably robust approach, although not tailored to the risk function considered, and because it is very fast to solve. The confidence intervals used by the robust approaches are percentile bootstrap 95 \% confidence intervals derived from resampling the available data with replacement. When solving the discretization schemes \eqref{eq-discretization-scheme-nominal} and \eqref{eq-discretization-scheme-robust}, ties in the argument of the maximum are broken in favor of the (estimated) least expected travel time to the destination. To solve the robust problems, we use the column generation scheme and the special-purpose procedure described in Section \ref{section-algorithm-robust} while we use the scheme based on fast Fourier transforms described in Section \ref{section-algorithm-nominal} for the nominal approach.

\subsection{Results}
\label{section-routingresults}

\begin{figure}[t]
		\centering 
		\subfloat[Average probability of on-time arrival.]{\label{fig-resultalpha001average}\includegraphics[scale=0.36]{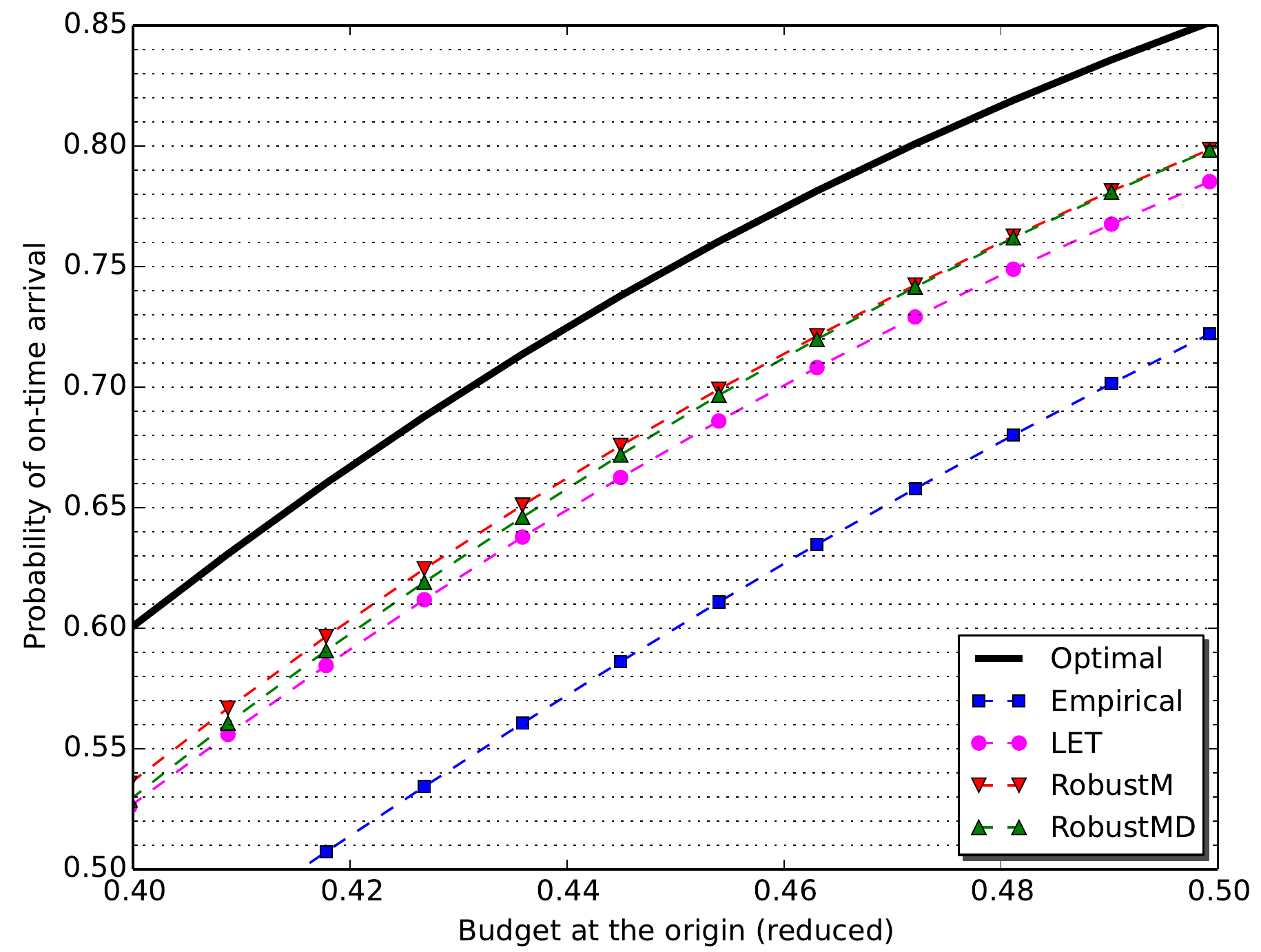}} 
		\subfloat[5\% worst-case probability of on-time arrival.]{\label{fig-resultalpha001worst}\includegraphics[scale=0.36]{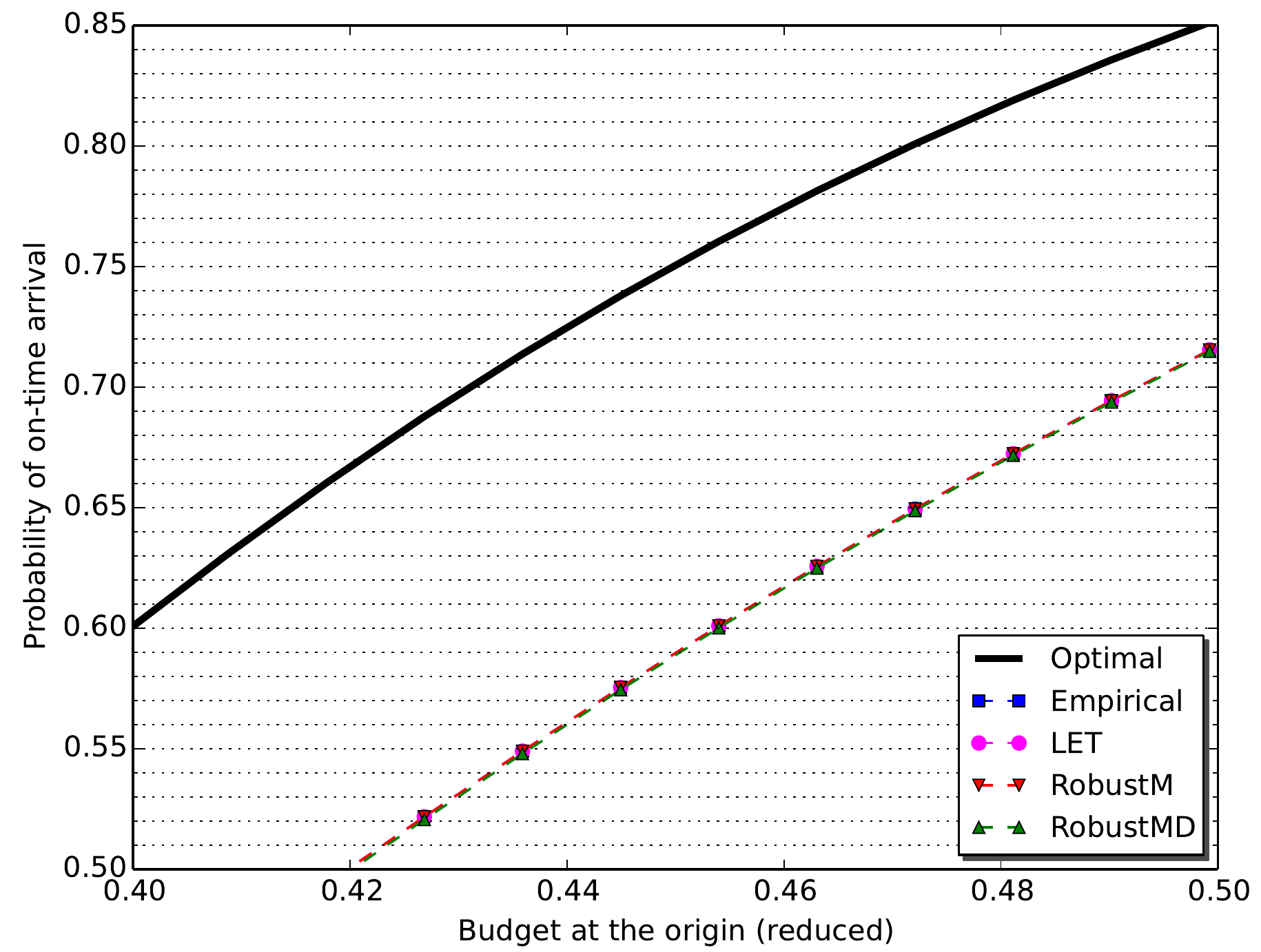}}
		\caption{$\lambda = 0.001$, average number of samples per link: $\sim$ 5.5.}	
		\label{fig-resultalpha001}
\end{figure}	
\begin{figure}[t]
		\centering 
		\subfloat[Average probability of on-time arrival.]{\label{fig-resultalpha002average}\includegraphics[scale=0.36]{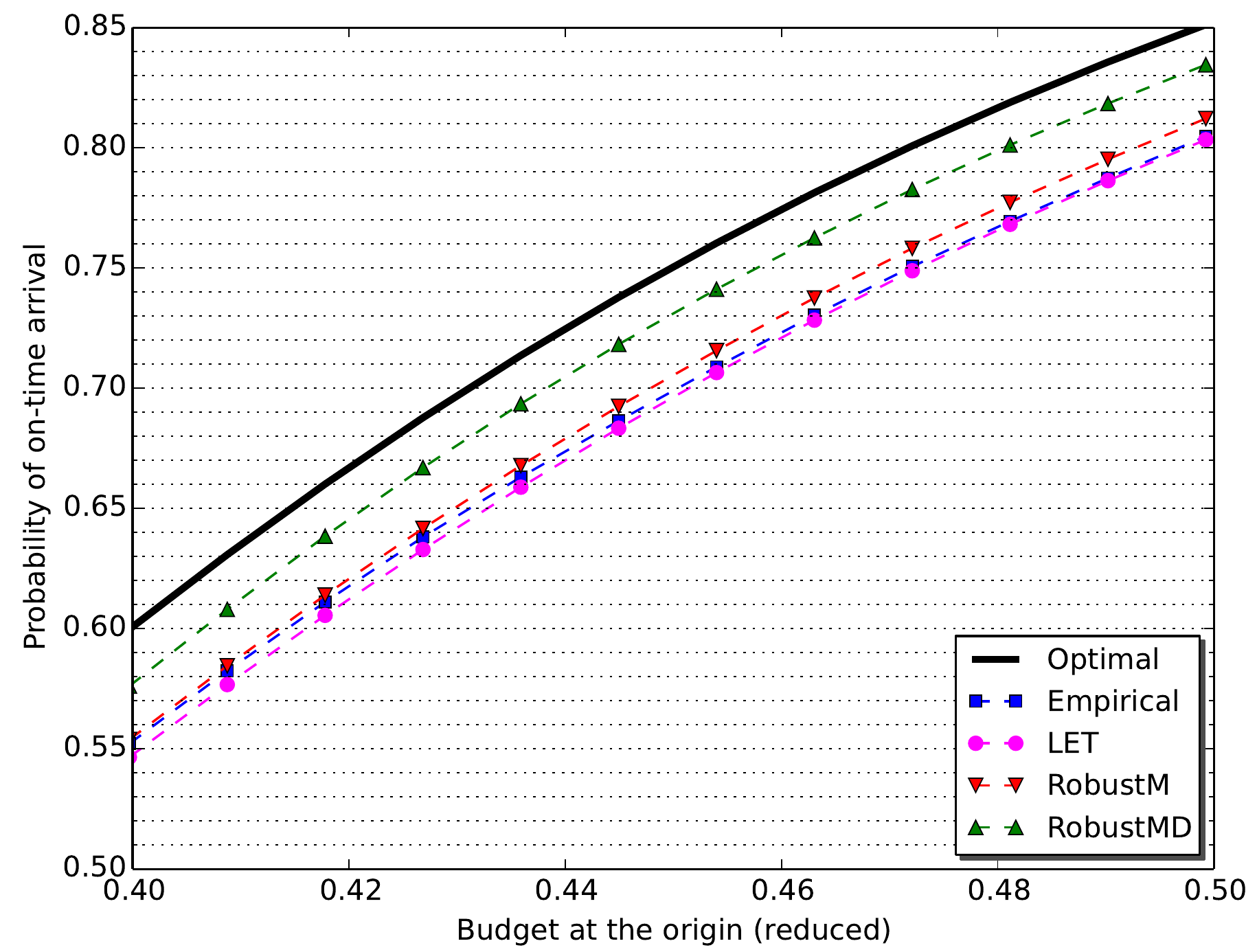}} 
		\subfloat[5\% worst-case probability of on-time arrival.]{\label{fig-resultalpha002worst}\includegraphics[scale=0.36]{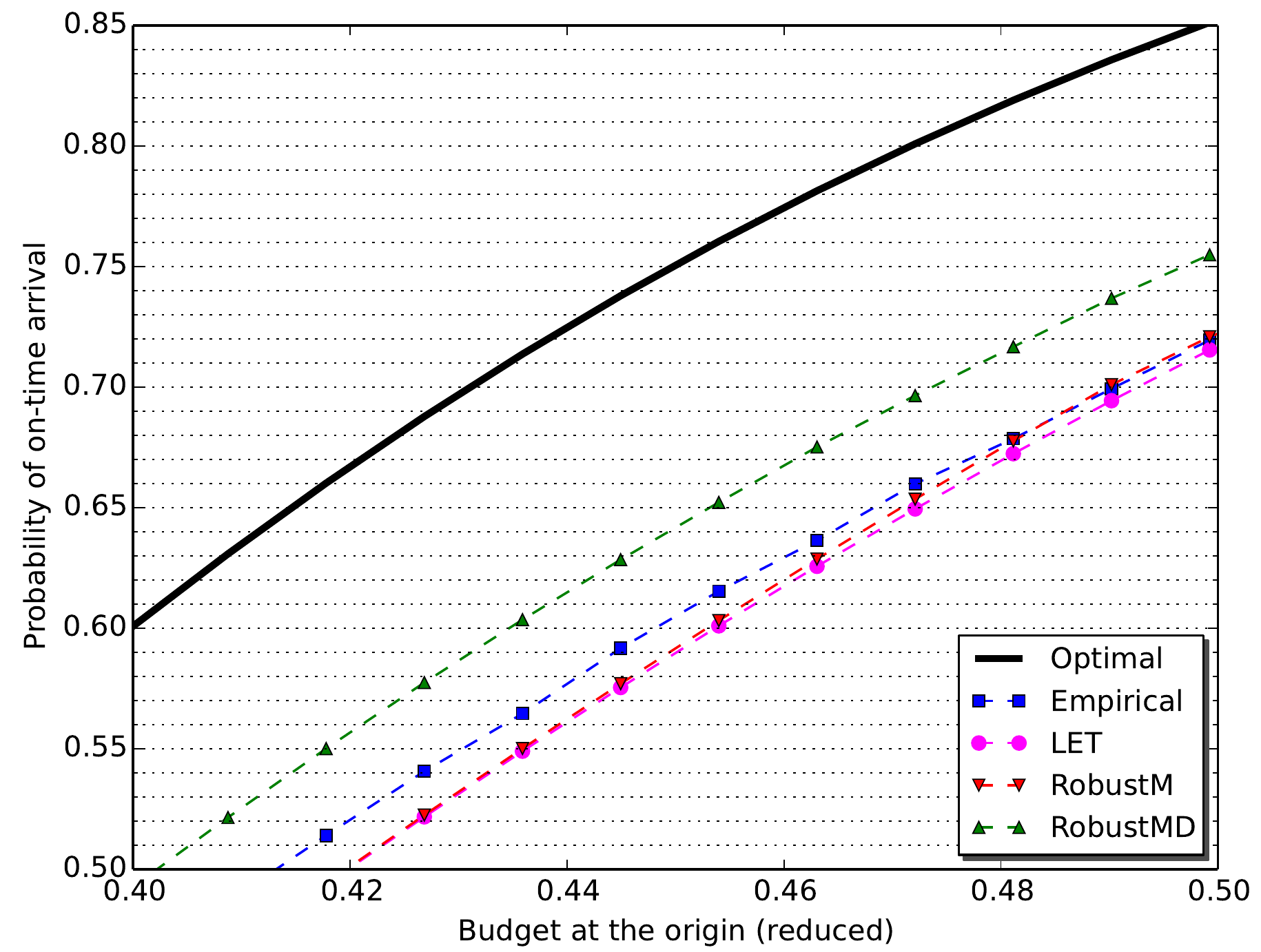}}
		\caption{$\lambda = 0.002$, average number of samples per link: $\sim$ 9.4.}	
		\label{fig-resultalpha002}
\end{figure}	
\begin{figure}[t]
		\centering 
		\subfloat[Average probability of on-time arrival.]{\label{fig-resultalpha005average}\includegraphics[scale=0.36]{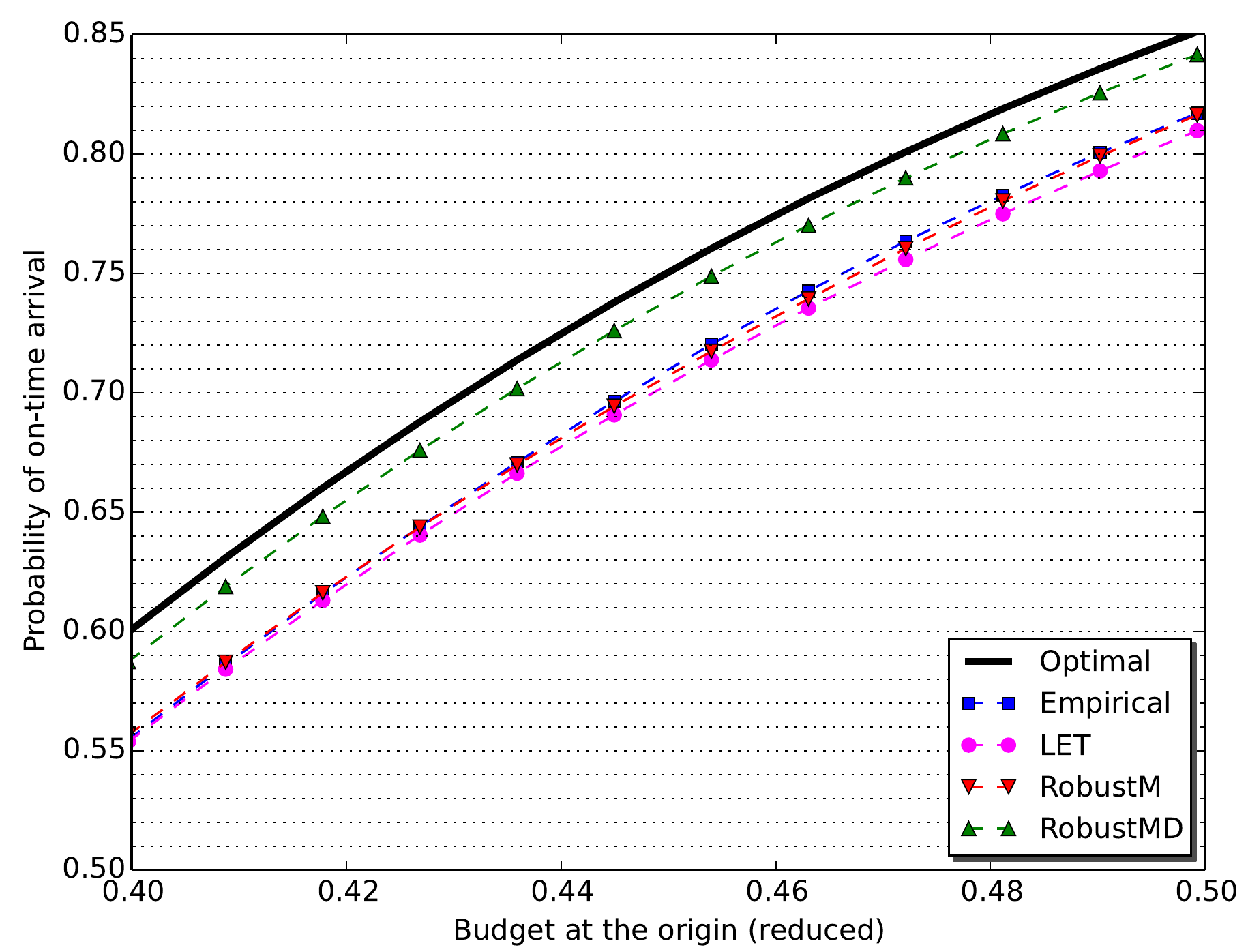}} 
		\subfloat[5\% worst-case probability of on-time arrival.]{\label{fig-resultalpha005worst}\includegraphics[scale=0.36]{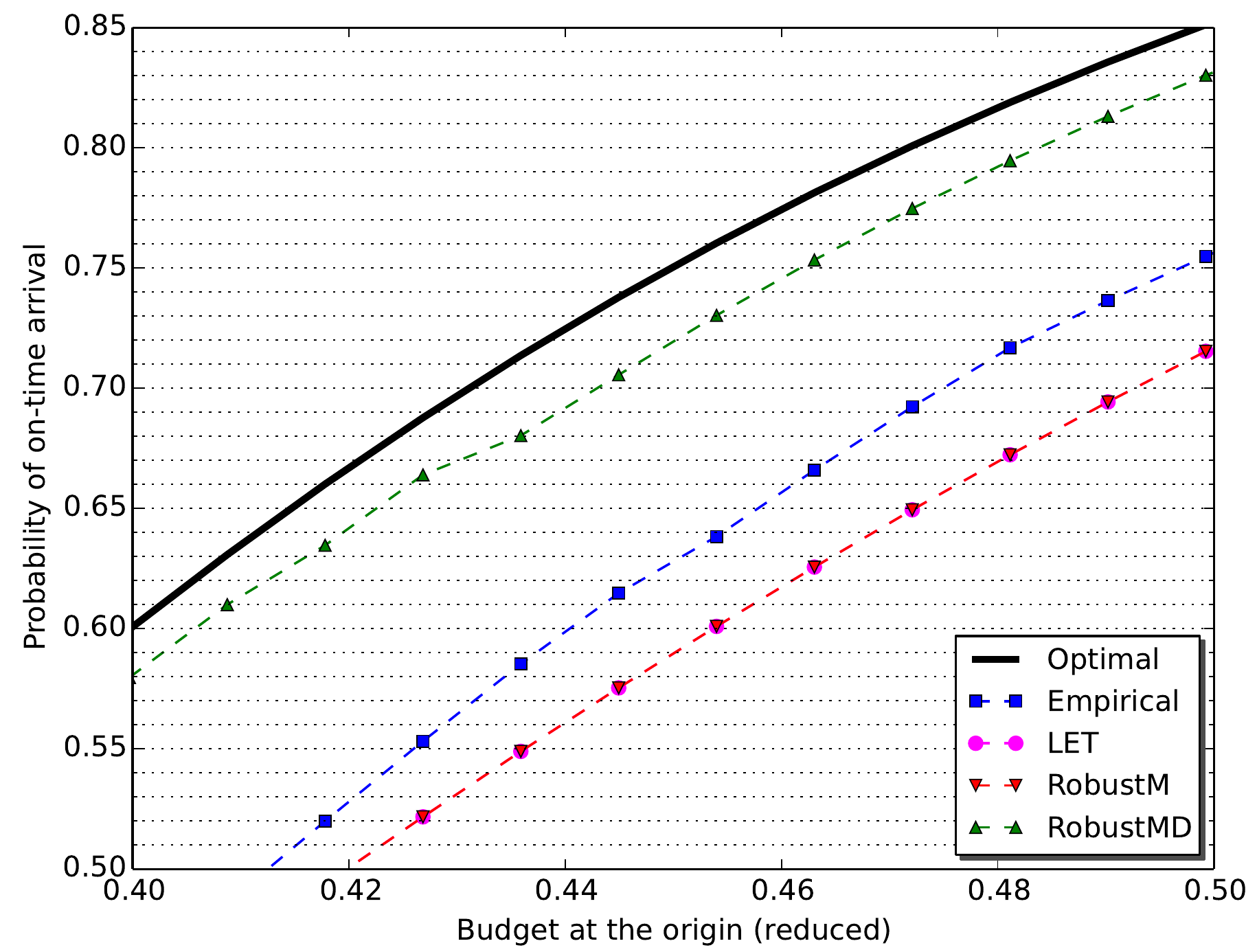}}
		\caption{$\lambda = 0.005$, average number of samples per link: $\sim$ 25.1.}	
		\label{fig-resultalpha005}
\end{figure}	

\begin{figure}[t]
	\centering
	\includegraphics[scale=0.35]{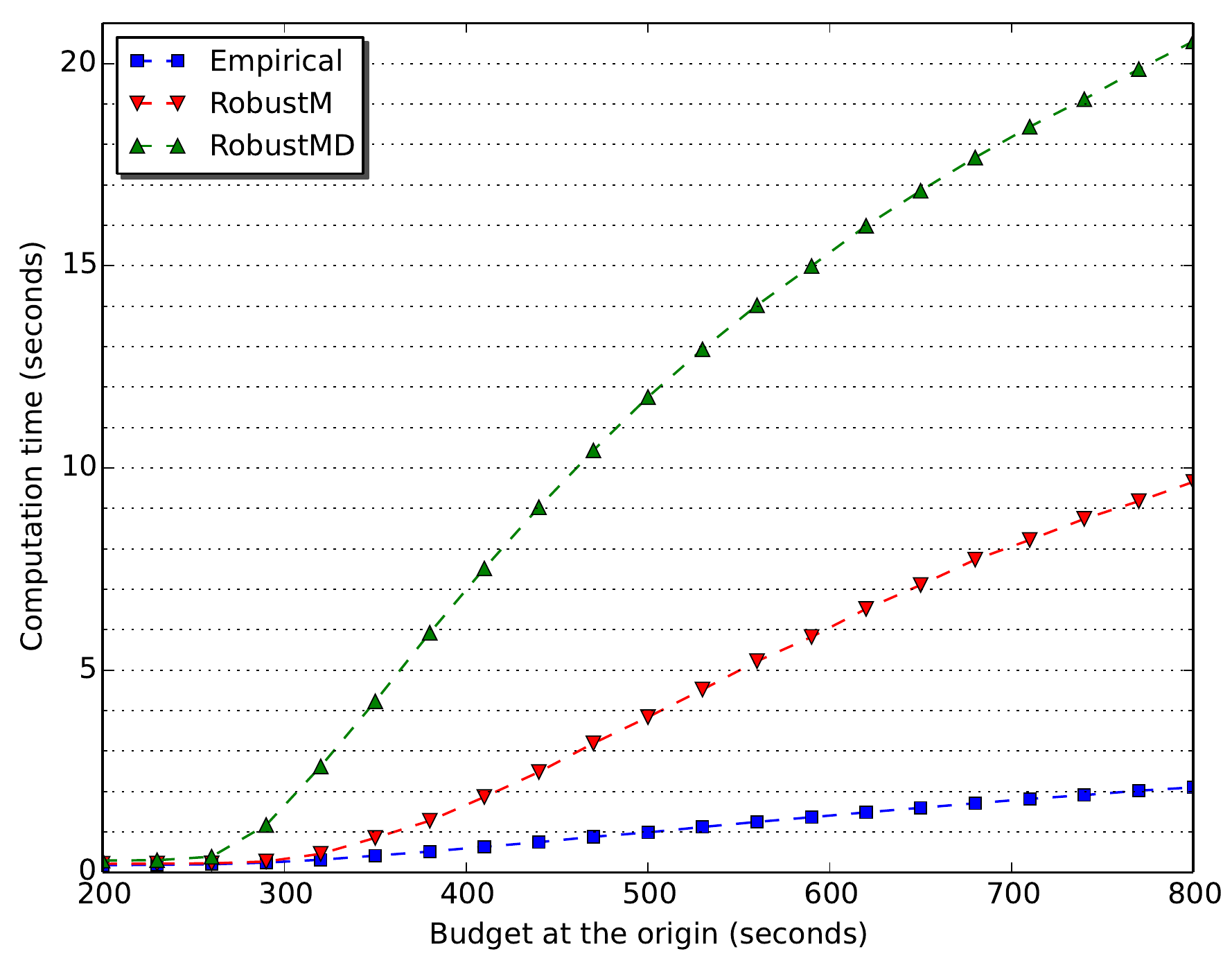}
	\caption{Average computation time as a function of the time budget for $\lambda = 0.001$.}
	\label{fig-runtimeres}
\end{figure}

The results are plotted in Figure \ref{fig-resultalpha001}, \ref{fig-resultalpha002}, and \ref{fig-resultalpha005}. Each of these figures corresponds to one of the fraction $\lambda \in \Lambda$ so as to see the impact of an increasing knowledge. The time budget is \enquote{normalized}: 0 (resp. 1) corresponds to the minimum (resp. maximum) amount of time it takes to reach $d$ from $s$. For each $\lambda$, for each method in Table \ref{tab-allmethods}, for each time budget $T$, and for each of the 100 simulations, we compute the actual probability of on-time arrival of the corresponding strategy. The average (resp. 5 \% worst-case)  probability of on-time arrival over the simulations is plotted on the figures labeled \enquote{a} (resp. \enquote{b}). The 5 \% worst-case measure, which corresponds to the average over the 5 simulations out 100 that yield the lowest probability of arriving on-time, is particularly relevant as commuters opting for this risk function would expect the approach to have good results even under bad scenarios. We also plot the average runtime for each of the method as a function of the time budget in Figure \ref{fig-runtimeres}.

\paragraph{Conclusions.}
As can be observed on the figures, Empirical is not competitive when only a few samples are available. To be specific, RobustM slightly outperforms the other methods when there are very few measurements, see Figure \ref{fig-resultalpha001}, while RobustMD is a clear winner when more samples are available, in terms of both average and worst-case performances, see Figures \ref{fig-resultalpha002} and \ref{fig-resultalpha005}. Observe that, as expected, the performance of Empirical improves as more samples get available and Empirical eventually outperforms RobustM, see Figure \ref{fig-resultalpha005}. Our interpretation of these results is that relying on quantities, either moments or distributions, that cannot be accurately estimated may be misleading even for robust strategies. On the other hand, failure to capture the increasing knowledge on the actual travel-time probability distributions (e.g. by estimating more moments) as the amount of available data increases may lead to poor performances.

\section{Extensions}
\label{sec-extensions}
In this section, we sketch how to extend the results derived in Sections \ref{section-analysis-nominal-problem} and \ref{section-analysis-robust-problem} when either Assumption \ref{assumption-independence} or Assumption \ref{assumption-time-dependence} is relaxed. Most of the results also extend when both assumptions are relaxed at the same time but we choose to discuss one assumption  at a time to highlight their respective implications.

\subsection{Relaxing the independence Assumption \ref{assumption-independence}: Markovian costs}
\label{sec-extensions-independence}
We consider here the case where the experienced costs of crossing arcs define a \emph{Markov} chain of finite order $m$. To simplify the presentation, we provide in details the extensions of our previous results to the case $m = 1$. Adapting these extensions to a general $m$ amounts to augmenting the state space of the underlying MDP by the costs of the last $m$ visited arcs. We emphasize that while \emph{Markov} chains can model the reality of the decision making process more accurately, this comes at a price: this requires an estimation of $m$-dimensional probability distributions, and the computational time needed to find an optimal strategy grows exponentially with $m$.
  
\paragraph{Extension for the nominal problem.}
A variant of Theorem \ref{lemma-LET-is-opt-nominal} can be shown to hold if the arc cost distributions are discrete. Under this assumption and as soon as the total cost spent so far is larger than $T - T_f$, the optimal strategy coincides with the strategy of minimizing the expected costs, which may no longer be a shortest path but can still be shown to be a solution without cycles. Under the same assumption, Proposition \ref{lemma-dp-optimal-nominal} remains valid under the following higher-dimensional dynamic program:
\begin{equation}
	\label{eq-dp-nominal-non-independent}
	\begin{aligned}
	 & u_d(t, z, \theta) = f(t) \quad  && t \leq T, z \in \mathcal{A}(d), \theta \in \Theta_{zd} \\
	 & u_i(t, z, \theta) = \max\limits_{j \in \mathcal{V}(i)} \int_0^{\infty} p_{ij}(\omega \; | \; z, \theta) \cdot u_j(t-\omega, i, \omega) \mathrm{d}\omega \quad && i \neq d, t \leq T, z \in \mathcal{A}(i), \theta \in \Theta_{zi} \\
	 & \pi^*_f(i, t, z, \theta) \in \argmax\limits_{j \in \mathcal{V}(i)} \int_0^{\infty} p_{ij}(\omega \; | \; z, \theta) \cdot u_j(t-\omega, i, \omega) \mathrm{d}\omega \quad && i \neq d, t \leq T, z \in \mathcal{A}(i), \theta \in \Theta_{zi},
	\end{aligned}
\end{equation}
where $\mathcal{A}(i)$ denotes the set of immediate antecedents of $i$ in $\mathcal{G}$, $\Theta_{zi}$ is the finite set of possible
values taken by $c_{zi}$ for $z \in \mathcal{A}(i)$, and $p_{ij}(\cdot \; | \; z, \theta)$ is the conditional distribution of $c_{ij}$ given that the last visited node is $z$ and that $c_{zi} = \theta$. The discretization scheme of Section \ref{section-discretization-nominal} can be adapted for this new dynamic equation and the approximation guarantees carry over. To solve this new discretization scheme, the label-setting approach from Section \ref{section-algorithm-nominal} can be adapted by observing that the functions $(u_i(\cdot, z, \theta))_{i \in \mathcal{V}, z \in \mathcal{A}(i), \theta \in \Theta_{zi}}$ can be computed block by block by interval increments of size $\deltainf$. However, the schemes based on fast Fourier transforms and the idea of zero-delay convolution do not apply anymore, and we need to use the pointwise definition of convolution products with computational complexity:
$$
	O(  \max\limits_{(i,j) \in \mathcal{A}} |\Theta_{ij}| \cdot \frac{ |\mathcal{A}| \cdot (T - T_f) + |\mathcal{V}|^2 \cdot \deltasup}{\Delta t}).
$$

\paragraph{Extension for the robust problem.}
For any $(i, j) \in \mathcal{A}$, $z \in \mathcal{A}(i)$, and $\theta \in \Theta_{zi}$, $p_{ij}( \cdot \; | \; z, \theta) $ is only known to lie in the ambiguity set $\mathcal{P}_{ij, z, \theta}$. If $\mathcal{P}_{ij, z, \theta}$ is only comprised of discrete distributions with finite support $\Theta_{ij}$, a variant of Theorem \ref{lemma-LET-is-opt-robust} can be shown to hold. Specifically, as soon as the total cost spent so far is larger than $T - T^r_f$, the optimal strategy coincides with the strategy of minimizing the worst-case expected costs, which can also be shown not to cycle. Under this assumption, Proposition \ref{lemma-dp-optimal-robust} remains valid under the following higher-dimensional dynamic program:
\begin{equation}
	\label{eq-dp-robust-non-independent}
	\begin{aligned}
	 & u_d(t, z, \theta) = f(t) \quad  && t \leq T, z \in \mathcal{A}(d), \theta \in \Theta_{zd} \\
	 & u_i(t, z, \theta) = \max\limits_{j \in \mathcal{V}(i)} \inf\limits_{p_{ij} \in \mathcal{P}_{ij, z, \theta}} \int_0^{\infty} p_{ij}(\omega) \cdot u_j(t-\omega, i, \omega) \mathrm{d}\omega \quad && i \neq d, t \leq T, z \in \mathcal{A}(i), \theta \in \Theta_{zi} \\
	 & \pi^*_{f, \mathcal{P}}(i, t, z, \theta) \in \argmax\limits_{j \in \mathcal{V}(i)} \inf\limits_{p_{ij} \in \mathcal{P}_{ij, z, \theta}} \int_0^{\infty} p_{ij}(\omega) \cdot u_j(t-\omega, i, \omega) \mathrm{d}\omega \quad && i \neq d, t \leq T, z \in \mathcal{A}(i), \theta \in \Theta_{zi}. \\
	\end{aligned}
\end{equation}
The discretization scheme of Section \ref{section-discretization-robust} can be adapted for this new set of equations and the approximation guarantees of Proposition \ref{lemma-quality-approx-robust} carry over. Moreover, the label-setting approach can also be adapted along the same lines as for the nominal problem. The ideas underlying the algorithmic developments of Section \ref{section-algorithm-robust} remain valid but we now have to recompute the convex hulls from scratch at each time step using Andrew's monotone chain convex hull algorithm, as opposed to using a dynamic convex hull algorithm, which leads to the computational complexity:
$$
	O(  \max\limits_{(i,j) \in \mathcal{A}} |\Theta_{ij}| \cdot \log(\max\limits_{(i,j) \in \mathcal{A}} |\Theta_{ij}|)  \cdot \frac{ |\mathcal{A}| \cdot (T - T^r_f) + |\mathcal{V}|^2 \cdot \deltasup }{ \Delta t}  \cdot \log(  \frac{ |\mathcal{V}| + \frac{T - T^r_f}{\deltainf}  }{ \epsilon }   )  ),
$$
when we want to compute an $\epsilon$-approximate strategy solution to the discretization scheme \eqref{eq-discretization-scheme-robust}.

\subsection{Relaxing Assumption \ref{assumption-time-dependence}: $\tau$-dependent arc cost probability distributions}
\label{sec-extensions-time-dependent-cost}

\paragraph{Extension for the nominal problem.}
For any $\tau \geq 0$ and $(i, j) \in \mathcal{A}$, we denote by $p^\tau_{ij}$ the distribution of $c^\tau_{ij}$ and by $m^\tau_{ij}$ the mean of $p^\tau_{ij}$. Theorem \ref{lemma-LET-is-opt-nominal} remains valid if, for any $(i, j) \in \mathcal{A}$, $m^\tau_{ij}$ converges as $\tau \rightarrow \infty$, in which case the shortest-path tree mentioned in the statement is defined with respect to the limits of the mean arc costs. For instance, this assumption is satisfied when the distributions are time-varying during a peak period and stationary anytime thereafter, see \cite{miller2000least}. Under this assumption, Proposition \ref{lemma-dp-optimal-nominal} also remains valid but for the slightly modified dynamic program:
\begin{equation}
	\label{eq-dp-nominal-time-dependent}
	\begin{aligned}
	 & u_d(t) = f(t) \quad  && t \leq T \\
	 & u_i(t) = \max\limits_{j \in \mathcal{V}(i)} \int_0^{\infty} p^{T-t}_{ij}(\omega) \cdot u_j(t-\omega) \mathrm{d}\omega \quad && i \neq d, t \leq T \\
	 & \pi^*_f(i, t) \in \argmax\limits_{j \in \mathcal{V}(i)} \int_0^{\infty} p^{T-t}_{ij}(\omega) \cdot u_j(t-\omega) \mathrm{d}\omega \quad && i \neq d, t \leq T.
	\end{aligned}
\end{equation}
The discretization scheme of Section \ref{section-discretization-nominal} can be trivially adapted for this new dynamic equation, although we may loose the approximation guarantees provided by Proposition \ref{lemma-quality-approx-nominal}. For them to carry over, we need additional assumptions. To be specific, one of the following properties must be satisfied: 
\begin{itemize}
	\item{the arc cost distributions vary smoothly, in the sense that, for any arc $(i, j) \in \mathcal{A}$, there exists $K $ such that the Kolmogorov distance between $p^{\tau_1}_{ij}$ and $p^{\tau_2}_{ij}$ is smaller than $K \cdot |\tau_1 - \tau_2|$ for any $\tau_1, \tau_2 \geq 0$,}
	\item{the arc cost distributions are discrete and the discretization length $\Delta t$ is chosen appropriately,}
	\item{the arc cost distributions change finitely many times and the discretization length $\Delta t$ is chosen appropriately.}
\end{itemize}
To solve the discretization scheme, the label-setting approach described in Section \ref{section-algorithm-nominal} remains relevant but we now have to apply the pointwise definition of convolution products, as opposed to using fast Fourier transforms and zero-delay convolutions, with computational complexity quadratic in $\frac{1}{\Delta t}$:
$$
	O( \frac{ |\mathcal{A}| \cdot (T - T_f) + |\mathcal{V}|^2 \cdot \deltasup}{\Delta t} \cdot  \frac{\deltasup - \deltainf}{\Delta t} ).
$$

\paragraph{Extension for the robust problem.}
For any $\tau \geq 0$ and $(i, j) \in \mathcal{A}$, $p^\tau_{ij}$ is only known to lie in the ambiguity set $\mathcal{P}^\tau_{ij}$. First observe that \eqref{eq-robust-problem} turns into:
$$
		\sup\limits_{\pi \in \Pi} \; \inf\limits_{\forall \tau, \forall (i, j) \in \mathcal{A}, \; p^\tau_{ij} \in \mathcal{P}^\tau_{ij}} \; \mathbb{E}_{ \mathbf{p^\tau} }[f(T - X_\pi)],
$$
which is exactly the robust counterpart of \eqref{eq-nominal-problem}, as opposed to a robust relaxation when the arc cost distributions are stationary. Theorem \ref{lemma-LET-is-opt-robust} remains valid if, for any $(i, j) \in \mathcal{A}$, $\max_{ p_{ij} \in \mathcal{P}^\tau_{ij} } \mathbb{E}_{X \sim p_{ij}}[X]$ converges as $\tau \rightarrow \infty$, in which case the shortest-path tree mentioned in the statement is defined with respect to the limits. Again, this assumption is, for instance, satisfied when the ambiguity sets are time-varying during a peak period and stationary anytime thereafter. Under this assumption, Proposition \ref{lemma-dp-optimal-robust} also remains valid but for the slightly modified dynamic program:
\begin{equation}
	\label{eq-dp-robust-time-dependent}
	\begin{aligned}
     & u_d(t) = f(t) \quad  && t \leq T \\
	 & u_i(t) = \max\limits_{j \in \mathcal{V}(i)} \inf\limits_{p_{ij} \in \mathcal{P}^{T-t}_{ij}} \int_0^{\infty} p_{ij}(\omega) \cdot u_j(t-\omega) \mathrm{d}\omega \quad && i \neq d, t \leq T \\
	 & \pi^*_{f, \mathcal{P}}(i, t) \in \argmax\limits_{j \in \mathcal{V}(i)} \inf\limits_{p_{ij} \in \mathcal{P}^{T-t}_{ij}} \int_0^{\infty} p_{ij}(\omega) \cdot u_j(t-\omega) \mathrm{d}\omega \quad && i \neq d, t \leq T.
	\end{aligned}
\end{equation}
Similarly as for the nominal problem, the discretization scheme can be trivially adapted but we may loose the approximation guarantees provided by Proposition \ref{lemma-quality-approx-robust}. For them to carry over, one of the following properties has to be satisfied:
\begin{itemize}
	\item{the ambiguity sets vary smoothly, in the sense that, for any arc $(i, j) \in \mathcal{A}$, there exists $K $ such that the Kolmogorov distance between $\mathcal{P}^{\tau_1}_{ij}$ and $\mathcal{P}^{\tau_2}_{ij}$ is smaller than $K \cdot |\tau_1 - \tau_2|$ for any $\tau_1, \tau_2 \geq 0$,}
	\item{the ambiguity sets are only comprised of discrete distributions and the discretization length $\Delta t$ is chosen appropriately,}
	\item{the ambiguity sets change finitely many times and the discretization length $\Delta t$ is chosen appropriately.}
\end{itemize}
In contrast to the nominal problem, all the algorithms developed in Section \ref{section-algorithm-robust} can still be used to solve the discretization sheme with the same computational complexity as long as the ambiguity sets are defined by confidence intervals on piecewise affine statistics, as precisely defined in Section \ref{section-ambiguity-set}.

\ACKNOWLEDGMENT{This research is supported in part by the National Research Foundation (NRF) Singapore through the Singapore MIT Alliance for Research and Technology (SMART) and its Future Urban Mobility (FM) Interdisciplinary Research Group. The authors would like to thank Chong Yang Goh from the Massachusetts Institute of Technology for his help in preprocessing the data.}

%%%%%%%%%%%%%%%%%%%%%%%%%%%%%%%%%%%%%%%%%%%%%%%%%%%%%%%%%%

\bibliographystyle{ormsv080} 
\bibliography{biblio}

%% Here starts the e-companion (EC)
%%%%%%%%%%%%%%%%%%%%%%%%%%%%%%%%%%%%%%%%%%%%%%%%%%%%%%%%%%

\ECSwitch

%\ECDisclaimer
%%%%%%%%%%%%%%%%%%%%%%%%%%%%%%%%%%%%%%%%%%%%%%%%%%%%%%%%%%

%%% Main head for the e-companion
\ECHead{Online Supplement}

\begin{APPENDICES}

\section{Tailored dynamic convex hull algorithm}
\label{section-tailored-dynamic-algorithm}
The fact that deletions and insertions always occur on the same side of the set allows us to deal with deletions in an indirect way, by building and merging upper convex hulls of partial input data. The only downside is that this requires an efficient merging procedure. In this respect, we state without proof a result derived from \cite{overmars1981maintenance}.

\begin{lemma}
\label{lemma-MergingConvexHull}
Consider a set $S$ of $N$ points in $\mathbb{R}^2$ partitioned into two sets of points $S_1$ and $S_2$ such that, for any two points $(x_1, y_1) \in S_1$ and $(x_2, y_2) \in S_2$ we have $x_1 < x_2$. Suppose that the extreme points of $\hat{S_1}$ (resp. $\hat{ S_2 }$) are stored in an array $A_1$ (resp. $A_2$) of size $N$ in ascending order of their first coordinates. We can find two indices $l_1$ and $l_2$ in $O(\log(N))$ time such that the set comprised of the points contained in $A_1$ with index smaller than $l_1$ and the points contained in $A_2$ with index larger than $l_2$ is precisely the set of extreme points of $\hat{ S }$.
\end{lemma}

\paragraph{Algorithm.} 
We use two arrays $A_{\text{left}}$ and $A_{\text{right}}$ along with a stack $\mathcal{S}$. The arrays $A_{\text{left}}$ and $A_{\text{right}}$ are of size $L' - L + 1$, indexed from $0$ to $L' - L$, and store points in $\mathbb{R}^2$ in ascending order of their first coordinates. The stack $\mathcal{S}$ stores stacks of points in $\mathbb{R}^2$. We keep track of two indices $l_{\text{left}}$ and $l_{\text{right}}$ such that, at any step $k = k^{r, \text{min}}_i + p \cdot (L' - L + 2) + r$ for some $p \in \mathbb{N}$ and $0 \leq r \leq L' - L + 1$, the following invariant holds:
\begin{itemize}
	\item{$\{ A_{\text{left}}[ l ], l = l_{\text{left}} + 1, \cdots, L' - L \}$ is the set of extreme points of the upper convex hull of $\{ (l \cdot \Delta t, u^{\Delta t}_j( l \cdot \Delta t)), \; l = k - L', \cdots, k - L - r \}$, }
	\item{$\{ A_{\text{right}}[ l ], l = 0, \cdots, l_{\text{right}} - 1\}$ is the set of extreme points of the upper convex hull of $\{ (l \cdot \Delta t, u^{\Delta t}_j( l \cdot \Delta t)), \; l = k - L - r + 1, \cdots, k - L \}$. }
\end{itemize}
Using the procedure of Lemma \ref{lemma-MergingConvexHull} and this invariant, we can find a pair of indices $(l_1, l_2)$ in $O(\log(L' - L))$ time such that $\{ A_{\text{left}}[ l ], l = l_{\text{left}} + 1, \cdots, l_1 \} \cup \{ A_{\text{right}}[ l ], l = l_2, \cdots, l_{\text{right}} - 1 \}$ is the set of extreme points of $\hat{ \mathcal{C}_k}$. Hence, all we have left to do is to provide a procedure to maintain $A_{\text{left}}$, $A_{\text{right}}, l_{\text{left}}$ and $l_{\text{right}}$, which we do next.
\\
\indent $A_{\text{left}}$, $A_{\text{right}}$, and $\mathcal{S}$ are initially empty. The algorithm proceeds in two phases and loops back to the first one every $L' - L + 2$ steps. For convenience, we define $\text{cross}$ as the function taking as an input three points $a, b, c$ in $\mathbb{R}^2$ and returning the cross product of the vector $\vec{ab}$ and $\vec{ac}$. 
\\
\textbf{Phase 1}: Suppose that the current step is $k$. Hence, the values $u^{\Delta t}_j( k - L'), \cdots, u^{\Delta t}_j( k - L)$ are available. This phase is based on Andrew's monotone chain convex hull algorithm to find the extreme points of $\hat{\mathcal{C}_k}$ with the difference that we store the points removed along the process in stacks for future use. Specifically, set $l_{\text{left}} = L' - L$ and $l_{\text{right}} = 0$ and for $l$ decreasing from $k - L$ to $k - L'$, do the following: 
\begin{enumerate}[(a)]
	\item Initialize a new stack $\mathcal{S}'$,
	\item While $l_{\text{left}} \leq L' - L - 2$ and $\text{cross}(A_{\text{left}}[l_{\text{left}} + 2], A_{\text{left}}[l_{\text{left}} + 1], (l \cdot \Delta t, u^{\Delta t}_j( l \cdot \Delta t))) \geq 0$:
		\begin{itemize}
			\item Push $A_{\text{left}}[l_{\text{left}} + 1]$ to $\mathcal{S}'$,
			\item Increment $l_{\text{left}}$,
		\end{itemize}
	\item Push $\mathcal{S}'$ to $\mathcal{S}$,
	\item{Set $A_{\text{left}}[l_{\text{left}}] = (l \cdot \Delta t, u^{\Delta t}_j( l \cdot \Delta t))$ and decrement $l_{\text{left}}$. At this point, $\{ A_{\text{left}}[l], l = l_{\text{left}} + 1, \cdots, L' - L \}$ is the set of extreme points of the upper convex hull of $\{ (m \cdot \Delta t, u^{\Delta t}_j( m \cdot \Delta t)), \; m = l, \cdots, k - L \}$.}
\end{enumerate}
\textbf{Phase 2}: At step $k + l$, for $l$ increasing from $1$ to $L' - L$, observe that the value $u^{\Delta t}_j( k + l - L)$ becomes available. To maintain $A_{\text{left}}$ and $l_{\text{left}}$, we remove the leftmost point $(x, y)$ and reinsert the points, stored in the topmost stack of $\mathcal{S}$, that were previously removed from $A_{\text{left}}$ when appending $(x, y)$ to $A_{\text{left}}$ in the course of running Andrew's monotone chain convex hull algorithm. Specifically:
\begin{enumerate}[(a)]
	\item{Increment $l_{\text{left}}$,}
	\item{Pop the topmost stack $\mathcal{S}'$ out of $\mathcal{S}$,}
	\item While $\mathcal{S}'$ is not empty:
		\begin{itemize}
			\item Pop the topmost point $(x, y)$ of $\mathcal{S}'$,
			\item Set $A_{\text{left}}[l_{\text{left}}] = (x, y)$,
			\item Decrement $l_{\text{left}}$.
		\end{itemize}
\end{enumerate}
To maintain $A_{\text{right}}$ and $l_{\text{right}}$, we run an iteration of Andrew's monotone chain convex hull algorithm. Specifically:
\begin{enumerate}[(a)]
	\item While $l_{\text{right}} \geq 2$ and $\text{cross}(A_{\text{right}}[l_{\text{right}} - 2], A_{\text{right}}[l_{\text{right}} - 1], ((k + l - L) \cdot \Delta t, u^{\Delta t}_j( (k + l - L) \cdot \Delta t))) \leq 0$: 
		\begin{itemize}
			\item Decrement $l_{\text{right}}$,
		\end{itemize}			
	\item{Set $A_{\text{right}}[l_{\text{right}}] = ( (k + l - L ) \cdot \Delta t, u^{\Delta t}_j( (k + l - L) \cdot \Delta t))$ and increment $l_{\text{right}}$.}
\end{enumerate}	

\paragraph{Complexity Analysis.}
Observe that any point added to $A_{\text{left}}$ can only be removed once, and the same holds for $A_{\text{right}}$. This means that Phase 1 and Phase 2 take $O( L' - L )$ computation time. These two phases are repeated $\ceil*{ \frac{ \floor*{ \frac{T}{\Delta t} } - k^{r, \text{min}}_i  }{ L' - L + 2 } }$ times leading to an overall complexity of $O( \floor*{ \frac{T}{\Delta t} } - k^{r, \text{min}}_i )$. Since the merging procedure outlined in Lemma \ref{lemma-MergingConvexHull} takes $O( \log(L' - L) )$ computation time at each step, the global complexity is $O( (\floor*{ \frac{T}{\Delta t} } - k^{r, \text{min}}_i )  \cdot \log(L' - L) )$.

\section{Omitted Proofs}

\subsection{Proof of Theorem \ref{lemma-LET-is-opt-nominal}}
\label{section-proofextendsota}

\proof{Proof of Theorem \ref{lemma-LET-is-opt-nominal}.} 
Let us start with the last part of the theorem. If the support of $f(\cdot)$ is included in $[T_f, \infty)$, any strategy is optimal when having already spent a budget of $T - T_f$ with an optimal objective function of $0$. \\
\indent Let us now focus on the first part of the theorem. Consider an optimal strategy $\pi^*_f$ solution to \eqref{eq-nominal-problem}. For a given history $h \in \mathcal{H}$, we define $t_h$ as the remaining budget, i.e. $T$ minus the total cost spent so far, and $i_h$ as the current location. The policy $\pi^*_f$ maps $h \in \mathcal{H}$ to a probability distribution over $\mathcal{V}(i_h)$. Observe that randomizing does not help because the costs are independent across time and arcs so that, without loss of generality, we can assume that $\pi^*_f$ actually maps $h$ to the node in $\mathcal{V}(i_h)$ minimizing the objective function given $h$. For $h \in \mathcal{H}$, we denote by $X_{\pi^*_f}^{h}$ the random cost-to-go incurred by following strategy $\pi^*_f$, i.e. not including the total cost spent up to this point of the history $T - t_h$. We define $(m_{ij})_{(i, j) \in \mathcal{A}}$ as the expected arc costs and by $M_i$ as the minimum expected cost to go from $i$ to $d$ for any $i \in \mathcal{V}$. We also define $\pi_s$ as a policy associated with an arbitrary shortest path from $i$ to $d$ with respect to the expected costs. Specifically, $\pi_s$ maps the current location $i_h$ to a node in $\mathcal{T}(i_h)$, irrespective of the history of the process. Similarly as for $\pi^*_f$, we denote by $X_{\pi_s}^{h}$ the random cost-to-go incurred by following strategy $\pi_s$ for $h \in \mathcal{H}$. We first show that there exists $T_f$ such that, for both cases (a) and (b):
\begin{equation}
	\label{eq-proof-LET-optimal}
	\mathbb{E}[X_{\pi^*_f}^{h}] - M_{i_h} < \min_{i \neq d} \min_{j \in \mathcal{V}(i), j \notin \mathcal{T}(i) } \{ m_{ij} + M_j - M_i \} \quad \forall h \in \mathcal{H} \; \text{ such that } \; t_h \leq T_f,
\end{equation}
with the convention that the minimum of an empty set is equal to infinity. Note that the right-hand side is always positive. Let $\alpha = |\mathcal{V}| \cdot \deltasup$. \\
(a) For $h \in \mathcal{H}$ such that $t_h < T_1$, we have, using a Taylor's series expansion:
$$
	f(t_h - X_{\pi^*_f}^{h}) = f(t_h - \alpha) + f'(t_h - \alpha) \cdot (\alpha - X_{\pi^*_f}^{h}) + \frac{1}{2} \cdot f''(\xi_h) \cdot (\alpha - X_{\pi^*_f}^{h})^2,
$$
where $\xi_h \in [\min(t_h - \alpha, t_h - X_{\pi^*_f}^{h}), \max(t_h - \alpha, t_h - X_{\pi^*_f}^{h})]$, and:
$$
	f(t_h - X_{\pi_s}^h) = f(t_h - \alpha) + f'(t_h - \alpha) \cdot (\alpha - X_{\pi_s}^h) + \frac{1}{2} \cdot f''(\zeta_h) \cdot (\alpha - X_{\pi_s}^h)^2,
$$
where $\zeta_h \in [\min(t_h - \alpha, t_h - X_{\pi_s}^h), \max(t_h - \alpha, t_h - X_{\pi_s}^h)]$. Using \emph{Bellman's Principle of Optimality} for $\pi^*_f$, we have:
$$
	\mathbb{E}[f(t_h - X_{\pi^*_f}^{h})] = \mathbb{E}[f(T - ( (T - t_h) + X_{\pi^*_f}^{h} )] \geq  \mathbb{E}[f(T - ( (T - t_h) + X_{\pi_s}^{h} )]  \geq \mathbb{E}[f(t_h - X_{\pi_s}^h)].
$$ 
Expanding and rearranging yields:
$$ 
	- f'(t_h - \alpha) \cdot ( \mathbb{E}[ X_{\pi^*_f}^{h} ] - \mathbb{E}[ X_{\pi_s}^{h} ] ) \geq \frac{1}{2} \cdot ( \mathbb{E}[ - f''(\xi_h) \cdot (\alpha - X_{\pi^*_f}^{h})^2 ] + \mathbb{E}[f''(\zeta_h) \cdot (\alpha - X_{\pi_s}^h)^2]).
$$
Since the costs are independent across time and arcs:
$$
	\mathbb{E}[ X_{\pi_s}^{h} ] = M_{i_h}.
$$
Concavity of $f(\cdot)$ implies that $f''(\xi_h) \cdot (\alpha - X_{\pi^*_f}^{h})^2 \leq 0$ almost surely. Since $f(\cdot)_{(-\infty, T_1)}$ is increasing, we obtain $\mathbb{E}[ X_{\pi^*_f}^{h} ] - M_{i_h} \leq \frac{ \mathbb{E}[- f''(\zeta_h) \cdot (\alpha - X_{\pi_s}^h)^2] }{2 \cdot f'(t_h - \alpha)}$. As $X_{\pi_s}^h$ is the cost of a path, Assumption \ref{assumption-compactsupport} implies $0 \leq X_{\pi_s}^h \leq \alpha$. We get that $\zeta_h \in [t_h - \alpha, t_h]$ and:
$$
	\mathbb{E}[ X_{\pi^*_f}^{h} ] - M_{i_h} \leq - \alpha^2 \cdot \frac{ \inf\limits_{[t_h - \alpha, t_h] } f'' } {2 \cdot f'(t_h - \alpha)}.
$$
As $f''(\cdot)$ is continuous, there exists $\alpha_{t_h} \in [0, \alpha]$ such that $\inf\limits_{[t_h - \alpha, t_h] } f''  = f''(t_h - \alpha_{t_h})$. Since $f'(\cdot)$ is non-increasing on $(-\infty, T_1)$, we derive:
$$
	\mathbb{E}[ X_{\pi^*_f}^{h} ] - M_{i_h} \leq - \alpha^2 \cdot \frac{ f''(t_h - \alpha_{t_h})} {2 \cdot f'(t_h - \alpha_{t_h})}.
$$
By assumption $\frac{ f'' } {f'}(\cdot)$ vanishes at $- \infty$ therefore we can pick $T_f$ small enough to get the desired inequality.
\\
(b) As $f' \rightarrow_{- \infty} a > 0$, we can find $T_f < T_1$ small enough such that:
$$
	|f'(t) - a| < \epsilon \quad \forall t \leq T_f,
$$
with $\epsilon = a \cdot \frac{\beta}{2 \alpha + \beta}$ and where $\beta$ is the right-hand side of the desired inequality. Consider $h \in \mathcal{H}$ such that $t_h \leq T_f$. Using \emph{Bellman's Principle of Optimality} for $\pi^*_f$, we have:
$$
	\mathbb{E}[f(t_h - X_{\pi^*_f}^{h})] = \mathbb{E}[f(T - ( (T - t_h) + X_{\pi^*_f}^{h} )] \geq  \mathbb{E}[f(T - ( (T - t_h) + X_{\pi_s}^{h} )]  \geq \mathbb{E}[f(t_h - X_{\pi_s}^h)].
$$ 
Since $f$ is $C^1$ on $(-\infty, T_f)$, this yields:
\begin{align*}
	0 & \leq \mathbb{E}[f(t_h - X_{\pi^*_f}^{h}) - f(t_h - X_{\pi_s}^h)] \\
	& \leq  \mathbb{E}[f(t_h - X_{\pi^*_f}^{h}) - f(t_h) + f(t_h) - f(t_h - X_{\pi_s}^h)] \\
	& \leq  \mathbb{E}[- \int_{t_h - X_{\pi^*_f}^{h}}^{t_h} f' + \int_{t_h - X_{\pi_s}^h}^{t_h} f' ] \\
	& \leq  \mathbb{E}[ - (a - \epsilon) \cdot X_{\pi^*_f}^{h} + (a + \epsilon) \cdot X_{\pi_s}^h ].
\end{align*}
Since the costs are independent across time and arcs:
$$
	\mathbb{E}[ X_{\pi_s}^{h} ] = M_{i_h}.
$$
Rearranging the last inequality, we derive:
\begin{align*}
	\mathbb{E}[ X_{\pi^*_f}^{h} ] - M_{i_h} & \leq \frac{2 \epsilon}{a - \epsilon} \cdot M_{i_h} \\
	& \leq \frac{2 \epsilon}{a - \epsilon} \cdot \alpha \\
	& < \beta,
\end{align*}
where we use the fact that $M_{i_h} \leq \alpha$ and the definition of $\epsilon$.
\\
\indent Starting from \eqref{eq-proof-LET-optimal}, consider $h \in \mathcal{H}$ such that $t_h \leq T_f$ and suppose by contradiction that $\pi^*_f(h) = j_h \notin \mathcal{T}(i_h)$. Even though the overall policy can be fairly complicated (history-dependent), the first action is deterministic and incurs an expected cost of $m_{i_h j_h}$ because the costs are independent across time and arcs. Moreover, when the objective is to minimize the average cost, the optimal strategy among all history-dependent rules is to follow the shortest path with respect to the mean arc costs (once again because the costs are independent across time and arcs). As a result:
$$
	\mathbb{E}[ X_{\pi^*_f}^{h} ] \geq  m_{i_h j_h } + M_{j_h},
$$
which implies:
$$
	\mathbb{E}[ X_{\pi^*_f}^{h} ] - M_{i_h} \geq  m_{i_h j_h } + M_{j_h} - M_{i_h},
$$
a contradiction.
\Halmos
\endproof

\subsection{Proof of Proposition \ref{lemma-dp-optimal-nominal}}
\label{section-proof-dp-optimal-nominal}

\proof{Proof of Proposition \ref{lemma-dp-optimal-nominal}.}
Using Theorem \ref{lemma-LET-is-opt-nominal}, the optimization problem \eqref{eq-nominal-problem} can be equivalently formulated as a discrete-time finite-horizon MDP in the extended space state $(i, t) \in \mathcal{V} \times  [T - \deltasup \cdot \ceil*{ \frac{T - T_f }{\deltainf} }, T]$ where $i$ is the current location and $t$ is the, possibly negative, remaining budget. Specifically:
\begin{itemize}
	\item{The time horizon is $\ceil*{ \frac{T - T_f }{\deltainf} }$,}
	\item{The initial state is $(s, T)$,}
	\item{The set of available actions at state $(i, t)$, for $i \neq d$, is taken as $\mathcal{V}(i)$. Picking $j \in \mathcal{V}(i)$ corresponds to crossing link $(i, j)$ and results in a transition to state $(j, t - \omega)$ with probability $p_{ij}(\omega) \mathrm{d}\omega$,}
	\item{The only available action at a state $(d, t)$ is to remain in this state,}
	\item{The transition rewards are all equal to $0$,}	
	\item{The final reward at the epoch $\ceil*{ \frac{T - T_f}{\deltainf} }$ for any state $(i, t)$ is equal to $f_i(t)$, which is the optimal expected objective-to-go when following the shortest path tree $\mathcal{T}$ starting at node $i$ with remaining budget $t$. Specifically, the collection of functions $(f_i(\cdot))_{i \in \mathcal{V}}$ is a solution to the following program:}
	\begin{align*}
     & f_d(t) = f(t), && t \leq T_f, \\
	 & f_i(t) = \max\limits_{j \in \mathcal{T}(i)} \int_0^{\infty} p_{ij}(\omega) \cdot f_j(t-\omega) \mathrm{d}\omega \quad && i \neq d, t \leq T_f.
	\end{align*}
\end{itemize}
Observe that Theorem \ref{lemma-LET-is-opt-nominal} is crucial to be able to define the final rewards. Proposition 4.4.3 of \cite{puterman2014markov} shows that any \emph{Markov} policy solution to \eqref{eq-dp-nominal} is an optimal solution to \eqref{eq-nominal-problem}.

\Halmos
\endproof

\subsection{Proof of Proposition \ref{lemma-quality-approx-nominal}}
\label{section-proof-quality-approx-nominal}

\proof{Proof of Proposition \ref{lemma-quality-approx-nominal}.}
For any node $i \in \mathcal{V}$, and $t \leq T$, we denote by $u^{\pi^{\Delta t}}_i(t)$ the expected risk function when following policy $\pi^{\Delta t}$ starting at $i$ with remaining budget $t$.  We deal with each case separately.

\paragraph{Case 1.}
We use the following useful facts:
\begin{itemize}
	\item{The functions $(u_i(\cdot))_{i \in \mathcal{V}}$ are non-decreasing,}
	\item{The functions $(u^{\Delta t}_i(\cdot))_{i \in \mathcal{V}}$ are non-decreasing,}
	\item{The functions $(u^{\Delta t}_i(\cdot))_{i \in \mathcal{V}}$ lower bound the functions $(u_i(\cdot))_{i \in \mathcal{V}}$.}
\end{itemize}
The main difficulty in proving convergence lies in the fact that the approximation $u_i^{\Delta t}(t)$ may not necessarily improve as $\Delta t$ decreases. However, this is the case for regular mesh size sequences such as $(\Delta t_p = \frac{1}{2^p})_{p \in \mathbb{N}}$. Hence, we first demonstrate convergence in that particular case in Lemma \ref{aux-lemma-regularmesh-nominal} and rely on this last result to prove pointwise convergence in general in Lemma \ref{aux-lemma-nonregularmesh-nominal}.
\begin{lemma}
\label{aux-lemma-regularmesh-nominal}
For the regular mesh $(\Delta t_p = \frac{1}{2^p})_{p \in \mathbb{N}}$, the sequence $(u_i^{\Delta t_p}(t))_{p \in \mathbb{N}}$ converges to $u_i(t)$ for almost every point $t$ in $[k^{\text{min}}_i \cdot \Delta t, T]$.
\proof{Proof}
	First observe that, for any $t$, the sequence $(u_i^{\Delta t_p}(t))_{p \in \mathbb{N}}$ is non-decreasing since (i) the discretization mesh used at step $p+1$ is strictly contained in the discretization mesh used at step $p$ and (ii) the functions $(u_i(\cdot))_{i \in \mathcal{V}}$ are non-decreasing. This shows that the functions $(u_i^{\Delta t_p}(\cdot))_{i \in \mathcal{V}}$ converge pointwise to some limits $(f_i(\cdot))_{i \in \mathcal{V}}$. Using the preliminary remarks, we get:
	$$	
		f_i(t) \leq u_i(t) \quad \forall t \in [k^{\text{min}}_i \cdot \Delta t, T], \forall i \in \mathcal{V}.
	$$
	Next, we establish that for any $i \in \mathcal{V}, t \in [k^{\text{min}}_i \cdot \Delta t, T]$ and $\epsilon >0$, $f_i(t) \geq u_i(t - \epsilon)$. This will enable us to squeeze $f_i(t)$ to finally derive $f_i(t) = u_i(t)$. We start with node $d$. Observe that, by construction of the approximation, $u_d^{\Delta t}(\cdot)$ converges pointwise to $f(\cdot)$ at every point of continuity of $f(\cdot)$. Furthermore, since $f_d(\cdot)$ and $u_d(\cdot)$ are non-decreasing, we have $f_d(t) \geq u_d(t - \epsilon)$ for all $t \in [k^{\text{min}}_d \cdot \Delta t, T]$ and for all $\epsilon >0$. Consider $\epsilon >0$ and a large enough $p$ such that $\epsilon > \frac{1}{2^p}$ which implies $\Delta t_p \cdot \lfloor \frac{t}{\Delta t_p} \rfloor \geq t - \epsilon$. We first show by induction on the level of the nodes in $\mathcal{T}$ that:
	$$
		f_i(t) \geq  u_i(t - \text{level}(i, \mathcal{T}) \cdot \epsilon) \quad \forall t \in [k^{\text{min}}_i \cdot \Delta t, \floor*{ \frac{T_f}{ \Delta t} } \cdot \Delta t), \forall i \in \mathcal{V}.
	$$
	The base case follows from the discussion above. Assume that the induction property holds for all nodes of level less than $l$ and consider a node $i \in \mathcal{V}$ of level $l +1$. We have, for $t \in [k^{\text{min}}_i \cdot \Delta t, \floor*{ \frac{T_f}{ \Delta t} } \cdot \Delta t)$:
	\begin{align*}
		u_i^{\Delta t_p}(t) 
			& = u_i^{\Delta t_p}( \floor*{\frac{t}{\Delta t_p}} \cdot \Delta t_p ) \\
			& \geq  \max\limits_{j \in \mathcal{T}(i)} \mathbb{E} [ u_{j}^{\Delta t_p}( \floor*{\frac{t}{\Delta t_p}} \cdot \Delta t_p - c_{ij}) ] \\
			& \geq \max\limits_{j \in \mathcal{T}(i)} \mathbb{E} [ u_{j}^{\Delta t_p}( t - \epsilon - c_{ij}) ].
	\end{align*}
	To take the limit $p \rightarrow \infty$ in the previous inequality, note that, for any $j \in \mathcal{T}(i)$:
	$$
		u_{j}^{\Delta t_p}(t - \epsilon - c_{ij}) \geq u_{j}^{\Delta t_1}(t - \epsilon - \deltasup),
	$$ 
	while $(u_{j}^{\Delta t_p}(t - \epsilon - c_{ij}))_{p \in \mathbb{N}}$ is non-decreasing and converges almost surely to $f_j(t - \epsilon - c_{ij})$ as $p \rightarrow \infty$. Therefore, we can apply the monotone convergence theorem and derive:
	$$
		f_i(t) \geq \mathbb{E}[ f_{j}(t - \epsilon -  c_{ij})].
	$$
	As the last inequality holds for any $j \in \mathcal{T}(i)$, we finally obtain:
	$$
		f_i(t) \geq \max_{j \in \mathcal{T}(i)} \mathbb{E} [ f_{j}(t - \epsilon - c_{ij})] \quad \forall t \in [k^{\text{min}}_i \cdot \Delta t, \floor*{ \frac{T_f}{ \Delta t} } \cdot \Delta t).
	$$
	 Using the induction property along with Theorem \ref{lemma-LET-is-opt-nominal}, we get: 
	\begin{align*}
		f_i(t) 
		& \geq \max_{j \in \mathcal{T}(i)} \mathbb{E} [ u_j(t - \text{level}(i, \mathcal{T}) \cdot \epsilon - c_{ij}) ] \\
		& \geq u_i(t - \text{level}(i, \mathcal{T}) \cdot \epsilon),
	\end{align*}
	for all $t \in [k^{\text{min}}_i \cdot \Delta t, \floor*{ \frac{T_f}{ \Delta t} } \cdot \Delta t)$, which concludes the induction. We can now prove by induction on $m$, along the same lines as above, that:
	$$
		f_i(t) \geq u_i(t - (|\mathcal{V}| + m) \cdot \epsilon) \quad \forall t \in [k^{\text{min}}_i \cdot \Delta t, \floor*{ \frac{T_f}{\Delta t} } \cdot \Delta t + m \cdot \deltainf), \forall i \in \mathcal{V},
	$$	 
	for all $m \in \mathbb{N}$. This last result can be reformulated as:
	$$
		f_i(t) \geq u_i(t - \epsilon) \quad \forall \epsilon >0, \forall t \in [k^{\text{min}}_i \cdot \Delta t, T], \forall i \in \mathcal{V}.
	$$
	Combining this lower bound with the upper bound previously derived, we get:
	$$
		u_i(t) \geq f_i(t) \geq u_i(t^-) \quad \forall t \in [k^{\text{min}}_i \cdot \Delta t, T], \forall i \in \mathcal{V},
	$$
	where $u_i(t^{-})$ refers to the left one-sided limit of $u_i(\cdot)$ at $t$. Since, $u_i(\cdot)$ is non-decreasing, it has countably many discontinuity points and the last inequality shows that $f_i(\cdot) = u_i(\cdot)$ almost everywhere on $[k^{\text{min}}_i \cdot \Delta t, T]$.
\Halmos
\endproof
\end{lemma}

\begin{lemma}
\label{aux-lemma-nonregularmesh-nominal}
For any sequence $(\Delta t_p)_{p \in \mathbb{N}}$ converging to $0$, the sequence $(u_i^{\Delta t_p}(t))_{p \in \mathbb{N}}$ converges to $u_i(t)$ for almost every point $t$ in $[k^{\text{min}}_i \cdot \Delta t, T]$.
\proof{Proof}
In contrast to the particular case handled by Lemma \ref{aux-lemma-regularmesh-nominal}, our approximation of $u_i(t)$ may not improve as $p$ increases. For that reason, there is no straightforward comparison between $(u_i^{\Delta t_p}(t))_p$ and $u_i(t)$. However, for a given $i \in \mathcal{V}$, $t \in [k^{\text{min}} \cdot \Delta t, T]$, $\epsilon >0$ and a large enough $p$, $(u_i^{\Delta t_p}(t))_p$ can be shown to be lower bounded by a subsequence of $(u_i^{\frac{1}{2^p}}(t - \epsilon))_p$. This is how we proceed to establish convergence.
	\\
	\indent Consider $i \in \mathcal{V}$, $t \in [k^{\text{min}} \cdot \Delta t, T]$, $\epsilon >0$ and $p \in \mathbb{N}$. Define $\sigma(p) \in \mathbb{N}$ as the unique integer satisfying $\frac{1}{2^{\sigma(p)-1}} < \Delta t_p \leq \frac{1}{2^{\sigma(p)}}$. Since $\lim_{p \rightarrow \infty} \Delta t_p = 0$, we necessarily have $\lim_{p \rightarrow \infty} \sigma(p) = \infty$. Remark that $u_i^{\frac{1}{2^{\sigma(p)}}}(\cdot)$ has steps of size $\frac{1}{2^{\sigma(p)}} \geq \Delta t_p$, i.e. $u_i^{\Delta t_p}(\cdot)$ is expected to be a tighter approximation of $u_i(\cdot)$ than $u_i^{\frac{1}{2^{\sigma(p)}}}(\cdot)$ is. However, the time steps do not overlap (multiples of either $\Delta t_p$ or $\frac{1}{2^p}$) making the two sequences impossible to compare. Nevertheless, the time steps differ by no more than $\Delta t_p$. Thus, if $p$ is large enough so that $\Delta t_p < \epsilon$, for each update needed to calculate $u_i^{\frac{1}{2^{\sigma(p)}}}(t - \epsilon)$, there is a corresponding update for a larger budget to compute $u_i^{\Delta t_p}(t)$. As a consequence, the sequence $(u_i^{\frac{1}{2^{\sigma(p)}}}(t- \epsilon))_p$ constitutes a lower bound on the sequence of interest $(u_i^{\Delta t_p}(t))_p$. Using the preliminary remarks, we are able to squeeze $(u_i^{\Delta t_p}(t))_p$:
	$$
		u_i^{\frac{1}{2^{\sigma(p)}}}(t - \epsilon) \leq u_i^{\Delta t_p}(t) \leq u_i(t),
	$$
	for all $i \in \mathcal{V}$, $t \in [k^{\text{min}} \cdot \Delta t, T]$, $\epsilon >0$ and for $p$ large enough. This can be proved first by induction on the level of the nodes in $\mathcal{T}$ and then by interval increments of size $\deltainf$ along the same lines as what is done in Lemma \ref{aux-lemma-regularmesh-nominal}. Yet, Lemma \ref{aux-lemma-regularmesh-nominal} shows that:
	$$
		\lim_{p \rightarrow \infty} u_i^{\frac{1}{2^{\sigma(p)}}}(t - \epsilon) = u_i(t-\epsilon),
	$$
	provided $t - \epsilon$ is a point of continuity for $u_i(\cdot)$. As $u_i(\cdot)$ has countably many discontinuity points (it is non-decreasing), the last inequality shows, by taking $p$ large enough and $\epsilon$ small enough, that $u_i^{\Delta t_p}(t) \rightarrow_{p \rightarrow \infty} u_i(t)$ for $t$ a point of continuity of $u_i(\cdot)$.
\Halmos
\endproof
\end{lemma}

\paragraph{Case 2.}
The first step consists in proving that the functions $(u_i(\cdot))_{i \in \mathcal{V}}$ are continuous on $(-\infty, T]$. By induction on $l$, we start by proving that $u_i(\cdot)$ is continuous on $(-\infty, T_f)$ for all nodes $i$ of level $l$ in $\mathcal{T}$. The base case follows from the continuity of $f(\cdot)$. Assuming the property holds for some $l \geq 1$, we consider a node $i$ of level $l+1$ in $\mathcal{T}$, $t < T_f$ and a sequence $t_n \rightarrow_{n \rightarrow \infty} t$. Using Theorem \ref{lemma-LET-is-opt-nominal}, we have:
$$
	|u_i(t) - u_i(t_n)|
		\leq \max_{j \in \mathcal{T}(i)} \mathbb{E}[| u_j(t - c_{ij}) - u_j(t_n - c_{ij}) |].
$$
For any $j \in \mathcal{T}(i)$, we can use the uniform continuity of $u_j(\cdot)$ on $[t - 2 \cdot \deltasup, t]$ to prove that this last term converges to $0$ as $n \rightarrow \infty$. We conclude that all the functions $(u_i(\cdot))_{i \in \mathcal{V}}$ are continuous on $(-\infty, T_f)$. By induction on $m$, we can then show that the functions $(u_i(\cdot))_{i \in \mathcal{V}}$ are continuous on $(-\infty, T_f + m \cdot \deltainf)$, to finally conclude that they are continuous on $(-\infty, T]$. We are now able to prove uniform convergence. Since $[T_f - |\mathcal{V}| \cdot \deltasup, T]$ is a compact set, the functions $(u_i(\cdot))_{i \in \mathcal{V}}$ are also uniformly continuous on this set. Take $\epsilon >0$, there exists $\alpha > 0$ such that:
$$
	\forall i \in \mathcal{V}, | u_i(\omega) - u_i(\omega') | \leq \epsilon, \; \forall (\omega, \omega') \in [T_f - |\mathcal{V}| \cdot \deltasup, T]^2 \; \text{with} \; | \omega - \omega'| \leq \alpha.
$$
Building on this, we can show, by induction on the level of the nodes in $\mathcal{T}$, that:
$$
	\sup_{\omega \in [k^{\text{min}}_i \cdot \Delta t,  \floor*{ \frac{T_f}{\Delta t} } \cdot \Delta t )} | u_i(\omega) -  u^{\Delta t}_i(\omega) | \leq \text{level}(i, \mathcal{T}) \cdot \epsilon, \quad \forall i \in \mathcal{V}.
$$
This follows from the sequence of inequalities:
\begin{align*}
	\sup_{\omega \in [k^{\text{min}}_i \cdot \Delta t, \floor*{ \frac{T_f}{\Delta t} } \cdot \Delta t )} | u^{\Delta t}_i(\omega) - u_i(\omega)| 
	& \leq \sup_{k \in \{ k^{\text{min}}_i, \cdots, \floor*{ \frac{T_f}{\Delta t}} - 1  \} } | u^{\Delta t}_i( k \cdot \Delta t ) - u_i(k \cdot \Delta t)| \\
	& + \sup_{\omega \in [k^{\text{min}}_i \cdot \Delta t, \floor*{ \frac{T_f}{\Delta t} } \cdot \Delta t ]} | u_i(\omega) - u_i( \floor*{ \frac{\omega}{\Delta t} } \cdot \Delta t)| \\
	& \leq  \sup_{k \in \{ k^{\text{min}}_i, \cdots, \floor*{ \frac{T_f}{\Delta t} } - 1 \} } \max_{j \in \mathcal{T}(i)} \mathbb{E}[| u^{\Delta t}_j( k \cdot \Delta t - c_{ij}) - u_j( k \cdot \Delta t - c_{ij})  |] \\
	& + \epsilon \\
	& \leq  (\text{level}(i, \mathcal{T})-1) \cdot \epsilon + \epsilon \\
	& \leq  \text{level}(i, \mathcal{T}) \cdot \epsilon.
\end{align*}
We conclude that:
$$
	\sup_{\omega \in [k^{\text{min}}_i \cdot \Delta t, \floor*{ \frac{T_f}{\Delta t} } \cdot \Delta t ) } | u_i(\omega) -  u^{\Delta t}_i(\omega) | \leq |\mathcal{V}| \cdot \epsilon, \forall i \in \mathcal{V}.	
$$
Along the same lines, we can show by induction on $m$ that:
$$
	\sup_{\omega \in [k^{\text{min}}_i \cdot \Delta t, \floor*{ \frac{T_f}{\Delta t} } \cdot \Delta t + m \cdot \deltainf)} | u^{\Delta t}_i(\omega) - u_i(\omega) | \leq (|\mathcal{V}| + m) \cdot \epsilon, \quad \forall i \in \mathcal{V}. 
$$
This implies:
$$
	\sup_{\omega \in [k^{\text{min}}_i \cdot \Delta t, T]} | u^{\Delta t}_i(\omega) - u_i(\omega) | \leq (|\mathcal{V}| + \ceil*{ \frac{T - T_f}{\deltainf} } + 1) \cdot \epsilon, \quad \forall i \in \mathcal{V},
$$
assuming $\Delta t \leq \deltainf$. In particular, this shows uniform convergence. To conclude the proof of Case 2, we show that $\pi^{\Delta t}$ is a $o(1)$-approximate optimal solution to \eqref{eq-nominal-problem} as $\Delta t \rightarrow 0$. Using the last set of inequalities derived in combination with the uniform continuity of the functions $(u_i(\cdot))_{i \in \mathcal{V}}$, we can show that:
\begin{equation}
	\label{eq-proof-uni-continuity-approx}
	\forall i \in \mathcal{V}, | u^{\Delta t}_i(\omega) - u^{\Delta t}_i(\omega') | \leq (2 \cdot |\mathcal{V}| + 2 \cdot \ceil*{ \frac{T - T_f}{\deltainf} } + 3) \cdot \epsilon, \; \forall (\omega, \omega') \in [k^{\text{min}}_i \cdot \Delta t, T]^2 \; \text{with} \; | \omega - \omega'| \leq \alpha,
\end{equation}
and:
\begin{equation}
	\label{eq-proof-cross-uni-continuity-approx}
	\forall i \in \mathcal{V}, | u^{\Delta t}_i(\omega) - u_i(\omega') | \leq (|\mathcal{V}| + \ceil*{ \frac{T - T_f}{\deltainf} } + 2) \cdot \epsilon, \; \forall (\omega, \omega') \in [k^{\text{min}}_i \cdot \Delta t, T]^2  \; \text{with} \; | \omega - \omega'| \leq \alpha.
\end{equation}
We can now prove, by induction on the level of the nodes in $\mathcal{T}$, that:
$$
	u^{\pi^{\Delta t}}_i(t) \geq u_i(t) - 3 \cdot \text{level}(i, \mathcal{T}) \cdot (|\mathcal{V}| + \ceil*{ \frac{T - T_f}{\deltainf} } + 2) \cdot \epsilon, \quad \forall t \in [k^{\text{min}}_i \cdot \Delta t, \floor*{ \frac{T_f}{\Delta t} } \cdot \Delta t), \forall i \in \mathcal{V}.
$$
This follows from the sequence of inequalities:
\begin{align*}
	u^{\pi^{\Delta t}}_i(t) 
	& = \int_0^\infty p_{i\pi^{\Delta t}(i, t)}(\omega) \cdot u^{\pi^{\Delta t}}_{ \pi^{\Delta t}(i, t) }(t - w) \mathrm{d}\omega \\
	& \geq  \int_0^\infty p_{i\pi^{\Delta t}(i, t)}(\omega) \cdot u_{ \pi^{\Delta t}(i, t) }(t - w) \mathrm{d}\omega - 3 \cdot ( \text{level}(i, \mathcal{T}) - 1) \cdot (|\mathcal{V}| + \ceil*{ \frac{T - T_f}{\deltainf} } + 2) \cdot \epsilon \\
	& \geq  \int_0^\infty p_{i\pi^{\Delta t}(i, t)}(\omega) \cdot u^{\Delta t}_{ \pi^{\Delta t}(i, t) }(t - w) \mathrm{d}\omega - \epsilon - 3 \cdot ( \text{level}(i, \mathcal{T}) - 1) \cdot (|\mathcal{V}| + \ceil*{ \frac{T - T_f}{\deltainf} } + 2) \cdot \epsilon \\
	& \geq  \int_0^\infty p_{i\pi^{\Delta t}(i, t)}(\omega) \cdot u^{\Delta t}_{ \pi^{\Delta t}(i, \floor*{ \frac{t}{\Delta t} } \cdot \Delta t ) }( \floor*{ \frac{t}{\Delta t} } \cdot \Delta t  - w) \mathrm{d}\omega - \epsilon - (2 \cdot |\mathcal{V}| + 2 \cdot \ceil*{ \frac{T - T_f}{\deltainf} } + 3) \cdot \epsilon \\
	& - 3 \cdot ( \text{level}(i, \mathcal{T}) - 1) \cdot (|\mathcal{V}| + \ceil*{ \frac{T - T_f}{\deltainf} } + 2) \cdot \epsilon  \\
	& \geq u^{\Delta t}_{ i }( \floor*{ \frac{t}{\Delta t} } \cdot \Delta t )  - 3 \cdot \text{level}(i, \mathcal{T}) \cdot (|\mathcal{V}| + \ceil*{ \frac{T - T_f}{\deltainf} } + 2) \cdot \epsilon,
\end{align*}
where we use the induction property for the first inequality, the uniform convergence for the second, \eqref{eq-proof-uni-continuity-approx} for the third, the definition of $\pi^{\Delta t}(i, t)$ for the fourth and finally \eqref{eq-proof-cross-uni-continuity-approx}. We conclude that:
$$
	u^{\pi^{\Delta t}}_i(t) \geq u_i(t) - 3 \cdot |\mathcal{V}| \cdot (|\mathcal{V}| + \ceil*{ \frac{T - T_f}{\deltainf} } + 2) \cdot \epsilon, \quad \forall t \in [k^{\text{min}}_i \cdot \Delta t, \floor*{ \frac{T_f}{\Delta t} } \cdot \Delta t), \forall i \in \mathcal{V}.
$$
We can then prove by induction on $m$, in the same fashion as above, that:
$$
	u^{\pi^{\Delta t}}_i(t) \geq u_i(t) - 3 \cdot (m + |\mathcal{V}|) \cdot (|\mathcal{V}| + \ceil*{ \frac{T - T_f}{\deltainf} } + 2) \cdot \epsilon, \forall t \in [k^{\text{min}}_i \cdot \Delta t, \floor*{ \frac{T_f}{\Delta t} } \cdot \Delta t + m \cdot \deltainf), \forall i \in \mathcal{V},
$$
for all $m$. We conclude that $u^{\pi^{\Delta t}}_s(T) \geq u_s(T) - 3 \cdot (|\mathcal{V}| + \ceil*{ \frac{T - T_f}{\deltainf} } + 2)^2 \cdot \epsilon$, which establishes the claim.

\paragraph{Case 3.}
The first step consists in showing that the functions $(u_i(\cdot))_{i \in \mathcal{V}}$ are Lipschitz on $[T_f - |\mathcal{V}| \cdot \deltasup, T]$. Take $K$ to be a Lipschitz constant for $f(\cdot)$ on $[T_f - ( \ceil*{ \frac{T - T_f}{\deltainf} } + 2 \cdot |\mathcal{V}| - 1) \cdot \deltasup, T]$. We first show by induction on $l$ that $u_i(\cdot)$ is $K$-Lipschitz on $[T_f - ( \ceil*{ \frac{T - T_f}{\deltainf} } + 2 \cdot |\mathcal{V}| - l) \cdot \deltasup, T_f]$ for all nodes $i$ of level $l$ in $\mathcal{T}$. The base case follows from the definition of $K$. Assuming the property holds for some $l \geq 1$, we consider a node $i$ of level $l+1$ in $\mathcal{T}$. Using Theorem \ref{lemma-LET-is-opt-nominal}, we have, for $(t, t') \in [T_f - ( \ceil*{ \frac{T - T_f}{\deltainf} } + 2 \cdot  |\mathcal{V}| - l - 1) \cdot \deltasup, T_f]^2$:
\begin{align*}
	|u_i(t) - u_i(t')|
		& \leq \max_{j \in \mathcal{T}(i)} \mathbb{E}[| u_j(t - \omega) - u_j(t' - \omega) |] \\
		& \leq K \cdot |t - t'|,
\end{align*}
where we use the induction property for $l$ (recall that $p_{ij}(\omega) = 0$ for $\omega \geq \deltasup$). We conclude that the functions $(u_i(\cdot))_{i \in \mathcal{V}}$ are all $K-$Lipschitz on $[T_f - ( \ceil*{ \frac{T - T_f}{\deltainf} } + |\mathcal{V}| ) \cdot \deltasup, T_f]$. We now prove, by induction on $m$, that the functions $(u_i(\cdot))_{i \in \mathcal{V}}$ are all $K$-Lipschitz on $[T_f - (\ceil*{ \frac{T - T_f}{\deltainf} } - m + |\mathcal{V}|) \cdot \deltasup, T_f + m \cdot \deltainf]$. The base case follows from the previous induction. Assuming the property holds for some $m$, we have for $i \in \mathcal{V}$ and for $(t, t') \in [T_f - (\ceil*{ \frac{T - T_f}{\deltainf} } - m - 1 + |\mathcal{V}|) \cdot \deltasup, T_f + (m+1) \cdot \deltainf]^2$:
\begin{align*}
	|u_i(t) - u_i(t')|
		& \leq \max_{j \in \mathcal{V}(i)} \mathbb{E}[| u_j(t - \omega) - u_j(t' - \omega) |]\\
		& \leq K \cdot |t - t'|,
\end{align*}
where we use the fact that $p_{ij}(\omega) = 0$ for $\omega \leq \deltainf$ or $\omega \geq \deltasup$ and the induction property. We conclude that the function $(u_i(\cdot))_{i \in \mathcal{V}}$ are all $K$-Lipschitz on $[T_f - |\mathcal{V}| \cdot \deltasup, T]$. Using this last fact, we can prove, by induction on the level of the nodes in $\mathcal{T}$, in a similar fashion as done for Case 2, that:
$$
	\sup_{\omega \in [k^{\text{min}}_i \cdot \Delta t, \floor*{ \frac{T_f}{\Delta t} } \cdot \Delta t)} | u^{\Delta t}_i(\omega) - u_i(\omega) | \leq \text{level}(i, \mathcal{T}) \cdot K \cdot \Delta t, \quad \forall i \in \mathcal{V}.
$$ 
By induction on $m$, we can then show that:
$$
	\sup_{\omega \in [k^{\text{min}}_i \cdot \Delta t, \floor*{ \frac{T_f}{\Delta t} } \cdot \Delta t + m \cdot \deltainf)} | u^{\Delta t}_i(\omega) - u_i(\omega) | \leq (|\mathcal{V}| + m) \cdot K \cdot \Delta t, \quad \forall i \in \mathcal{V}, \forall m \in \mathbb{N}.
$$
This implies:
$$
	\sup_{\omega \in [k^{\text{min}}_i \cdot \Delta t, T]} | u^{\Delta t}_i(\omega) - u_i(\omega) | \leq (|\mathcal{V}| + \ceil*{ \frac{T - T_f}{\deltainf} } + 1 ) \cdot K \cdot \Delta t, \quad \forall i \in \mathcal{V},
$$
assuming $\Delta t \leq \deltainf$. This shows uniform convergence at speed $\Delta t$. To conclude the proof of Case 3, we show that $\pi^{\Delta t}$ is a $O(\Delta t)$-approximate optimal solution to \eqref{eq-nominal-problem} as $\Delta t \rightarrow 0$. We can show, using the last inequality derived along with the same sequence of inequalities as in Case 2, by induction on the level of the nodes in $\mathcal{T}$ that:
$$
	u^{\pi^{\Delta t}}_i(t) \geq u_i(t) - 6 \cdot \text{level}(i, \mathcal{T}) \cdot (|\mathcal{V}| + \ceil*{ \frac{T - T_f}{\deltainf} } + 1) \cdot K \cdot \Delta t, \quad \forall t \in [k^{\text{min}}_i \cdot \Delta t, \floor*{ \frac{T_f}{\Delta t} } \cdot \Delta t), \forall i \in \mathcal{V}.
$$
We can then prove by induction on $m$, in the same fashion as in Case 2, that:
$$
	u^{\pi^{\Delta t}}_i(t) \geq u_i(t) - 6 \cdot (m + |\mathcal{V}|) \cdot (|\mathcal{V}| + \ceil*{ \frac{T - T_f}{\deltainf} } + 1) \cdot K \cdot \Delta t, \forall t \in [k^{\text{min}}_i \cdot \Delta t, \floor*{ \frac{T_f}{\Delta t} } \cdot \Delta t + m \cdot \deltainf), \forall i \in \mathcal{V},
$$
for all $m \in \mathbb{N}$. We conclude that $u^{\pi^{\Delta t}}_s(T) \geq u_s(T) - 6 \cdot (|\mathcal{V}| + \ceil*{ \frac{T - T_f}{\deltainf} } + 1 )^2 \cdot K \cdot \Delta t$, which establishes the claim.

\paragraph{Case 4.}
In this situation, we can show by induction on $m$ that the functions $(u_i(\cdot))_{i \in \mathcal{V}}$ are continuous on $[0, m \cdot \deltainf]$ and conclude that the functions $(u_i(\cdot))_{i \in \mathcal{V}}$ are continuous on $[0, T]$. Since $[0, T]$ is a compact set, these functions are also uniformly continuous on $[0, T]$. Moreover, $u^{\Delta t}_i(t) = u_i(t) = 0$ for all $t \leq 0$ and $i \in \mathcal{V}$. Using these two observations, we can apply the same techniques as in Case 2 to obtain the same results.
\Halmos
\endproof

\subsection{Proof of Theorem \ref{lemma-LET-is-opt-robust}}
\label{section-proof-lemma-LET-is-opt-robust}

\proof{Proof of Theorem \ref{lemma-LET-is-opt-robust}.} 
As in Theorem \ref{lemma-LET-is-opt-nominal}, the last part of the theorem is trivial because any strategy is optimal when having already spent a budget of $T - T^r_f$. 
\\
\indent The proof for the first part is an extension of Theorem \ref{lemma-LET-is-opt-nominal} and follows the same steps. We denote by $(m_{ij})_{(i, j) \in \mathcal{A}}$ the worst-case expected costs, i.e.:
$$
	m_{ij} = \sup_{p_{ij} \in \mathcal{P}_{ij}} \mathbb{E}_{X \sim p_{ij}}[X] \quad \forall (i, j) \in \mathcal{A}.
$$
Observe that these quantities are well-defined as $\mathcal{P}_{ij}$ is not empty and $\mathbb{E}_{X \sim p_{ij}}[X] \leq \deltasup$ for any $p_{ij} \in \mathcal{P}_{ij}$. Furthermore, there exists $p^*_{ij} \in \mathcal{P}_{ij}$ such that $m_{ij} = \mathbb{E}_{X \sim p^*_{ij}}[X]$ as $\mathcal{P}_{ij}$ is compact for the weak topology. For any node $i \neq d$, we define $M_i$ as the length of a shortest path from $i$ to $d$ in $\mathcal{G}$ when the arc cots are taken as $(m_{ij})_{(i, j) \in \mathcal{A}}$. Just like in Theorem \ref{lemma-LET-is-opt-nominal}, we consider an optimal strategy $\pi^*_{f, \mathcal{P}}$ solution to \eqref{eq-robust-problem}. For a given history $h \in \mathcal{H}$, we define $t_h$ as the remaining budget, i.e. $T$ minus the total cost spent so far, and $i_h$ as the current location. The policy $\pi^*_{f, \mathcal{P}}$ maps $h \in \mathcal{H}$ to a probability distribution over $\mathcal{V}(i_h)$. Observe that randomizing does not help because (i) the costs are independent across time and arcs and (ii) the ambiguity set is rectangular. Hence, without loss of generality, we may assume that $\pi^*_{f, \mathcal{P}}$ actually maps $h$ to the node in $\mathcal{V}(i_h)$ minimizing the worst-case objective function given $h$. For $h \in \mathcal{H}$, we denote by $X_{\pi^*_{f, \mathcal{P}}}^{h}$ the random cost-to-go incurred by following strategy $\pi^*_{f, \mathcal{P}}$, i.e. not including the total cost spent up to this point of the history $T - t_h$. We define $\pi_s$ as a policy associated with an arbitrary shortest path from $i$ to $d$ with respect to $(m_{ij})_{(i, j) \in \mathcal{A}}$. Specifically, $\pi_s$ maps the current location $i_h$ to a node in $\mathcal{T}^r(i_h)$, irrespective of the history of the process. Similarly as for $\pi^*_{f, \mathcal{P}}$, we denote by $X_{\pi_s}^{h}$ the random cost-to-go incurred by following strategy $\pi_s$ for $h \in \mathcal{H}$. Using \emph{Bellman's Principle of Optimality} for $\pi^*_{f, \mathcal{P}}$, we have:
\begin{align*}
	\mathbb{E}_{ \mathbf{p^*} } \; [f(t_h - X^{h}_{\pi^*_{f, \mathcal{P}}})] 
		& \geq \inf_{ \forall \tau \geq T - t_h, \forall (i, j) \in \mathcal{A}, \; p^\tau_{ij} \in \mathcal{P}_{ij} } \; \mathbb{E}_{ \mathbf{p^\tau} } [f(t - X^{h}_{\pi^*_{f, \mathcal{P}}})]  \\
		& \geq \sup_{\pi \in \Pi} \; \inf_{ \forall \tau \geq T - t_h, \forall (i, j) \in \mathcal{A}, \; p^\tau_{ij} \in \mathcal{P}_{ij} } \; \mathbb{E}_{ \mathbf{p^\tau} } [f(t - X^{h}_{\pi})]  \\
		& \geq \inf_{ \forall \tau \geq T - t_h, \forall (i, j) \in \mathcal{A}, \; p^\tau_{ij} \in \mathcal{P}_{ij} } \; \mathbb{E}_{ \mathbf{p^\tau} } [f(t - X^{h}_{\pi_s})]  \\
		& \geq \mathbb{E}_{ \mathbf{q^\tau}} [f(t - X^{h}_{\pi_s})]  ,
\end{align*}
where $(q^\tau_{ij})_{(i, j) \in \mathcal{A}, \tau \geq T - t_h}$ is given by the worst-case scenario in the ambiguity sets, i.e.:
$$
	(q^\tau_{ij})_{(i, j) \in \mathcal{A}, \tau \geq T - t_h} \in \argmin_{ \forall \tau \geq T - t_h, \forall (i, j) \in \mathcal{A}, \; p^\tau_{ij} \in \mathcal{P}_{ij} } \mathbb{E}_{ \mathbf{p^\tau} } [f(t - X^{h}_{\pi_s})],
$$
which can be shown to exist because the ambiguity sets are compact. Using the last inequality derived, we can prove, using the exact same sequence of inequalities as in Theorem \ref{lemma-LET-is-opt-nominal}, that there exists $T^r_f$ such that, for both cases (a) and (b):
\begin{equation}
	\label{eq-proof-worst-case-LET-optimal}
	\mathbb{E}_{ \mathbf{p^*} }[ X^{h}_{\pi^*_{f, \mathcal{P}}} ] - \mathbb{E}_{ \mathbf{q^\tau} }[ X^{h}_{\pi_s} ] < \min_{i \neq d} \min_{j \in \mathcal{V}(i), j \notin \mathcal{T}^r(i) } \{ m_{ij} + M_j - M_i \} \quad \forall h \in \mathcal{H} \; \text{ such that } \; t_h \leq T^r_f,
\end{equation}
with the convention that the minimum of an empty set is equal to infinity. Starting from \eqref{eq-proof-worst-case-LET-optimal}, consider $h \in \mathcal{H}$ such that $t_h \leq T^r_f$ and suppose by contradiction that $\pi^*_{f, \mathcal{P}}(h) = j_h \notin \mathcal{T}^r(i_h)$. As mentioned in Theorem \ref{lemma-LET-is-opt-nominal}, even though $\pi^*_{f, \mathcal{P}}$ can be fairly complicated, the first action is deterministic and incurs an expected cost of $m_{i_h j_h}$ because the costs are independent across time and arcs. Moreover, when the objective is to minimize the average cost, the optimal strategy among all history-dependent rules is to follow the shortest path with respect to the mean arc costs (once again because the costs are independent across time and arcs). As a result:
$$
	\mathbb{E}_{ \mathbf{p^*} }[ X^{h}_{\pi^*_{f, \mathcal{P}}} ] \geq  m_{i_h j_h} + M_{j_h}.
$$
Additionally, by definition of $(p^*_{ij})_{(i, j) \in \mathcal{A}}$:
\begin{align*}
	\mathbb{E}_{ \mathbf{q^\tau} }[ X^{h}_{\pi_s} ]
		& \leq \mathbb{E}_{ \mathbf{p^*} }[ X^{h}_{\pi_{s}} ]\\
		& \leq M_{i_h}.
\end{align*}
This implies:
$$
	\mathbb{E}_{ \mathbf{p^*} }[ X^{h}_{\pi^*_{f, \mathcal{P}}} ] - \mathbb{E}_{ \mathbf{q^\tau} }[ X^{h}_{\pi_s} ] \geq m_{i_h j_h } + M_{j_h} - M_{i_h},
$$
a contradiction. We conclude that:
$$
	\pi^*_{f, \mathcal{P}}(h) \in \mathcal{T}^r(i_h) \quad \forall h \in \mathcal{H} \; \text{ such that } t_h \leq T^r_f.
$$
\Halmos
\endproof

\subsection{Proof of Proposition \ref{lemma-dp-optimal-robust}.}
\label{section-proof-lemma-dp-optimal-robust}

\proof{Proof of Proposition \ref{lemma-dp-optimal-robust}.} 
The proof uses a reduction to distributionally robust finite-horizon MDPs in a similar fashion as in Proposition \ref{lemma-dp-optimal-nominal}. Using Theorem \ref{lemma-LET-is-opt-robust}, the optimization problem \eqref{eq-robust-problem} can be equivalently formulated as a discrete-time finite-horizon distributionally robust MDP in the extended space state $(i, t) \in \mathcal{V} \times  [T - \deltasup \cdot \ceil*{ \frac{T - T^r_f }{\deltainf} }, T]$ where $i$ is the current location and $t$ is the, possibly negative, remaining budget. Specifically:
\begin{itemize}
	\item{The time horizon is $\ceil*{ \frac{T - T^r_f }{\deltainf} }$,}
	\item{The initial state is $(s, T)$,}
	\item{The set of available actions at state $(i, t)$, for $i \neq d$, is taken as $\mathcal{V}(i)$. Picking $j \in \mathcal{V}(i)$ corresponds to crossing link $(i, j)$ and results in a transition to state $(j, t - \omega)$ with probability $p_{ij}(\omega) \mathrm{d}\omega$,}
	\item{The probability of transitions are only known to lie in the rectangular ambiguity set:}
		$$\prod\limits_{\substack{ (i, j) \in \mathcal{A}  \\ t \in [T - \deltasup \cdot \ceil*{ \frac{T - T^r_f }{\deltainf} }, T] }} \mathcal{P}_{ij},
		$$
	\item{The only available action at a state $(d, t)$ is to remain in this state,}
	\item{The transition rewards are all equal to $0$,}	
	\item{The final reward at the epoch $\ceil*{ \frac{T - T^r_f}{\deltainf} }$ for any state $(i, t)$ is equal to $f_i(t)$, which is the optimal worst-case expected objective-to-go when following the shortest path tree $\mathcal{T}^r$ starting at node $i$ with remaining budget $t$. Specifically, the collection of functions $(f_i(\cdot))_{i \in \mathcal{V}}$ is a solution to the following program:}
	\begin{align*}
     	& f_d(t) = f(t), && t \leq T^r_f \\
	 	& f_i(t) = \max\limits_{j \in \mathcal{T}^r(i)} \inf\limits_{p_{ij} \in \mathcal{P}_{ij}} \int_0^{\infty} p_{ij}(\omega) \cdot f_j(t-\omega) \mathrm{d}\omega \quad && i \neq d, t \leq T^r_f.
	\end{align*}
\end{itemize}
As a consequence, we can conclude the proof with Theorem 2.2 of \cite{iyengar2005robust} (or equivalently Theorem 1 of \cite{RobustMDPElGahoui}).
\Halmos
\endproof

\subsection{Proof of Proposition \ref{lemma-quality-approx-robust}}
\label{section-proof-lemma-quality-approx-robust}
The proofs are along the same lines as for Proposition \ref{lemma-quality-approx-nominal}.
\proof{Proof of Proposition \ref{lemma-quality-approx-robust}.}
We deal with each case separately.

\paragraph{Case 1.}
We make use the following facts:
\begin{itemize}
	\item{The functions $(u^{\Delta t}_i(\cdot))_{i \in \mathcal{V}}$ are non-decreasing,}
	\item{The functions $(u_i(\cdot))_{i \in \mathcal{V}}$ are non-decreasing,}
	\item{The functions $(u^{\Delta t}_i(\cdot))_{i \in \mathcal{V}}$ lower bound the functions $(u_i(\cdot))_{i \in \mathcal{V}}$.}
\end{itemize}
We follow the same recipe as in Proposition \ref{lemma-quality-approx-nominal}. We start by proving convergence for the discretizattion sequence $(\Delta t_p = \frac{1}{2^p})_{p \in \mathbb{N}}$. Then, we conclude the general study with the exact same argument as in Lemma \ref{aux-lemma-nonregularmesh-nominal}.

\begin{lemma}
\label{aux-lemma-regularmesh-robust-sota}
For the regular mesh $(\Delta t_p = \frac{1}{2^p})_{p \in \mathbb{N}}$, the sequence $(u_i^{\Delta t_p}(t))_{p \in \mathbb{N}}$ converges to $u_i(t)$ for almost every point $t$ in $[k^{r, \text{min}}_i \cdot \Delta t, T]$.
\proof{Proof}
	Just like in Lemma \ref{aux-lemma-regularmesh-nominal} we can prove that the sequence $(u_i^{\Delta t_p}(t))_{p \in \mathbb{N}}$ is non-decreasing for any $t$ and $i \in \mathcal{V}$. Hence, the functions $(u_i^{\Delta t_p}(\cdot))_{i \in \mathcal{V}}$ converge pointwise to some limits $(f_i(\cdot))_{i \in \mathcal{V}}$. Using the preliminary remarks, we get:
	$$	
		f_i(t) \leq u_i(t) \quad \forall t \in [k^{r, \text{min}}_i \cdot \Delta t, T], \forall i \in \mathcal{V}.
	$$
	Next, we establish that for any $t \in [k^{r, \text{min}}_i \cdot \Delta t, T]$ and for any $\epsilon >0$, $f_i(t) \geq u_i(t - \epsilon)$. This will enable us to squeeze $f_i(t)$ to finally derive $f_i(t) = u_i(t)$. We start with node $d$. Observe that, by construction of the approximation, $u_d^{\Delta t}(\cdot)$ converges pointwise to $f(\cdot)$ at every point of continuity of $f(\cdot)$. Furthermore, since $f_d(\cdot)$ and $u_d(\cdot)$ are non-decreasing, we have $f_d(t) \geq u_d(t - \epsilon)$ for all $t \in [k^{r, \text{min}}_d \cdot \Delta t, T]$ and for all $\epsilon >0$. Consider $\epsilon >0$ and a large enough $p$ such that $\epsilon > \frac{1}{2^p}$ which implies $\Delta t_p \cdot \lfloor \frac{t}{\Delta t_p} \rfloor \geq t - \epsilon$. We first show by induction on the level of the nodes in $\mathcal{T}^r$ that:
	$$
		f_i(t) \geq  u_i(t - \text{level}(i, \mathcal{T}^r) \cdot \epsilon) \quad \forall t \in [k^{r, \text{min}}_i \cdot \Delta t, \floor*{ \frac{T^r_f}{ \Delta t} } \cdot \Delta t], \forall i \in \mathcal{V}.
	$$
	The base case follows from the discussion above. Assume that the induction property holds for all nodes of level less than $l$ and consider a node $i \in \mathcal{V}$ of level $l +1$. We have, for $t \in [k^{r, \text{min}}_i \cdot \Delta t, \floor*{ \frac{T^r_f}{ \Delta t} } \cdot \Delta t]$:
	\begin{align*}
		u_i^{\Delta t_p}(t) 
			& \geq u_i^{\Delta t_p}( \floor*{\frac{t}{\Delta t_p}} \cdot \Delta t_p ) \\
			& \geq  \max\limits_{j \in \mathcal{T}^r(i)} \inf_{p_{ij} \in \mathcal{P}_{ij}}  \mathbb{E}_{X \sim p_{ij} }[ u_{j}^{\Delta t_p}( \floor*{\frac{t}{\Delta t_p}} \cdot \Delta t_p - X) ] \\
			& \geq \max\limits_{j \in \mathcal{T}^r(i)} \inf_{p_{ij} \in \mathcal{P}_{ij}}  \mathbb{E}_{X \sim p_{ij} }[ u_{j}^{\Delta t_p}(t - \epsilon - X) ].
	\end{align*}
	Take $j \in \mathcal{T}^r(i)$. Since $u_{j}^{\Delta t_p}(\cdot)$ is continuous and $\mathcal{P}_{ij}$ is compact, the infimum in the previous inequality is attained for some $p_{ij}^p \in \mathcal{P}_{ij}$ which gives:
	$$
		u_i^{\Delta t_p}(t) \geq \mathbb{E}_{X \sim p_{ij}^p} [ u_{j}^{\Delta t_p}(t - \epsilon - X)].
	$$
	As the sequence $(u_{j}^{\Delta t_p}(t - \epsilon - \omega))_p$ is non-decreasing for any $\omega$, we have, for any $m \leq p$:
	$$
		u_i^{\Delta t_p}(t) \geq  \mathbb{E}_{X \sim p_{ij}^p} [ u_{j}^{\Delta t_m}(t - \epsilon - X)].
	$$
	Because $\mathcal{P}_{ij}$ is a compact set for the weak topology, there exists a subsequence of $(p_{ij}^p)_p$ converging weakly in $\mathcal{P}_{ij}$ to some probability measure $p_{ij}$. Without loss of generality, we continue to refer to this subsequence as $(p_{ij}^p)_p$. We can now take the limit $p \rightarrow \infty$ in the previous inequality which yields:
	$$
		f_i(t) \geq \mathbb{E}_{X \sim p_{ij}} [ u_{j}^{\Delta t_m}(t - \epsilon - X)],
	$$
	since $u_{j}^{\Delta t_m}(\cdot)$ is continuous. To take the limit $m \rightarrow \infty$, note that:
	$$
		u_{j}^{\Delta t_m}(t - \epsilon - X) \geq u_{j}^{\Delta t_1}(t - \epsilon - \deltasup),
	$$ 
	while $(u_{j}^{\Delta t_m}(t - \epsilon - X))_{m \in \mathbb{N}}$ is non-decreasing and converges almost surely to $f_j(t - \epsilon - X)$ as $m \rightarrow \infty$. Therefore, we can apply the monotone convergence theorem and derive:
	$$
		f_i(t) \geq \mathbb{E}_{X \sim p_{ij}} [ f_{j}(t - \epsilon -  X)],
	$$
	which further implies 
	$$
		f_i(t) \geq \inf\limits_{p_{ij} \in \mathcal{P}_{ij}} \mathbb{E}_{X \sim p_{ij}} [ f_{j}(t - \epsilon - X)].
	$$
	As the last inequality holds for any $j \in \mathcal{T}^r(i)$, we finally obtain:
	$$
		f_i(t) \geq \max_{j \in \mathcal{T}^r(i)} \inf\limits_{p_{ij} \in \mathcal{P}_{ij}} \mathbb{E}_{X \sim p_{ij}} [ f_{j}(t - \epsilon - X)] \quad \forall t \in [k^{r, \text{min}}_i \cdot \Delta t, \floor*{ \frac{T^r_f}{ \Delta t} } \cdot \Delta t], \forall i \in \mathcal{V}.
	$$
	 Using the induction property along with Theorem \ref{lemma-LET-is-opt-robust}, we get: 
	\begin{align*}
		f_i(t) 
		& \geq \max_{j \in \mathcal{T}^r(i)} \inf\limits_{p_{ij} \in \mathcal{P}_{ij}} \mathbb{E}_{X \sim p_{ij}} [ u_j(t - \text{level}(i, \mathcal{T}^r) \cdot \epsilon - X) ] \\
		& \geq u_i(t - \text{level}(i, \mathcal{T}^r) \cdot \epsilon),
	\end{align*}
	for all $t \in [k^{r, \text{min}}_i \cdot \Delta t, \floor*{ \frac{T^r_f}{ \Delta t} } \cdot \Delta t]$, which concludes the induction. We can now prove by induction on $m$, along the same lines as above, that:
	$$
		f_i(t) \geq u_i(t - (|\mathcal{V}| + m) \cdot \epsilon) \quad \forall t \in [k^{r, \text{min}}_i \cdot \Delta t, \floor*{ \frac{T^r_f}{\Delta t} } \cdot \Delta t + m \cdot \deltainf], \forall i \in \mathcal{V},
	$$	 
	for all $m \in \mathbb{N}$. This last result can be reformulated as:
	$$
		f_i(t) \geq u_i(t - \epsilon) \quad \forall \epsilon >0, \forall t \in [k^{r, \text{min}}_i \cdot \Delta t, T], \forall i \in \mathcal{V}.
	$$
	Combining this lower bound with the upper bound previously derived, we get:
	$$
		u_i(t) \geq f_i(t) \geq u_i(t^-) \quad \forall t \in [k^{r, \text{min}}_i \cdot \Delta t, T], \forall i \in \mathcal{V},
	$$
	where $u_i(t^{-})$ refers to the left one-sided limit of $u_i(\cdot)$ at $t$. Since, $u_i(\cdot)$ is non-decreasing, it has countably many discontinuity points and the last inequality shows that $f_i(\cdot) = u_i(\cdot)$ almost everywhere on $[k^{r, \text{min}}_i \cdot \Delta t, T]$.
\Halmos
\endproof
\end{lemma}

\paragraph{Case 2.}
The first step consists in proving that the functions $(u_i(\cdot))_{i \in \mathcal{V}}$ are continuous on $(-\infty, T]$. By induction on $l$, we start by proving that $u_i(\cdot)$ is continuous on $(-\infty, T^r_f]$ for all nodes $i$ of level $l$ in $\mathcal{T}^r$. The base case follows from the continuity of $f(\cdot)$. Assuming the property holds for some $l \geq 1$, we consider a node $i$ of level $l+1$ in $\mathcal{T}^r$, $t \leq T^r_f$ and a sequence $t_n \rightarrow_{n \rightarrow \infty} t$. Using Theorem \ref{lemma-LET-is-opt-robust}, we have:
$$
	|u_i(t) - u_i(t_n)|
		\leq \max_{j \in \mathcal{T}^r(i)} \sup\limits_{p_{ij} \in \mathcal{P}_{ij}} \int_0^\infty p_{ij}(\omega) \cdot | u_j(t - \omega) - u_j(t_n - \omega) | \mathrm{d}\omega. 
$$
For any $j \in \mathcal{T}^r(i)$, we can use the uniform continuity of $u_j(\cdot)$ on $[t - 2 \cdot \deltasup, t]$ to prove that this last term converges to $0$ as $n \rightarrow \infty$. We conclude that all the functions $(u_i(\cdot))_{i \in \mathcal{V}}$ are continuous on $(-\infty, T^r_f]$. By induction on $m$, we can then show that the functions $(u_i(\cdot))_{i \in \mathcal{V}}$ are continuous on $(-\infty, T^r_f + m \cdot \deltainf]$, to finally conclude that they are continuous on $(-\infty, T]$. We are now able to prove uniform convergence. Since $[T^r_f - |\mathcal{V}| \cdot \deltasup, T]$ is a compact set, the functions $(u_i(\cdot))_{i \in \mathcal{V}}$ are also uniformly continuous on this set. Take $\epsilon >0$, there exists $\alpha > 0$ such that:
$$
	\forall i \in \mathcal{V}, | u_i(\omega) - u_i(\omega') | \leq \epsilon, \; \forall (\omega, \omega') \in [T^r_f - |\mathcal{V}| \cdot \deltasup, T]^2 \; \text{with} \; | \omega - \omega'| \leq \alpha.
$$
Building on this, we can show, by induction on the level of the nodes in $\mathcal{T}^r$, that:
$$
	\sup_{\omega \in [k^{r, \text{min}}_i \cdot \Delta t,  \floor*{ \frac{T^r_f}{\Delta t} } \cdot \Delta t ]} | u_i(\omega) -  u^{\Delta t}_i(\omega) | \leq 2 \cdot \text{level}(i, \mathcal{T}^r) \cdot \epsilon, \quad \forall i \in \mathcal{V}.
$$
This follows from the sequence of inequalities:
\begin{align*}
	\sup_{\omega \in [k^{r, \text{min}}_i \cdot \Delta t, \floor*{ \frac{T^r_f}{\Delta t} } \cdot \Delta t ]} | u^{\Delta t}_i(\omega) - u_i(\omega)| 
	& \leq \sup_{k \in \{ k^{r, \text{min}}_i, \cdots, \floor*{ \frac{T^r_f}{\Delta t} }  \} } | u^{\Delta t}_i( k \cdot \Delta t ) - u_i(k \cdot \Delta t)| \\
	& + \sup_{\omega \in [k^{r, \text{min}}_i \cdot \Delta t, \floor*{ \frac{T^r_f}{\Delta t} } \cdot \Delta t ]} | u_i(\omega) - u_i( \floor*{ \frac{\omega}{\Delta t} } \cdot \Delta t)| \\
	& + \sup_{\omega \in [k^{r, \text{min}}_i \cdot \Delta t, \floor*{ \frac{T^r_f}{\Delta t} } \cdot \Delta t ]} | u_i(\omega) - u_i( \ceil*{ \frac{\omega}{\Delta t} } \cdot \Delta t)| \\
	& \leq  \sup_{k \in \{ k^{r, \text{min}}_i, \cdots, \floor*{ \frac{T^r_f}{\Delta t} } \} }\max_{j \in \mathcal{T}(i)} \sup_{p_{ij} \in \mathcal{P}_{ij}} \\
	& \{ \int_0^\infty p_{ij}(\omega) \cdot | u^{\Delta t}_j( k \cdot \Delta t - \omega) - u_j( k \cdot \Delta t - \omega)  | \mathrm{d}\omega \} \\
	& + 2 \cdot \epsilon \\
	& \leq  2 \cdot (\text{level}(i, \mathcal{T}^r)-1) \cdot \epsilon + 2 \cdot \epsilon \\
	& \leq  2 \cdot \text{level}(i, \mathcal{T}^r) \cdot \epsilon.
\end{align*}
We conclude that:
$$
	\sup_{\omega \in [k^{r, \text{min}}_i \cdot \Delta t, \floor*{ \frac{T^r_f}{\Delta t} } \cdot \Delta t ] } | u^{\Delta t}_i(\omega) - u_i(\omega) | \leq 2 \cdot |\mathcal{V}| \cdot \epsilon, \quad \forall i \in \mathcal{V}.	
$$
Along the same lines, we can show by induction on $m$ that:
$$
	\sup_{\omega \in [k^{r, \text{min}}_i \cdot \Delta t, \floor*{ \frac{T^r_f}{\Delta t} } \cdot \Delta t + m \cdot \deltainf]} | u^{\Delta t}_i(\omega) - u_i(\omega) | \leq 2 \cdot (|\mathcal{V}| + m) \cdot \epsilon, \quad \forall i \in \mathcal{V}. 
$$
This implies:
$$
	\sup_{\omega \in [k^{r, \text{min}}_i \cdot \Delta t, T]} | u^{\Delta t}_i(\omega) - u_i(\omega) | \leq 2 \cdot (|\mathcal{V}| + \ceil*{ \frac{T - T^r_f}{\deltainf} }) \cdot \epsilon, \quad \forall i \in \mathcal{V},
$$
assuming $\Delta t \leq \deltainf$. In particular, this shows uniform convergence. To conclude the proof of Case 2, we show that $\pi^{\Delta t}$ is a $o(1)$-approximate optimal solution to \eqref{eq-robust-problem} as $\Delta t \rightarrow 0$. We denote by $u^{\pi^{\Delta t}}_i(t)$ the worst-case expected risk function when following policy $\pi^{\Delta t}$ starting at $i$ with remaining budget $t$. We can show, by induction on the level of the nodes in $\mathcal{T}^r$, that :
$$
	u^{\pi^{\Delta t}}_i(t) \geq u_{ i }(t) - 12 \cdot \text{level}(i, \mathcal{T}^r) \cdot (|\mathcal{V}| + \ceil*{ \frac{T - T^r_f}{\deltainf} } ) \cdot \epsilon, \quad \forall t \in [k^{r, \text{min}}_i \cdot \Delta t, T^r_f], \forall i \in \mathcal{V}.
$$
To do so, we can use the same sequence of inequalities as in Case 2 of Proposition \ref{lemma-quality-approx-nominal}, except that we also take the infimum over $p_{i\pi^{\Delta t}(i, t)} \in \mathcal{P}_{i\pi^{\Delta t}(i, t)}$. We derive:
$$
	u^{\pi^{\Delta t}}_i(t) \geq u_{ i }(t) - 12 \cdot |\mathcal{V}| \cdot (|\mathcal{V}| + \ceil*{ \frac{T - T^r_f}{\deltainf} } ) \cdot \epsilon, \; \forall t \in [k^{r, \text{min}}_i \cdot \Delta t, T^r_f], \forall i \in \mathcal{V}.
$$
Along the same lines, we can show by induction on $m$ that:
$$
	u^{\pi^{\Delta t}}_i(t) \geq u_{ i }(t) - 12 \cdot ( |\mathcal{V}| + m ) \cdot (|\mathcal{V}| + \ceil*{ \frac{T - T^r_f}{\deltainf} }) \cdot \epsilon, \; \forall t \in [k^{r, \text{min}}_i \cdot \Delta t, T^r_f + m \cdot \deltainf], \forall i \in \mathcal{V}.
$$
We conclude that $u^{\pi^{\Delta t}}_s(T)  \geq u_s(T) - 12 \cdot (|\mathcal{V}| + \ceil*{ \frac{T - T^r_f}{\deltainf} })^2 \cdot \epsilon$, which establishes the claim.

\paragraph{Case 3.}
This case is essentially identical to Case 2 substituting uniform continuity for Lipschitz continuity and the proof mirrors the proof of Case 3 of Proposition \ref{lemma-quality-approx-nominal}.

\Halmos
\endproof

\subsection{Proof of Lemma \ref{lemma-tight-relaxation-if-convex}}
\label{section-proof-lemma-tight-relaxation-if-convex}

\proof{Proof of Lemma \ref{lemma-tight-relaxation-if-convex}.}
For a real value $x$, $\delta_x$ refers to the Dirac distribution at $x$. We denote by $(m_{ij})_{(i, j) \in \mathcal{A}}$ the worst-case expected costs, i.e.:
$$
	m_{ij} = \sup_{p_{ij} \in \mathcal{P}_{ij}} \mathbb{E}_{X \sim p_{ij}}[X] \quad \forall (i, j) \in \mathcal{A}.
$$
We define $M_i$ as the length of a shortest path from $i$ to $d$ in $\mathcal{G}$ when the arc costs are taken as $(m_{ij})_{(i, j) \in \mathcal{A}}$. Observe that $f(\cdot)$ is increasing since $f(\cdot)$ is convex and $f' \rightarrow_{- \infty} a >0$. We use Proposition \ref{lemma-dp-optimal-robust} and consider a solution $(\pi^*_{f, \mathcal{P}}, (u_i(\cdot))_{i \in \mathcal{V}})$ to the dynamic programming equation \eqref{eq-dp-robust}. We first prove by induction on the level of the nodes in $\mathcal{T}^r$ that:
$$
	u_i(t) = f(t - M_i) \quad \forall t \in [T^r_f - ( |\mathcal{V}| - \text{level}(i, \mathcal{T}^r) + 1) \cdot \deltasup, T^r_f],
$$ 
for all nodes $i \in \mathcal{V}$. The base case is trivial. Assume that the property holds for all nodes of level less than $l$ and consider a node $i \in \mathcal{V}$ of level $l + 1$. Take $t \in [T^r_f - ( |\mathcal{V}| - \text{level}(i, \mathcal{T}^r) + 1) \cdot \deltasup, T^r_f]$. Using Theorem  \ref{lemma-LET-is-opt-robust}, we have:
\begin{align*}
	u_i(t) 
		& = \max_{j \in \mathcal{T}^r(i)} \inf_{p_{ij} \in \mathcal{P}_{ij}} \mathbb{E}_{X \sim p_{ij} }[ u_{j}(t - X) ] \\
		& = \max_{j \in \mathcal{T}^r(i)} \inf_{p_{ij} \in \mathcal{P}_{ij}} \mathbb{E}_{X \sim p_{ij} }[ f(t - X - M_j) ]  \\
		& \geq \max_{j \in \mathcal{T}^r(i)} \inf_{p_{ij} \in \mathcal{P}_{ij}} f(t -  \mathbb{E}_{X \sim p_{ij} }[X] - M_j) \\
		& \geq \max_{j \in \mathcal{T}^r(i)} f(t -  m_{ij} - M_j ) \\
		& \geq f(t -  \min_{j \in \mathcal{T}^r(i)} m_{ij} - M_j ) \\
		& \geq f(t -  M_i ) \\
		& \geq \max_{j \in \mathcal{T}^r(i)}  \mathbb{E}_{X \sim \delta_{ m_{ij} } }[ f(t - X - M_j) ],
\end{align*}
where the first inequality results from the convexity of $f(\cdot)$ and the third inequality is a consequence of the monotonicity of $f(\cdot)$. Since $\delta_{m_{ij}} \in \mathcal{P}_{ij}$, the last inequality shows that $u_i(t) = f(t -  M_i )$. This concludes the induction. We move on to prove by induction on $m$ that:
$$
	u_i(t) = f(t - M_i) \quad \forall t \in [T^r_f - ( |\mathcal{V}| - \text{level}(i, \mathcal{T}^r) + 1) \cdot \deltasup, T^r_f + m \cdot \deltainf], \forall i \in \mathcal{V}.
$$
Assume that the inductive property holds for some $m \in \mathbb{N}$. Consider $i \neq d$. We have, for $t \in [T^r_f + m \cdot \deltainf, T^r_f + (m+1) \cdot \deltainf]$:
\begin{align*}
	u_i(t) 
		& = \max_{j \in \mathcal{V}(i)} \inf_{p_{ij} \in \mathcal{P}_{ij}} \mathbb{E}_{X \sim p_{ij} }[ u_{j}(t - X) ] \\
		& = \max_{j \in \mathcal{V}(i)} \inf_{p_{ij} \in \mathcal{P}_{ij}} \mathbb{E}_{X \sim p_{ij} }[ f(t - X - M_j) ]  \\
		& \geq \max_{j \in \mathcal{V}(i)} \inf_{p_{ij} \in \mathcal{P}_{ij}} f(t -  \mathbb{E}_{X \sim p_{ij} }[X] - M_j) \\
		& \geq \max_{j \in \mathcal{V}(i)} f(t -  m_{ij} - M_j ) \\
		& \geq f(t -  \min_{j \in \mathcal{V}(i)} m_{ij} - M_j ) \\
		& \geq f(t -  M_i ) \\
		& \geq \max_{j \in \mathcal{V}(i)}  \mathbb{E}_{X \sim \delta_{ m_{ij} } }[ f(t - X - M_j) ],
\end{align*}
using the convexity and the monotonicity of $f(\cdot)$. The last inequality shows that $u_i(t) = f(t -  M_i )$. This concludes the induction. Hence:
$$
	u_i(t) = f(t - M_i) \quad \forall t \in [T^r_f - ( |\mathcal{V}| - \text{level}(i, \mathcal{T}^r) + 1) \cdot \deltasup, T], \forall i \in \mathcal{V}.	
$$
Using Theorem \ref{lemma-LET-is-opt-robust} and plugging this last expression back into \eqref{eq-dp-robust}, we conclude that:
$$
	\pi^*_{f, \mathcal{P}}(i, t) \in \mathcal{T}^r(i) \quad \forall t \leq T, \forall i \neq d.
$$
Moreover, for any arc $(i, j) \in \mathcal{A}$, we have proved that the infimum appearing in \eqref{eq-dp-robust} is attained at $\delta_{m_{ij}}$ irrespective of the remaining budget $t$. This shows:
\begin{align*}
		\sup\limits_{\pi \in \Pi} \; \inf\limits_{\forall \tau, \forall (i, j) \in \mathcal{A}, \; p^\tau_{ij} \in \mathcal{P}_{ij}} \; \mathbb{E}_{ \mathbf{p^\tau} }[f(T - X_\pi)]
		& \geq \sup\limits_{\pi \in \Pi} \mathbb{E}_{ \mathbf{\delta} }[f(T - X_\pi)] \\
		& \geq	\sup\limits_{\pi \in \Pi} \; \inf\limits_{\forall (i, j) \in \mathcal{A}, \; p_{ij} \in \mathcal{P}_{ij}} \; \mathbb{E}_{ \mathbf{p}  } [f(T - X_\pi)],
\end{align*}
where the notation $\mathbf{\delta}$ refers to the fact the costs $(c_{ij})_{(i, j) \in \mathcal{A}}$ are independent and distributed according to $(\delta_{ m_{ij} })_{(i, j) \in \mathcal{A}}$. 
Since \eqref{eq-robust-problem} is  a relaxation of \eqref{eq-ideal-robust-problem}, we get:
$$
	\sup\limits_{\pi \in \Pi} \; \inf\limits_{\forall (i, j) \in \mathcal{A}, \; p_{ij} \in \mathcal{P}_{ij}} \; \mathbb{E}_{ \mathbf{p} }[f(T - X_\pi)] = \sup\limits_{\pi \in \Pi} \; \inf\limits_{\forall \tau, \forall (i, j) \in \mathcal{A}, \; p^\tau_{ij} \in \mathcal{P}_{ij}} \; \mathbb{E}_{ \mathbf{p^\tau}  }[f(T - X_\pi)],
$$
and an optimal strategy for both problems is to always follow the shortest-path tree $\mathcal{T}^r$.
\Halmos
\endproof

\subsection{Proof of Lemma \ref{lemma-tight-relaxation-if-higher-order-convex}}
\label{section-proof-lemma-tight-relaxation-if-higher-order-convex}

\proof{Proof of Lemma \ref{lemma-tight-relaxation-if-higher-order-convex}.}
Without loss of generality, we assume that $f^{(K+1)} > 0$. The proof is almost identical in the converse situation. We use Proposition \ref{lemma-dp-optimal-robust} and consider a solution $(\pi^*_{f, \mathcal{P}}, (u_i(\cdot))_{i \in \mathcal{V}})$ to the dynamic programming equation \eqref{eq-dp-robust}. We first prove by induction on the level of the nodes $i$ in $\mathcal{G}$ that $u_i^{(K+1)} > 0$ and that, for $j$ the immediate successor of $i$ in $\mathcal{G}$, there exists $p_{ij} \in \mathcal{P}_{ij}$ such that: 
\begin{equation}
	\label{eq-proof-tightness-conclusion}
	u_i(t) = \mathbb{E}_{X \sim p_{ij}}[ u_j( t - X) ] \quad \forall t \leq T \; \text{ if } i \neq d.
\end{equation}
Assume that the property holds for all nodes of level less than $l$ and consider a node $i \in \mathcal{V}$ of level $l + 1$. Let $j$ be the immediate successor of $i$ in $\mathcal{G}$. As $\mathcal{P}_{ij}$ is not empty, Lemma 3.1 from \cite{Shapiro00onduality} shows that $\mathcal{P}_{ij}$ contains a discrete distribution whose support is a subset of $\{\delta_0, \cdots, \delta_{K+2} \}$ with $\delta_0 = \deltainf_{ij} < \delta_1 < \cdots < \delta_{K+2} = \deltasup_{ij}$. For any $n \in \mathbb{N}$, we define the ambiguity set:
$$
	\mathcal{P}^n_{ij} = \{ p \in \mathcal{P}_{ij} \; | \; \supp(p) \subset \{ \delta_0, \delta_0 + \frac{ \delta_1  - \delta_0}{n}, \delta_0 + 2 \cdot \frac{ \delta_1  - \delta_0}{n}, \cdots, \delta_1, \delta_1 + \frac{ \delta_2  - \delta_1}{n}, \cdots,  \delta_{K+1} \} \},
$$
which can be interpreted as a discretization of $\mathcal{P}_{ij}$. Observe that, by design, $\mathcal{P}^n_{ij}$ is not empty. Additionally, we define the sequence of functions $(f^n_i(\cdot))_{n \in \mathbb{N}}$ by:
\begin{equation}
	\label{eq-proof-tightness-sequence}
	f^n_i(t) = \inf_{p \in \mathcal{P}^n_{ij}} \mathbb{E}_{X \sim p }[ u_{j}(t - X) ] \quad \forall t \leq T.
\end{equation}
Since $u^{(K+1)}_j > 0$, \cite{prekopa1990discrete} shows that there exists $p^n_{ij} \in \mathcal{P}^n_{ij}$ such that:
$$
	f^n_i(t) = \mathbb{E}_{X \sim p^n_{ij} }[ u_{j}(t - X) ] \quad \forall t \leq T.
$$
Because $\mathcal{P}_{ij}$ is compact with respect to the weak topology and since $\mathcal{P}^n_{ij} \subset \mathcal{P}_{ij}$, we can take a subsequence of $(p^n_{ij})_{n \in \mathbb{N}}$ such that $p^n_{ij} \rightarrow p_{ij} \in \mathcal{P}_{ij}$ as $n \rightarrow \infty$ for the weak topology. Without loss of generality, we continue to denote this sequence $(p^n_{ij})_{n \in \mathbb{N}}$. Since $u_j(\cdot)$ is continuous, we derive that the sequence of functions $(f^n_i(\cdot))_{n \in \mathbb{N}}$ converges simply to a function $f_i(\cdot)$ which satisfies:
\begin{equation}
	\label{eq-proof-tightness-equality}
		f_i(t) = \mathbb{E}_{ X \sim p_{ij} }[ u_{j}(t - X) ] \quad \forall t \leq T.
\end{equation}
We now move on to show that $f_i(t) = u_i(t)$ for all $t \leq T$. This will conclude the induction because we can take the $(K+1)$th derivative in \eqref{eq-proof-tightness-equality} since $p_{ij}$ has compact support. Take $t \leq T$ and $\epsilon >0$. The function $u_j(\cdot)$ is continuous on $[t-\deltasup, t-\deltainf]$ hence, by uniform continuity, there exists $\alpha > 0$ such that:
$$
	|u_j(t - \omega) - u_j(t - \omega') | \leq \epsilon
$$
as soon as $|\omega - \omega'| \leq \alpha$ and $(\omega, \omega') \in [\deltainf_{ij}, \deltasup_{ij}]^2$. Consider $n > \frac{\deltasup_{ij} - \deltainf_{ij}}{\alpha}$. Using conic duality, Corollary 3.1 of \cite{Shapiro00onduality} shows that $u_i(t)$ is the optimal value of the infinite linear program:
\begin{equation}
	\label{eq-proof-dual-exact}
	\begin{aligned}
	& \sup_{(a_1, \cdots, a_K, b) \in \mathbb{R}^{K+1}} 
	& & \sum_{k=1}^K a_k \cdot m^k_{ij} + b \\
	& \text{subject to}
	& & \sum_{k=1}^K a_k \cdot \omega^k + b \leq u_j(t - \omega) \quad \forall \omega \in [\deltainf_{ij}, \deltasup_{ij}].
	\end{aligned}
\end{equation}	
Using strong linear programming duality, we also have that $f^n_i(t)$ is the optimal value of the finite linear program:
\begin{equation}
	\label{eq-proof-dual-approx}
	\begin{aligned}
	& \sup_{(a_1, \cdots, a_K, b) \in \mathbb{R}^{K+1}} 
	& & \sum_{k=1}^K a_k \cdot m^k_{ij} + b \\
	& \text{subject to}
	& & \sum_{k=1}^K a_k \cdot \omega^k + b \leq u_j(t - \omega) \quad \forall \omega \in \{ \delta_0, \delta_0 + \frac{ \delta_1  - \delta_0}{n}, \delta_0 + 2 \cdot \frac{ \delta_1  - \delta_0}{n}, \cdots, \delta_{K+1} \}.
	\end{aligned}
\end{equation}	
Take $(a^n_1, \cdots, a^n_K, b^n)$ an optimal basic feasible solution to \eqref{eq-proof-dual-approx}. By a standard linear programming argument:
$$
	\max(\max_{k=1, \cdots, K} |a^n_k|, |b^n|) \leq U,
$$
where $U = ((K+1) \cdot \max(1, u_j(t - \deltainf), (\deltasup_{ij})^K ))^{(K+1)}$ does not depend on $n$. Let us use the shorthand:
$$
	V = U \cdot (\deltasup_{ij} - \deltainf_{ij}) \cdot \sum_{k=1}^K k \cdot (\deltasup_{ij})^{(k-1)},
$$
and define $b = b^n - \frac{V}{n} - \epsilon$. We show that $(a^n_1, \cdots, a^n_K, b)$ is feasible for \eqref{eq-proof-dual-exact}. For any $w \in [\deltainf_{ij}, \deltasup_{ij}]$, take $w' \in \{ \delta_0, \delta_0 + \frac{ \delta_1  - \delta_0}{n}, \delta_0 + 2 \cdot \frac{ \delta_1  - \delta_0}{n}, \cdots, \delta_{K+1} \}$ such that $|w - w'| \leq \frac{\deltasup_{ij} - \deltainf_{ij}}{n}$. We have:
\begin{align*}
	\sum_{k=1}^K a^n_k \cdot \omega^k + b 
		& = \sum_{k=1}^K a^n_k \cdot (\omega')^k + b^n + \sum_{k=1}^K a^n_k \cdot ( \omega^k - (\omega')^k) - \frac{V}{n} - \epsilon\\
		& \leq u_j(t- \omega') + \sum_{k=1}^K | a^n_k | \cdot | \omega^k - (\omega')^k | - \frac{V}{n} - \epsilon \\
		& \leq u_j(t- \omega) + \sum_{k=1}^K U \cdot k \cdot (\deltasup_{ij})^{(k-1)} \cdot | \omega - \omega'| - \frac{V}{n} \\
		& \leq u_j(t- \omega),
\end{align*}
where we use the fact that $(a^n_1, \cdots, a^n_K, b^n)$ is feasible for \eqref{eq-proof-dual-approx} in the first inequality, the uniform continuity of $u_j(\cdot)$ in the second and the definition of $V$ in the last one. We derive:
$$
	f_i(t) - \frac{V}{n} - \epsilon \leq u_i(t) \leq f_i(t).
$$
Taking $n \rightarrow \infty$ and $\epsilon \rightarrow 0$, we obtain $f_i(t) = u_i(t)$. This concludes the induction.
\\
\indent As a consequence of \eqref{eq-proof-tightness-conclusion}, the infimum in \eqref{eq-dp-robust} is always attain for $p_{ij}$, irrespective of the remaining budget $t$, so we can conclude that \eqref{eq-ideal-robust-problem} and \eqref{eq-robust-problem} are equivalent.
\Halmos
\endproof

\subsection{Proof of Lemma \ref{lemma-bound-gap-robust-relaxation}}
\label{section-proof-lemma-bound-gap-robust-relaxation}
This result is a direct consequence of the following observations:
\begin{itemize}
	\item{when the risk function is $f(t) = t$, following the shortest path with respect to $(\max_{p \in \mathcal{P}_{ij}} \mathbb{E}_{X \sim p}[ X ] )_{(i,j) \in \mathcal{A}}$ is an optimal strategy for \eqref{eq-robust-problem},}
	\item{when the risk function is $f(t) = \exp(t)$, following the shortest path with respect to $(\max_{p \in \mathcal{P}_{ij}} - \log( \mathbb{E}_{X \sim p}[ \exp(- X ) ] ))_{(i,j) \in \mathcal{A}}$ is an optimal strategy for \eqref{eq-robust-problem},}
	\item{when the risk function is $f(t) = - \exp( - t)$, following the shortest path with respect to $(\max_{p \in \mathcal{P}_{ij}} \log( \mathbb{E}_{X \sim p}[ \exp( X ) ] ))_{(i,j) \in \mathcal{A}}$  is an optimal strategy for \eqref{eq-robust-problem}.}
\end{itemize}
As a consequence, for any of these risk functions, \eqref{eq-ideal-robust-problem} and \eqref{eq-robust-problem} are equivalent. Define $g(\cdot)$ as any of these risk functions. Assuming that $\gamma \cdot g(t) + \beta \geq f(t) \geq a \cdot g(t) + b, \forall t \leq T$, we get:
\begin{align*}
	\sup\limits_{\pi \in \Pi} \; \inf\limits_{\forall (i, j) \in \mathcal{A}, \; p_{ij} \in \mathcal{P}_{ij}} \; \mathbb{E}_{ \mathbf{p} }[f(T - X_\pi)] 
		& \leq \sup\limits_{\pi \in \Pi} \; \inf\limits_{\forall (i, j) \in \mathcal{A}, \; p_{ij} \in \mathcal{P}_{ij}} \; \mathbb{E}_{ \mathbf{p} }[  \gamma \cdot g(T - X_\pi) + \beta ] \\
		& \leq \beta + \gamma \cdot \sup\limits_{\pi \in \Pi} \; \inf\limits_{\forall (i, j) \in \mathcal{A}, \; p_{ij} \in \mathcal{P}_{ij}} \; \mathbb{E}_{ \mathbf{p} }[ g(T - X_\pi)] \\
		& \leq \beta + \gamma \cdot \sup\limits_{\pi \in \Pi} \; \inf\limits_{\forall \tau, \forall (i, j) \in \mathcal{A}, \; p^\tau_{ij} \in \mathcal{P}_{ij}} \; \mathbb{E}_{ \mathbf{p^\tau} }[g(T - X_\pi)] \\
		& \leq \beta - \frac{\gamma}{a} \cdot b + \frac{\gamma}{a} \cdot \sup\limits_{\pi \in \Pi} \; \inf\limits_{\forall \tau, \forall (i, j) \in \mathcal{A}, \; p^\tau_{ij} \in \mathcal{P}_{ij}} \; \mathbb{E}_{ \mathbf{p^\tau} }[a \cdot g(T - X_\pi) + b] \\
		& \leq \beta - \frac{\gamma}{a} \cdot b + \frac{\gamma}{a} \cdot \sup\limits_{\pi \in \Pi} \; \inf\limits_{\forall \tau, \forall (i, j) \in \mathcal{A}, \; p^\tau_{ij} \in \mathcal{P}_{ij}} \; \mathbb{E}_{ \mathbf{p^\tau} }[f(T - X_\pi)].
\end{align*}
This last inequality along with:
$$
	\sup\limits_{\pi \in \Pi} \; \inf\limits_{\forall (i, j) \in \mathcal{A}, \; p_{ij} \in \mathcal{P}_{ij}} \; \mathbb{E}_{ \mathbf{p} }[f(T - X_\pi)] \geq \sup\limits_{\pi \in \Pi} \; \inf\limits_{\forall \tau, \forall (i, j) \in \mathcal{A}, \; p^\tau_{ij} \in \mathcal{P}_{ij}} \; \mathbb{E}_{ \mathbf{p^\tau} }[f(T - X_\pi)]
$$
yields the claim with some basic algebra.

\subsection{Proof of Lemma \ref{lemma-convergence-more-moments}}
\label{section-proof-lemma-convergence-more-moments}

\proof{Proof of Lemma \ref{lemma-convergence-more-moments}.}
For any $k \in \mathbb{N}$, we define $(\pi^k, (u^k_i(\cdot))_{i \in \mathcal{V}})$ as a solution to the dynamic program \eqref{eq-dp-robust} when the ambiguity sets are taken as $(\mathcal{P}^k_{ij})_{(i, j) \in \mathcal{A}}$. Similarly, we define $(\pi^\infty, (u^\infty_i(\cdot))_{i \in \mathcal{V}})$ as a solution to the dynamic program \eqref{eq-dp-robust} when the ambiguity sets are taken as $(\cap_{ k \in \mathbb{N} } \mathcal{P}^k_{ij})_{(i, j) \in \mathcal{A}}$. Along the sames lines as what is done in the proof of Proposition \ref{lemma-quality-approx-robust}, we can show that the functions $(u^k_i(\cdot))_{k \in \mathbb{N}}$ and $u^\infty_i(\cdot)$ are continuous for any $i \in \mathcal{V}$. Because the ambiguity sets are nested, observe that the sequence $(u^k_i(t))_{k \in \mathbb{N}}$ is non-decreasing for any $t \leq T$, hence it converges to a limit $f_i(t) \leq u^\infty_i(t)$. Moreover, $f_d(t) = f(t)$ for all $t \leq T$. Take $i \neq d$ and $t \leq T$. We have, for any $k \in \mathbb{N}$ and $m \leq k$:
\begin{align*}
	f_i(t) 
		& \geq  u^k_i(t) \\
		& \geq \max\limits_{j \in \mathcal{V}(i)} \inf\limits_{p \in \mathcal{P}^k_{ij}} \int_0^\infty p(\omega) \cdot u^k_j(t - \omega) \mathrm{d}\omega \\
				& \geq \max\limits_{j \in \mathcal{V}(i)} \int_0^\infty p^k_{ij}(\omega) \cdot u^k_j(t - \omega) \mathrm{d}\omega \\		
		& \geq \max\limits_{j \in \mathcal{V}(i)} \int_0^\infty p^k_{ij}(\omega) \cdot u^m_j(t - \omega) \mathrm{d}\omega,
\end{align*}
where $p^k_{ij} \in \mathcal{P}^k_{ij}$ achieves the minimum for any $j \in \mathcal{V}(i)$, which can be shown to exist since $\mathcal{P}^k_{ij}$ is compact and $u^k_j(\cdot)$ is continuous. Because $\mathcal{P}^k_{ij}$ is compact for the weak topology, we can take a subsequence of $(p^k_{ij})_{k \in \mathbb{N}}$ that converges to a distribution $p^\infty_{ij}$ in $\cap_{ k \in \mathbb{N} } \mathcal{P}^k_{ij}$. Without loss of generality we continue to refer to this sequence as $(p^k_{ij})_{k \in \mathbb{N}}$. Taking the limit $k \rightarrow \infty$ in the last inequality derived yields:
$$
	f_i(t) \geq \max\limits_{j \in \mathcal{V}(i)} \int_0^\infty p^\infty_{ij}(\omega) \cdot u^m_j(t - \omega) \mathrm{d}\omega.
$$
Observing that $u^m_j(t - \omega) \geq u^1_j(t - \omega)$, we can use the monotone convergence theorem for $m \rightarrow \infty$ and conclude that:
\begin{align*}
		f_i(t) 
			& \geq \max\limits_{j \in \mathcal{V}(i)} \int_0^\infty p^\infty_{ij}(\omega) \cdot f_j(t - \omega) \mathrm{d}\omega \\
			& \geq \max\limits_{j \in \mathcal{V}(i)} \inf_{ p \in \cap_{k \in \mathbb{N} } \mathcal{P}^k_{ij} } \int_0^\infty p(\omega) \cdot f_j(t - \omega) \mathrm{d}\omega.
\end{align*}
We use Theorem \ref{lemma-LET-is-opt-robust} for the ambiguity sets $(\cap_{ k \in \mathbb{N} } \mathcal{P}^k_{ij})_{(i, j) \in \mathcal{A}}$ and denote by $T^r_f$ (resp. $\mathcal{T}^r$) the time budget (resp. the tree) put forth in the statement of the theorem. Using the last sequence of inequalities derived, we can prove, by induction on the levels of the nodes in $\mathcal{T}^r$ that: 
$$
	 f_i(t) \geq u^\infty_i(t) \quad \forall t \in [T^r_f - ( |\mathcal{V}| - \text{level}(i, \mathcal{T}^r) + 1) \cdot \deltasup, T^r_f], \forall i \in \mathcal{V},
$$
and then by induction on $m \in \mathbb{N}$ that:
$$
	f_i(t) \geq u^\infty_i(t) \quad \forall t \in [T^r_f - ( |\mathcal{V}| - \text{level}(i, \mathcal{T}^r) + 1) \cdot \deltasup, T^r_f + m \cdot \deltainf], \forall i \in \mathcal{V}.
$$
We finally obtain $f_s(T) \geq u^\infty_s(T)$ which concludes the proof.

\Halmos
\endproof

\subsection{Proof of Lemma \ref{lemma-optimal-basis-inner-problem-only-mean}}
\label{section-proof-lemma-optimal-basis-inner-problem-only-mean}

\proof{Proof of Lemma \ref{lemma-optimal-basis-inner-problem-only-mean}.}
First observe that, along the sames lines as in general case, the constraint
$$
	z + ( x  -  y ) \cdot l \cdot \Delta t \leq u^{\Delta t}_j( (k - l) \cdot \Delta t)
$$
does not limit the feasible region if $(l \cdot \Delta t, u^{\Delta t}_j( (k - l) \cdot \Delta t) )$ is not an extreme point of the upper convex hull of $\{ ( l \cdot \Delta t, u^{\Delta t}_j( l \cdot \Delta t)), \; l = k - \ceil*{ \frac{ \deltasup_{ij} }{\Delta t} }, \cdots, k - \floor*{ \frac{ \deltainf_{ij} }{\Delta t} } \} \cup \{ (\deltasup_{ij}, u^{\Delta t}_j( k \cdot \Delta t - \deltasup_{ij})), (\deltainf_{ij}, u^{\Delta t}_j( k \cdot \Delta t - \deltainf_{ij}))  \}$. Hence, we can discard the constraints that do no satisfy this property from \eqref{eq-primal-inner-problem-only-mean}. We denote by $S$ the sorted projection of the set of extreme points onto the first coordinate. Observe that the feasible region is pointed as the polyhedron described by the inequality constraints does not contain any line, therefore there exists a basic optimal feasible solution for which at least three inequality constraints are binding. By definition of $S$, only two of the constraints 
$$
	z + ( x  -  y ) \cdot \omega \leq u^{\Delta t}_j(\omega) \quad \omega \in S
$$
can be binding which further implies that at least one of the constraints $x \geq 0$ and $y \geq 0$ must be binding. There are three types of feasible basis depending on whether these last two constraints are binding or if only one of them is. We show that, for each type, we can identify an optimal basis among the basis of the same type by binary search on the first coordinate of the extreme points. This will conclude the proof as it takes constant time to compare the objective function achieved by each of the three potentially optimal basis. Since, by definition of $S$, $u^{\Delta t}_j(\cdot)$ is convex on $S$, we can partition $S$ into $S_1$ and $S_2$ such that $u^{\Delta t}_j(\cdot)$ is non-increasing on $S_1$ and non-decreasing on $S_2$ with $\max(S_1) = \min(S_2)$.
\\
\indent If $x \geq 0$ and $y \geq 0$ are binding then $z$ is the only non-zero variable and the objective is to maximize $z$. Hence, the optimal basis of this type is given by $x = 0$, $y = 0$ and $z = \min\limits_{\omega \in S} u^{\Delta t}_j(\omega)$ which can be computed by binary search since $u^{\Delta t}_j(\cdot)$ is convex on $S$.
\\
\indent If only $x \geq 0$ is binding, then the line $\omega \rightarrow z - y \cdot \omega$ must be joining two consecutive points in $S_1$. Since the objective function is precisely the value taken by the line $\omega \rightarrow z - y \cdot \omega$ at $\beta^{ij}$, the optimal straight line joins two consecutive points in $S_1$, $\omega_1$ and $\omega_2$, that satisfy $\omega_1 \leq \beta^{ij} \leq \omega_2$ assuming $\max(S_1) \geq \beta^{ij}$. If $\max(S_1) < \beta^{ij}$, the feasible basis of this type are dominated by the optimal basis of the first type. Computing $\omega_1$ and $\omega_2$ or showing that they do not exist can be done with a single binary search on $S$.
\\
\indent The discussion is analogous if only $y \geq 0$ is binding instead. The line $\omega \rightarrow z + x \cdot \omega$ must be joining two consecutive points in $S_2$. Since the objective function is precisely the value taken by this line at $\alpha^{ij}$, the optimal straight line joins two consecutive points in $S_2$, $\omega_1$ and $\omega_2$, that satisfy $\omega_1 \leq \alpha^{ij} \leq \omega_2$ assuming $\alpha^{ij} \geq \min(S_2)$. If $\min(S_2) > \alpha^{ij}$, the feasible basis of this type are dominated by the optimal basis of the first type. Computing $\omega_1$ and $\omega_2$ or showing that they do not exist can be done with a single binary search on $S$.

\Halmos
\endproof

\end{APPENDICES}

%%%%%%%%%%%%%%%%%
\end{document}